\documentclass[reprint,
superscriptaddress,
amsmath,amssymb,
aps,
prapplied,
]{revtex4-2}

\usepackage{graphicx}
\usepackage{dcolumn}
\usepackage{bm}
\usepackage{tensor}
\usepackage[dvipsnames]{xcolor}
\usepackage{soul}
\usepackage{svg}
\usepackage{subcaption}
\usepackage{tabularray}
\usepackage{physics}
\usepackage[utf8]{inputenc} 
\usepackage[T1]{fontenc}
\usepackage{dsfont}
\usepackage[symbol]{footmisc}
\usepackage{makecell}
\usepackage{braket} 
\usepackage{siunitx}
\usepackage{color}
\usepackage{xr}
\usepackage{upgreek}
\usepackage{multirow} 

\usepackage[colorlinks]{hyperref}
\hypersetup{colorlinks=true, citecolor=blue, urlcolor=blue, linkcolor=blue}
\usepackage{ragged2e}

\begin{document}
\setstcolor{red}
\newcommand{\HY}[1]{{\color{blue}{[HY: #1]}}}
\newcommand{\XW}[1]{{\color{orange}{[XW: #1]}}}
\newcommand{\ANC}[1]{{\color{red}{[ANC: #1]}}}

\title{Tunable Hybrid-Mode Coupler Enabling Strong Interactions between Transmons at a Centimeter-Scale Distance}

\author{Jianwen Xu}
\thanks{These authors contributed equally to this work.}
\affiliation{National Laboratory of Solid State Microstructures, School of Physics, Nanjing University, Nanjing 210093, China}
\affiliation{Shishan Laboratory, Nanjing University, Suzhou 215163, China}
\affiliation{Jiangsu Key Laboratory of Quantum Information Science and Technology, Nanjing University, Suzhou 215163, China}
\author{Xiang Deng}
\thanks{These authors contributed equally to this work.}
\affiliation{National Laboratory of Solid State Microstructures, School of Physics, Nanjing University, Nanjing 210093, China}
\affiliation{Shishan Laboratory, Nanjing University, Suzhou 215163, China}
\affiliation{Jiangsu Key Laboratory of Quantum Information Science and Technology, Nanjing University, Suzhou 215163, China}
\author{Wen Zheng}
\thanks{These authors contributed equally to this work.}
\email{Contact author: zhengwen@nju.edu.cn}
\affiliation{National Laboratory of Solid State Microstructures, School of Physics, Nanjing University, Nanjing 210093, China}
\affiliation{Shishan Laboratory, Nanjing University, Suzhou 215163, China}
\affiliation{Jiangsu Key Laboratory of Quantum Information Science and Technology, Nanjing University, Suzhou 215163, China}

\author{Wenchang Yan}
\affiliation{National Laboratory of Solid State Microstructures, School of Physics, Nanjing University, Nanjing 210093, China}
\affiliation{Shishan Laboratory, Nanjing University, Suzhou 215163, China}
\affiliation{Jiangsu Key Laboratory of Quantum Information Science and Technology, Nanjing University, Suzhou 215163, China}
\author{Tao Zhang}
\affiliation{National Laboratory of Solid State Microstructures, School of Physics, Nanjing University, Nanjing 210093, China}
\affiliation{Shishan Laboratory, Nanjing University, Suzhou 215163, China}
\affiliation{Jiangsu Key Laboratory of Quantum Information Science and Technology, Nanjing University, Suzhou 215163, China}
\author{Zhenchuan Zhang}
\affiliation{Shishan Laboratory, Nanjing University, Suzhou 215163, China}
\affiliation{Jiangsu Key Laboratory of Quantum Information Science and Technology, Nanjing University, Suzhou 215163, China}
\author{Wanli Huang}
\affiliation{Shishan Laboratory, Nanjing University, Suzhou 215163, China}
\affiliation{Jiangsu Key Laboratory of Quantum Information Science and Technology, Nanjing University, Suzhou 215163, China}
\author{Xiaoyu Xia}
\affiliation{National Laboratory of Solid State Microstructures, School of Physics, Nanjing University, Nanjing 210093, China}
\affiliation{Shishan Laboratory, Nanjing University, Suzhou 215163, China}
\affiliation{Jiangsu Key Laboratory of Quantum Information Science and Technology, Nanjing University, Suzhou 215163, China}
\author{Xudong Liao}
\affiliation{National Laboratory of Solid State Microstructures, School of Physics, Nanjing University, Nanjing 210093, China}
\affiliation{Shishan Laboratory, Nanjing University, Suzhou 215163, China}
\affiliation{Jiangsu Key Laboratory of Quantum Information Science and Technology, Nanjing University, Suzhou 215163, China}

\author{Yu Zhang}
\affiliation{National Laboratory of Solid State Microstructures, School of Physics, Nanjing University, Nanjing 210093, China}
\affiliation{Shishan Laboratory, Nanjing University, Suzhou 215163, China}
\affiliation{Jiangsu Key Laboratory of Quantum Information Science and Technology, Nanjing University, Suzhou 215163, China}
\author{Jie Zhao}
\affiliation{National Laboratory of Solid State Microstructures, School of Physics, Nanjing University, Nanjing 210093, China}
\affiliation{Shishan Laboratory, Nanjing University, Suzhou 215163, China}
\affiliation{Jiangsu Key Laboratory of Quantum Information Science and Technology, Nanjing University, Suzhou 215163, China}
\author{Shaoxiong Li}
\affiliation{National Laboratory of Solid State Microstructures, School of Physics, Nanjing University, Nanjing 210093, China}
\affiliation{Shishan Laboratory, Nanjing University, Suzhou 215163, China}
\affiliation{Jiangsu Key Laboratory of Quantum Information Science and Technology, Nanjing University, Suzhou 215163, China}
\affiliation{Synergetic Innovation Center of Quantum Information and Quantum Physics, University of Science and Technology of China, Hefei, Anhui 230026, China}
\affiliation{Hefei National Laboratory, Hefei 230088, China}
\author{Xinsheng Tan}
\affiliation{National Laboratory of Solid State Microstructures, School of Physics, Nanjing University, Nanjing 210093, China}
\affiliation{Shishan Laboratory, Nanjing University, Suzhou 215163, China}
\affiliation{Jiangsu Key Laboratory of Quantum Information Science and Technology, Nanjing University, Suzhou 215163, China}
\affiliation{Synergetic Innovation Center of Quantum Information and Quantum Physics, University of Science and Technology of China, Hefei, Anhui 230026, China}
\affiliation{Hefei National Laboratory, Hefei 230088, China}

\author{Dong Lan}
\email{Contact author: land@nju.edu.cn}
\affiliation{National Laboratory of Solid State Microstructures, School of Physics, Nanjing University, Nanjing 210093, China}
\affiliation{Shishan Laboratory, Nanjing University, Suzhou 215163, China}
\affiliation{Jiangsu Key Laboratory of Quantum Information Science and Technology, Nanjing University, Suzhou 215163, China}
\affiliation{Synergetic Innovation Center of Quantum Information and Quantum Physics, University of Science and Technology of China, Hefei, Anhui 230026, China}
\affiliation{Hefei National Laboratory, Hefei 230088, China}

\author{Yang Yu}
\affiliation{National Laboratory of Solid State Microstructures, School of Physics, Nanjing University, Nanjing 210093, China}
\affiliation{Shishan Laboratory, Nanjing University, Suzhou 215163, China}
\affiliation{Jiangsu Key Laboratory of Quantum Information Science and Technology, Nanjing University, Suzhou 215163, China}
\affiliation{Synergetic Innovation Center of Quantum Information and Quantum Physics, University of Science and Technology of China, Hefei, Anhui 230026, China}
\affiliation{Hefei National Laboratory, Hefei 230088, China}

\date{\today}

\begin{abstract}
    The transmon, a fabrication-friendly superconducting qubit, remains a leading candidate for scalable quantum computing.
    Recent advances in tunable couplers have accelerated progress toward high-performance quantum processors.
    However, extending coherent interactions beyond millimeter scales to enhance quantum connectivity presents a critical challenge.
    Here, we introduce a hybrid-mode coupler exploiting resonator-transmon hybridization to simultaneously engineer the two lowest-frequency modes, enabling high-contrast coupling between transmons spaced at a centimeter-scale distance.
    For a 1-cm coupler, we experimentally demonstrate flux-tunable hybrid modes and measure strong $XX$ and $ZZ$ couplings between qubits exceeding $23$ MHz and $10$ MHz, respectively, under current conditions, in agreement with an effective two-channel model.
	Our theoretical model further suggests $ZZ$ strengths reaching 100 MHz, with modulation contrasts exceeding $10^4$.
	This work provides an efficient pathway to mitigate the inherent connectivity constraints imposed by short-range interactions, enabling transmon-based architectures compatible with hardware-efficient quantum tasks.
\end{abstract}

\maketitle

\section{Introduction}
The advent of superconducting qubits \cite{you2011,krantz2019,Blais2021}, particularly the transmon \cite{Koch2007}, known for its fabrication-friendly nature, has enabled diverse architectures for qubit connectivity \cite{chen2014,kurpiers2018,zhou2023,abrams2020,nguyen2024,qiu2025a,zhao2020,yan2018,Sete2021,li2024}.
Specifically, the introduction of tunable couplers \cite{yan2018,Sete2021,li2024,Mundada2019,Xu2020,Sung2021,Stehlik2021}, a strategy now integral to high-performance superconducting processors \cite{gong2021,chu2023,zhang2023,arute2019,Wu2021,Google2025,Gao2025}, enables sophisticated control over coupling strengths \cite{McKay2016,Foxen2020,chen2025}, making significant progresses toward practical quantum processors \cite{Google2025,Gao2025}.
These systems have also shown quantum advantages in specialized tasks \cite{arute2019,Wu2021,Google2025,Gao2025} and enabled scalability to hundreds of qubits \cite{Google2025,Gao2025}.
Nevertheless, the short-range nature of nearest-neighbor connection presents challenges for achieving large-scale systems \cite{Jiang2007,gambetta2017,bravyi2022,ang2022,field2024} and implementing efficient error correction protocols \cite{cohen2022,Breuckmann2021,bravyi2024,gidney2025,wang2025}.

In recent years, research on quantum connection capabilities has increased.
For example, quantum teleportation gates \cite{gottesman1999} theoretically enable quantum entanglement operations over arbitrary distances for distributing quantum conputing \cite{main2025}, but due to limitations in measurement systems, local entangling gate speeds, measurement performance, and coherence times, quantum teleportation gates still cannot fully exploit their advantages \cite{chou2018,qiu2025}.
It is worth noting that significant progress has been made in achieving long-range high-fidelity state transfer using transmission lines \cite{zhong2019}, coaxial cables \cite{kurpiers2018,niu2023,zhong2021,Campagne2018}, and cryogenic waveguides \cite{Magnard2020,kannan2023,storz2023}.
However, how to realize high-fidelity entangling gates to enhance the connectivity remains an ongoing area of exploration \cite{zhao2022,mollenhauer2024,song2024,deng2025,xiong2025,heya2025}.
Therefore, establishing efficient remote connections between qubits remains one of the core challenges for superconducting qubits toward hardware-efficient quantum manipulation.

Resonators, which are foundational yet widely adopted due to their practical advantages in fabrication and manipulation, efficiently mediate inter-qubit coupling.
A paradigmatic example is the coplanar waveguide (CPW) resonator, demonstrated in experimental realizations of quantum gates between qubits \cite{majer2007,sillanpaa2007}, which has since become a ubiquitous building block in superconducting quantum processors.
Beyond basic coupling functionalities, these resonators unlock enhanced connectivity paradigms, such as all-to-all coupling architectures \cite{Hazra2021,wu2024}, which are critical for quantum simulation \cite{Zhong2016,wang2019} and entangled state preparation \cite{song2017,song2019}.
Specifically, the resonator exhibits a free spectral range (FSR), defined as $\mathrm{FSR}=\pi \nu/d$, corresponding to the frequency spacing between adjacent modes, where $\nu$ is the photon phase velocity and $d$ denotes the resonator length.
Their intrinsic length $d$ can further address quantum chip design challenges by enabling distant qubit coupling, thereby enhancing quantum connectivity and reducing control-line density.

Therefore, a natural consideration is the development of a modified resonator that enables both tunable and extended long-range coupling architectures.
Previous innovations include Josephson-junction embedded transmission-line resonators \cite{palacios2008, bourassa2012, mallet2009, leib2012,sandberg2008,wang2013} and Unimon qubits \cite{Tuohino2024,duda2025,hyyppa2022}, which integrate Josephson junctions directly into CPW resonators.
These approaches achieve hybridization between the non-linear modes provided by the Josephson energy and the harmonic modes provided by the resonators, enabling tunable mode operation tailored to manipulation requirements \cite{bourassa2012, hyyppa2022,Bao2021,marinelli2023}.

Building on these advances, we introduce a coupler architecture engineered for strong coupling between transmons spaced several centimeters apart.
We establish a theoretical framework for this flux-tunable hybrid-mode coupler and analyze its dynamics response to external flux,
including the evolution of electric field profiles.
The two-qubit coupling mediated by a 1-cm coupler decomposes into two distinct interaction channels.
Experimentally, we characterize the full spectroscopy of the coupler and its anharmonicity, with results showing excellent agreement with theoretical predictions.
We observe coupling strengths exceeding $23$ MHz for $XX$ interactions and $10$ MHz for $ZZ$ interactions.
Furthermore, theoretical analysis suggests that the $ZZ$ coupling can reach up to $100$ MHz with contrasts exceeding $10^4$ under optimal system parameters, highlighting the potential of the coupler for future applications.

This paper is organized as follows.
Section~\ref{sec2a} analyzes the hybrid-mode structure,
and Section~\ref{sec2b} investigates its capacitive coupling to transmons spaced several centimeters apart, quantifying the dependence of inter-qubit coupling strength on external flux.
Section~\ref{sec3} presents experimental results on two-qubit couplings,
followed by a discussion in Section~\ref{sec4} on the achievement of strong $ZZ$ interactions and an evaluation of the coupler's potential based on designed paramters.
Finally, Section~\ref{sec5} makes conclusions.

\section{Exploration of the system Hamiltonian}
\begin{figure}
	\centering
	\includegraphics[width=0.45\textwidth]{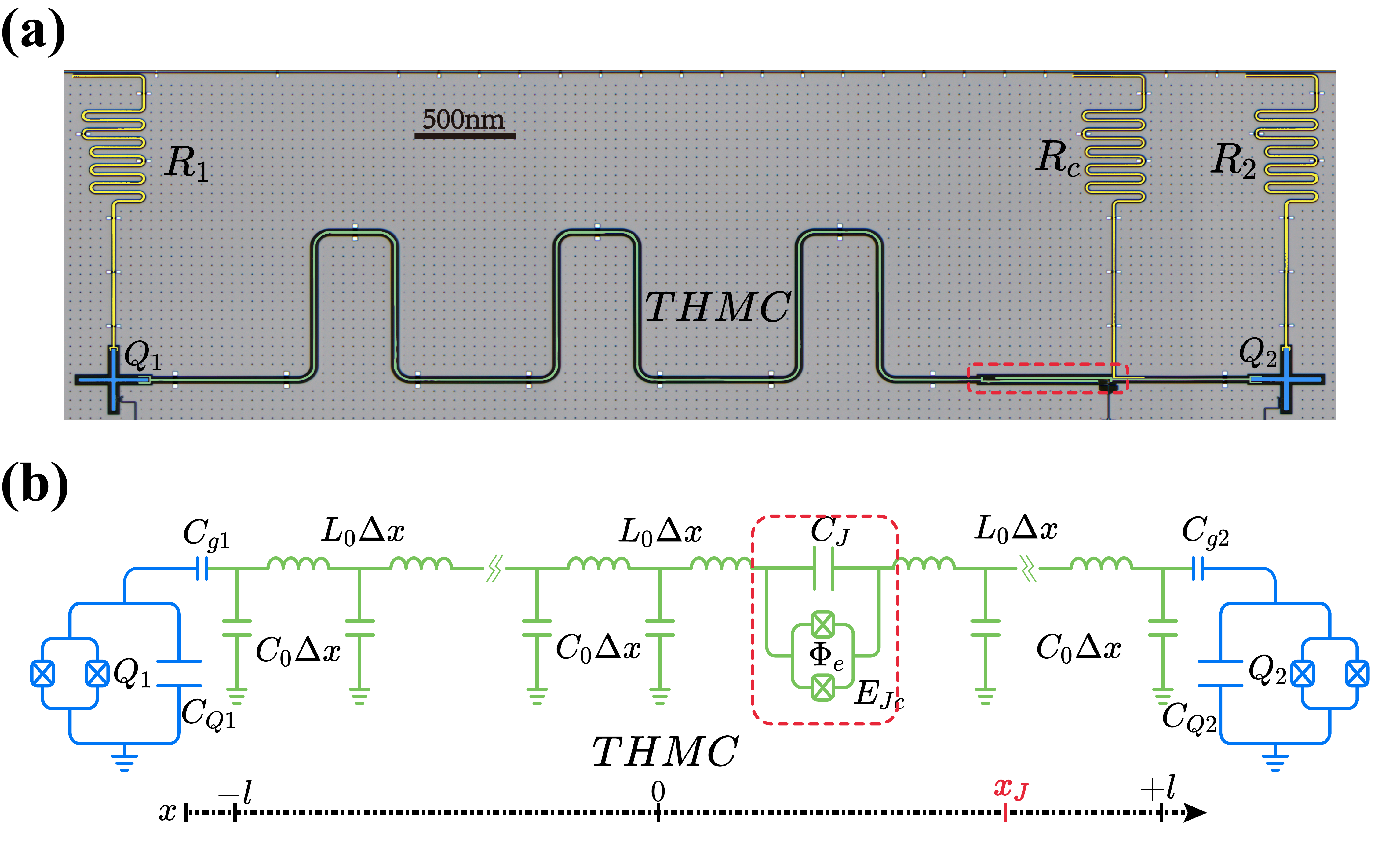}
    \caption{\justifying
      (a) False color sample image. Optical micrograph shows two grounded transmons ($Q_{1,2}$, blue), coupled via a flux-tunable hybrid-mode coupler (THMC, green), with an effective seperation of approximately 1~cm.
      The states of the qubits and coupler are read out through dispersive measurements using individual resonators ($R_{1,2,c}$, yellow).
      (b) Distributed-element circuit model. The THMC couples to both transmons via lumped elements characterized by per-unit-length parameters $C_0$ and $L_0$, defined in the continuum limit $\Delta x \to 0$.
      The embedded transmon at position $x_J$ partitions the coupler into left ($-l \leq x < x_J$) and right ($x_J < x \leq l$) segments.
      The embedded transmon contains a shunt capacitance $C_J$ and two Josephson junctions, with energy $E_{Jc}$, threaded by external flux $\Phi$.
    }
    \label{fig:figure1}
\end{figure}

Without loss of generality, as shown in Fig. \ref{fig:figure1}(a), we experimentally implement a CPW resonator of length $2l \approx 1\,\text{cm}$ (green) that couples two grounded transmons \cite{Barends2013} ($Q_1$ and $Q_2$, blue).
The fabrication procedure follows a standard and simple approach, suitable for application in prevalent superconducting chips, with details provided in our previous work \cite{zheng2022}.
To enable flux-tunable operation, a transmon is integrated at the resonator center conductor, hence can be thought of as a CPW resonator, open at both ends, connected to a transmon at $x_J$ of the central conductor. 
This architecture functions as a tunable bus module toward a modular quantum device~\cite{zhao2022}. 
Unlike the Unimon design~\cite{hyyppa2022} with its central junction and closed boundaries, our coupler features open ends and an off-center junction at $x_J \approx 0.75l$ on the biased side.
This distinction arises from differing boundary conditions and the necessity to modify anharmonicity.

Conventional circuit quantization of tunable couplers typically employs lumped-element models, neglecting the impact of coupler length on device parameters.
Yet achieving long-distance coupling, whether via grounded \cite{yan2018}, floating \cite{Sete2021}, or double-transmon designs \cite{li2024,Goto2022,Campbell2023,li2025}, requires extending the coupler resulting in distributed elements.
At centimeter scales, this finite length critically modifies coupler behavior.
We thus derive the Hamiltonian using a distributed-element model as shown in Fig. \ref{fig:figure1}(b), accounting for the resonator per-unit-length capacitance $C_0$ and inductance $L_0$.
The total Lagrangian of the system is given by
$\mathcal{L} = \mathcal{L}_{Q}+\mathcal{L}_{C}+\mathcal{L}_g$,
where $\mathcal{L}_{Q}$, $\mathcal{L}_{C}$, and $\mathcal{L}_{g}$ correspond to the Lagrangians of the qubits, coupler, and coupling terms, respectively. 
The qubit Lagrangian takes the form
\begin{equation}
	\label{eq:lagrangian_Q}
	\begin{aligned}
		\mathcal{L}_{Q}  = \sum_{j=1,2}  \frac{\left ( C_{Qj} \! +\! C_{gj} \right)}{2} \dot{\psi}_{Qj}^2(t) \! + \! E_{Jj} \cos\left(\frac{2\pi {\psi}_{Qj}}{\Phi_0}\right),         
	\end{aligned}
\end{equation}
The coupler Lagrangian comprises distributed and junction terms
\begin{equation}
	\label{eq:lagrangian_C}
	\begin{aligned}
			\mathcal{L}_{C} =& \int_{-l}^{+l}\left(\frac{C_0}{2} \dot{\psi}^2(x,t)-\frac{{[\partial_x \psi(x,t)]}^2}{2L_0}\right)dx\\
			                 &+\frac{C_{g1}}{2} \dot{\psi}^2(-l,t) +\frac{C_{g2}}{2} \dot{\psi}^2(+l,t)\\
			                 &+\frac{C_J}{2}{\dot{\delta}^2_C}+E_{Jc}(\Phi) \cos \left(\frac{2\pi {\delta_{C}}}{\Phi_0}\right),                       
	\end{aligned}
\end{equation}
where the integral captures CPW dynamics.
The capacitive coupling Lagrangian is given by
\begin{equation}
	\label{eq:lagrangian_g}
	\begin{aligned}
		\mathcal{L}_{g}=-C_{g1} \dot{\psi}(-l,t) \dot{\psi}_{Q1} - C_{g2} \dot{\psi}(+l,t) \dot{\psi}_{Q2}.                     
	\end{aligned}
\end{equation}
Here, $\psi$ represents the node flux variable at the circuit model, with $\dot{\psi}=\partial \psi/\partial t$ (proportional to voltage).
The capacitances $C_{Qj}$ and $C_{gj}$ denote the shunt capacitance of qubit $j$ and the coupling capacitance between the coupler and qubit $j$, respectively.
The phase difference across the Josephson junction is defined as $\delta_{C} \equiv \psi(x_J^{+},t) - \psi(x_J^{-},t)$, where $x_J^{\pm}$ mark the junction boundaries.
The flux-tunable Josephson energy follows  
$
E_{Jc}(\Phi) \approx E_{J}^{\text{max}} \left| \cos\left(2\pi \Phi / \Phi_0\right) \right|\!,
$
with $\Phi$ being the external flux bias and $\Phi_0$ the flux quantum.

\subsection{Hamiltonian of the coupler}\label{sec2a}

We first analyze the coupler Lagrangian.
Outside the boundaries and Josephson junction, the Euler-Lagrange equation
\begin{equation}
\frac{d}{dt} \frac{\partial \mathcal{L}}{\partial \dot{\psi}(x,t)} = \frac{\partial \mathcal{L}}{\partial \psi(x,t)}
\end{equation}
for the linearized coupler action yields solutions satisfying the wave equation
\begin{equation}
\ddot{\psi}(x,t) = v_p^2  \partial_{xx} \psi(x,t),
\end{equation}
where $ v_p = 1/\sqrt{L_0 C_0} $ is the phase velocity, and $ \omega_m = k_m v_p $ gives the mode frequency with wave number $k_m$.

The general solution decomposes into spatial and temporal components
\begin{equation}
\label{eq:wave equation_solution1}
\psi(x,t) = \sum_{m} u_m(x) \phi_m(t),
\end{equation}
where $ \phi_m(t) \sim e^{-i \omega_m t} $.
The spatial function $u_m(x)$ takes the piecewise form
\begin{equation}
\label{eq:wave equation_solution2}
u_m(x) = A_m \begin{cases} 
\sin [\alpha_1(x)] & -l \leq x < x_J \\
B_m \sin [\alpha_2(x)] & x_J < x \leq +l
\end{cases}
\end{equation}
with $\alpha_1(x) = k_m (x + l) + \theta_{m,1}$,
$\alpha_2(x) = k_m (x - l) + \theta_{m,2}$, wavenumber $k_m$, phase parameters $\theta_{m,1}$, $\theta_{m,2}$, and dimensionless amplitudes $A_m$, $B_m$ determined by boundary conditions and orthonormality.

The amplitude $B_m$ is constrained by current continuity at the junction
\begin{equation}
\label{eq:boundary_condition1}
\frac{\partial_{x} \psi(x_J^{-}, t)}{L_0} = \frac{\partial_{x} \psi(x_J^{+}, t)}{L_0} = C_J\ddot{\delta}_C + I_c \sin\left( \frac{2\pi\delta_C}{\Phi_0} \right).
\end{equation}
Substituting Eq. \eqref{eq:wave equation_solution1} into the Eq. \eqref{eq:boundary_condition1} yields
\begin{equation}
\label{eq:boundary_condition1_1}
\sum_{m} \phi_m(t)  \partial_x u_m(x_J^{-}) = \sum_{m} \phi_m(t)  \partial_x u_m(x_J^{+}),
\end{equation}
which, combined with  Eq. \eqref{eq:wave equation_solution2}, gives
\begin{equation}
\label{eq:Bm}
B_m = \frac{ \cos [\alpha_1(x_J)] }{ \cos [\alpha_2(x_J)]}.
\end{equation}
The phases $\theta_{m,1}$, $\theta_{m,2}$ are determined by resonator boundary conditions
\begin{equation}
\label{eq:boundary_condition2}
\begin{aligned}
\ddot{\psi}(-l,t) - \frac{1}{C_{g1} L_0}  \partial_x \psi(-l,t) &= 0 \\
\ddot{\psi}(+l,t) + \frac{1}{C_{g2} L_0}  \partial_x \psi(+l,t) &= 0
\end{aligned}
\end{equation}
Substituting the spatial envelope $u_m(x)$ in Eq. \eqref{eq:wave equation_solution2} into Eq. \eqref{eq:boundary_condition2} yields the relations
\begin{equation}
	\label{eq:boundary_condition2_1}
	\begin{aligned}
		\tan \theta_{m,1}=-\frac{1}{\omega_{m}^{2}} \frac{k_{m}}{C_{g1}L_0}=-\frac{C_0}{C_{g1}}\frac{1}{k_{m}}\\ 
		\tan \theta_{m,2}=\frac{1}{\omega_{m}^{2}} \frac{k_{m}}{C_{g2}L_0}=\frac{C_0}{C_{g2}}\frac{1}{k_{m}}
	\end{aligned}
\end{equation}
where we use $\omega_m=k_m v_p$ and $v_p=1/\sqrt{L_0 C_0}$.
The wave number $k_m$ satisfies the transcendental equation
\begin{equation}
	\label{eq:transcendental}
	\begin{aligned}
		\left ( \frac{L_0}{L_J}-\frac{C_J }{C_0}k_m^2\right)\left [ \tan \alpha_2(x_J)-\tan \alpha_1(x_J)\right] = k_m,
	\end{aligned}
\end{equation}
Note that the junction current in Eq. \eqref{eq:boundary_condition1} is linearized as $I_c \sin(\frac{2\pi}{\Phi_0}{\delta}_{C}) \approx \frac{2\pi I_c{\delta}_{C}}{\Phi_0}$, valid for small phase variations, and $L_J = \Phi_0/2\pi I_c$ is Josephson junction inductance.
Solving Eq. \eqref{eq:transcendental} numerically yields mode frequencies $\omega_m = k_m v_p$.

\begin{figure*}
	\centering
	\includegraphics[width=1\textwidth]{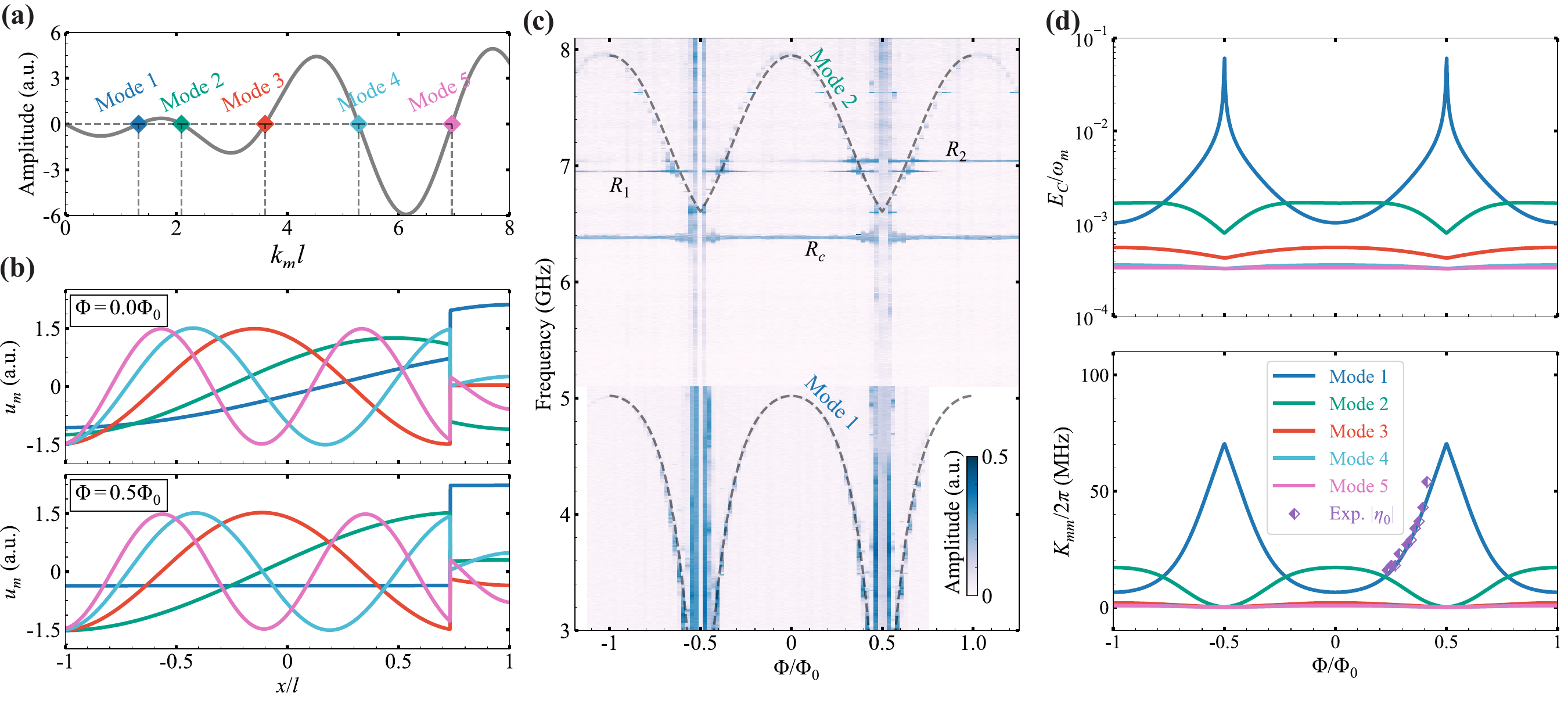}
    \caption{\justifying
	(a) Wave number solutions.
	Spatial mode characterization showing the calculated wave number $k_m$ with a geometric length $l$.
    The condition, amplitude at 0, corresponds to the solution of the transcendental equation Eq. \eqref{eq:transcendental}, which determines the wavenumber.
    Dashed lines represent analytic predictions.  
	(b) Flux-dependent mode envelopes.
		Flux envelope functions for the five lowest-frequency normal modes at a flux bias of $\Phi/\Phi_0 = 0$ (upper panel) and $\Phi/\Phi_0 = 0.5$ (lower panel).
		The results were obtained using the parameters provided in Table. \ref{table:designed_params}.
		The shifting of antinode values is attributed to the modulation of the Josephson inductance.
	(c) Experimentally measured flux-tunable spectroscopy.
		Transmission magnitude as a function of flux bias and driving frequency.
		Dashed lines correspond to the full-circuit model fits based on Eq. \eqref{eq:H_QM2}.
		As $\Phi/\Phi_0$ increases from 0 to 0.5, the first mode shifts from 5.02 GHz to below 3 GHz, while the second mode shifts from 7.95 GHz to 6.622 GHz.
		The vertical feature near $\Phi/\Phi_0=\pm0.5$ may result from the second mode approaching the readout resonator.
		The horizontal resonances at 6.4 GHz and 7.2 GHz corresponds to the fundamental modes of readout resonators $R_{1,c,2}$.
	(d) Anharmonicity.
    Theoretically calculated charge energy $E_{C,m}$ as a function of flux, normalized by the mode frequency $\omega_m$ (upper panel). 
    Theoretically calculated Kerr coefficient $\mathcal{K}_{mm} = \frac{E_{Jc}}{E_{L,m}}E_{C,m}$ (lower panel) is shown alongside the experimentally measured first-mode anharmonicity $\eta_0 = \omega_0^{21}-\omega_0^ {10}$ (dimond markers), which agrees well with the theoretical predictions.
		The results show that the tunable coupler exhibits weak anharmonicity in the lower-order modes, which can be adjusted by the external flux bias.
	}
  \label{fig:figure2}
\end{figure*}

The parameter $A_m$ follows from orthonormality conditions
\begin{equation}
	\label{eq:orthogonality relation1}
	\begin{aligned}
		&\left\langle u_{m}, u_{n}\right\rangle =  \int_{-l}^{+l} d x C_0 u_{m}(x) u_{n}(x)+C_{g1} u_{m}(-l) u_{n}(-l) \\
		&+C_{g2} u_{m}(+l) u_{n}(+l)+C_{J} \Delta u_{m} \Delta u_{n} \equiv C_{\Sigma} \delta_{mn},
	\end{aligned}
\end{equation}
\begin{equation}
	\label{eq:orthogonality relation2}
	\begin{aligned}
	\left\langle\partial_{x} u_{m}, \partial_{x} u_{n}\right\rangle &=\int_{-l}^{+l} \frac{dx}{L_0} \partial_{x} u_{m}(x) \partial_{x} u_{n}(x)\\ 
	&+\frac{ \Delta u_{m} \Delta u_{n}}{L_J} \equiv \frac{\delta_{m n}}{L_{m}},
	\end{aligned}
\end{equation}
where $C_{\Sigma} = \int_{-l}^{+l} C_{0} d x+C_{g1}+C_{g2}+C_{J}$ is the total capacitance and $L_m^{-1} = C_{\Sigma} \omega_m^2 $ is the effective inductance corresponding to the mode frequency of the coupler.
The junction discontinuity is $\Delta u_{m}=u_{m}(x_J^{+})-u_{m}(x_J^{-})$, changing in the envelope function $u_m(x)$ across the  Josephson junction.
Evaluating the integrals gives
\begin{equation}
	\label{eq:Am}
	\begin{aligned}
		A_m = \sqrt{\frac{C_{\Sigma}}{I_1+I_2+I_3+I_4}},
	\end{aligned}
\end{equation}
with
\begin{equation}
	\label{eq:A1}
	\begin{aligned}
		I_1 &= C_0\left[\frac{x_J+l}{2}-\frac{\sin 2\alpha_{1}(x_J)-\sin 2\theta_{m,1} }{4k_m}\right],\\
		I_2 &= C_0 B^2_m\left[\frac{l-x_J}{2} + \frac{\sin 2\alpha_{2}(x_J)-\sin 2\theta_{m,2}}{4k_m}\right],\\
		I_3 &=  C_{g1}{\sin}^2\theta_{m,1}+C_{g2}B_m^2{\sin}^2\theta_{m,2},\\
		I_4 &=  C_J{\left[ B_m\sin \alpha_{2}(x_J)-\sin \alpha_{1}(x_J)\right]}^2.
	\end{aligned}
\end{equation}
Meanwhile, the parameter $\Delta u_m$ quantifies the junction's influence on the CPW spatial modes, crucial for Hamiltonian quantization.
The coupler Lagrangian then becomes
\begin{equation}
	\label{eq:L_C1}
	\begin{aligned}
		\mathcal{L}_C 
		= &\sum_{m}  \frac{C_m^{\prime}}{2}\dot{\phi}_m^2-\frac{{\phi}_m^2(t)}{2L_m^{\prime}}-U_{nl}(\delta_C),
	\end{aligned}
\end{equation}
where $\phi_m = \psi_m \Delta u_m$ is the flux across the junction, with renormalized parameters
$C_m^{\prime}=C_{\Sigma}/ {(\Delta u_m)}^2$ and $L_m^{\prime}= L_m{(\Delta u_m)}^2$.
The nonlinear potential
$$
U_{nl}(\delta_C)=\sum_{i>1} \frac{(-1)^{i+1}}{(2 i)!}E_{Jc}\left(\frac{2 \pi}{\Phi_{0}}\sum_{m} \phi_{m} \right)^{2 i}
$$
captures higher-order junction effects beyond the linear approximation.

\begin{table}
\centering
\caption{\justifying
Sample parameters} 
\label{table:designed_params}
\begin{tabular*}{0.85\linewidth}{@{\extracolsep{\fill}}lc@{}}
\hline
\hline
Junction position, $x_J$ & 0.395 cm \\
Coupler length, $2l$ & 1.05 cm\\
Coupling capacitance, $C_{g1}$ & 9 fF\\
Coupling capacitance, $C_{g2}$ & 9 fF\\
$Q_1$ shunt capacitance, $C_{Q_1}$ & 80 fF\\
$Q_2$ shunt capacitance, $C_{Q_2}$ & 85 fF\\
Coupler shunt capacitance, $C_J$ & 30 fF\\
Linear capacitance, $C_0$ & 85.644 pF/m\\
Linear inductance, $L_0$ & 0.744 $\mu$H/m\\
Maximum Josephson energy, $E_J^{\mathrm{max}}/2\pi$ & 34.186 GHz\\
\hline
\hline
\end{tabular*}
\end{table}

The coupler Hamiltonian is derived via Legendre transformation $\mathcal{H} = \sum_m \frac{\partial \mathcal{L}}{\partial \dot{\phi}_m}\dot{\phi}_m - \mathcal{L}$. Retaining the leading nonlinear junction term $\frac{1}{24}E_{Jc}\left(\frac{2\pi}{\Phi_0}\sum_{m}{\phi}_m \right)^4$ while neglecting the coupling effect between different modes yields the classical Hamiltonian
\begin{equation}
	\label{eq:H_FTR}
	\begin{aligned}
		\mathcal{H}_C &\approx \sum_{m} \mathcal{H}_{C,m},\\ 
		\mathcal{H}_{C,m} &\approx \frac{Q_m^2}{2C_m^{\prime}}+\frac{{\phi}_m^2}{2L_m^{\prime}}-\frac{1}{24}E_{Jc}\left(\frac{2\pi}{\Phi_0}{\phi}_m \right)^4,
	\end{aligned}
\end{equation}
where the canonical momentum $Q_m \equiv \partial \mathcal{L} / \partial \dot{\phi}_{m}$ satisfies the canonical commutation relation $[\hat{\phi}_m,\hat{Q}_m] = i\hbar$ upon quantization.
Introducing the particle number operator $\hat{n}_m = \hat{Q}_m/2e$ and the phase operator $\hat{\varphi}_m = {2\pi} \hat{\phi}_m/{\Phi_0}$, and similarly $[\hat{\varphi}_m,\hat{n}_m]=i$, we obtain the quantum Hamiltonian
\begin{equation}
	\label{eq:H_QM1}
	\begin{aligned}
		\hat{\mathcal{H}}_{C,m} = 4E_{C,m}\hat{n}_m^2+\frac{1}{2}E_{L,m}\hat{\varphi}_m^2 -\frac{1}{24}E_{Jc}\hat{\varphi}_m^4
	\end{aligned}
\end{equation}
with $E_{C,m}= e^2/(2C_m^{\prime})$ and $E_{L,m}=({\Phi_0}/{2\pi})^2/L_m^{\prime}$.

\begin{figure*}
	\centering
	\includegraphics[width=1\textwidth]{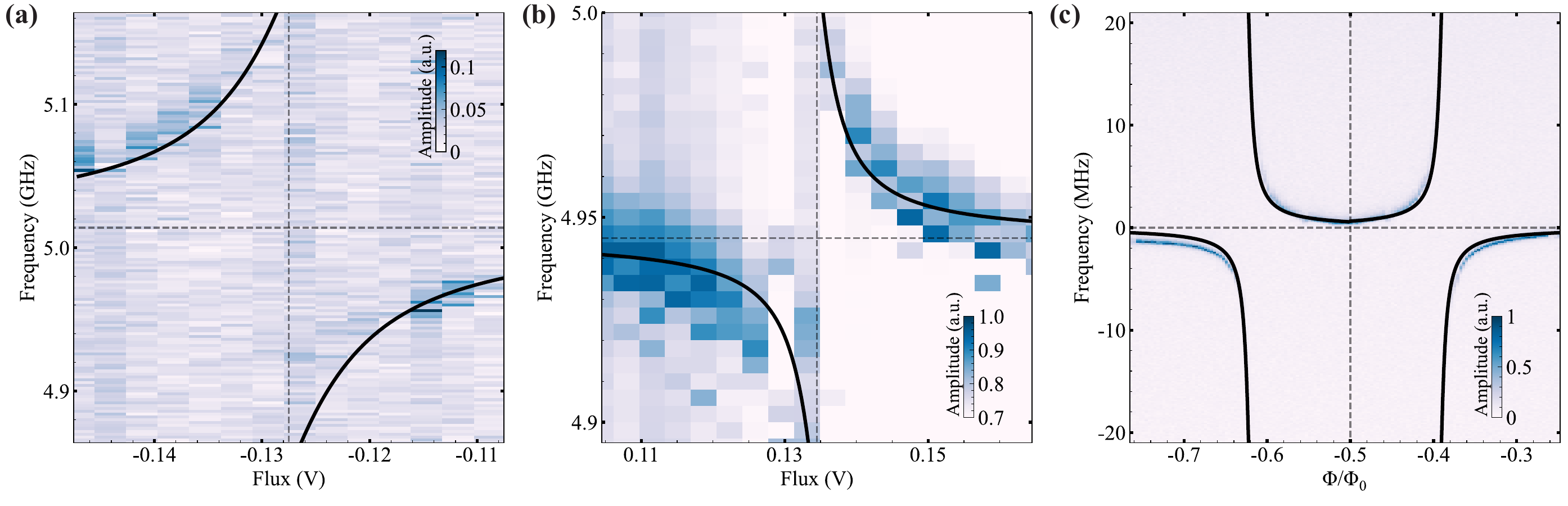}
    \caption{\justifying
    Experimentally measured avoided crossings.
    (a) At zero coupler bias, $\Phi/\Phi_0=0$, sweeping the $Q_2$ flux reveals an avoided crossing with coupling strength $|g_{21}|/2\pi=160$ MHz.
    (b) Sweeping the flux bias of $Q_1$ yields a coupling strength $|g_{11}|/2\pi=50$ MHz.
    (c) With $Q_1$ fixed at $\omega_1/2\pi=6.504$ GHz and $\Phi/\Phi_0=0$, sweeping the coupler flux $\Phi$ reveals an avoided crossing between the coupler mode $\nu_2$ and the $Q_1$ readout resonator with a dispersive frequency 6.954 GHz, indicating $g_{r1m2}/2\pi=21.04$ MHz.
    Considering the dispersive shift of $10.9$ MHz, the coupling strength is inferred to be $|g_{12}|/2\pi=135.18$ MHz.
    }
    \label{fig:figure3}
\end{figure*}

Routinely, expressing $\hat{n}_m$ and $\hat{\varphi}_m$ in harmonic oscillator basis
\begin{equation}
	\label{eq:a+a_operator}
	\begin{aligned}
		\hat{n}_m = \frac{i}{\sqrt{2}} \left(\frac{E_{L,m}}{8E_{C,m}}\right)^{\frac{1}{4}} (\hat{a}^{\dagger}_m-\hat{a}_m),\\
		\hat{\varphi}_m = \frac{1}{\sqrt{2}} \left(\frac{8E_{C,m}}{E_{L,m}}\right)^{\frac{1}{4}} (\hat{a}^{\dagger}_m+\hat{a}_m).
	\end{aligned}
\end{equation}
and applying the rotating wave approximation, we obtain
\begin{equation}
	\label{eq:H_QM2}
	\begin{aligned}
		\hat{\mathcal{H}}_{C,m}   \approx  \hbar \omega_{C,m} \hat{a}^{\dagger}_m \hat{a}_m -  \frac{\mathcal{K}_{mm}}{2}\hat{a}^{\dagger}_m     \hat{a}^{\dagger}_m \hat{a}_m \hat{a}_m,
        \end{aligned}
\end{equation}
where $\mathcal{K}_{mm} = E_{C,m}E_{Jc}/E_{L,m}$ is the self-Kerr coefficient and $\hbar \omega_{C,m} = \sqrt{8 E_{C,m}E_{L,m}} \!-\! \mathcal{K}_{mm}$ is  transition frequency of the coupler on $m$th mode.
The resulting energy spectrum
$$
		E_m^n = n \sqrt{8 E_{C,m}E_{L,m}}-\frac{\mathcal{K}_{mm}}{2}\left( n^2+n \right),
$$
exhibits transition frequencies
\begin{equation}
	\label{eq:Delta_E_m}
	\begin{aligned}
		\Delta E_m^{n+1,n} = \sqrt{8 E_{C,m}E_{L,m}}-\mathcal{K}_{mm}\left( n+1 \right),
	\end{aligned}
\end{equation}
with anharmonicity 
\begin{equation}\label{eq:anharmocity}
\eta_m = E_m^{21}-E_m^{10}.
\end{equation}

It is noteworthy that this anharmonicity $\eta_m=-E_CE_{Jc}/E_{L,m}$ differs from the transmon case ($\eta \simeq -E_C$) by the inductive ratio $E_{Jc}/E_{L,m}$, reflecting the coupler's hybrid Josephson-linear inductance character.
Here, we emphasize that this single-mode approximation neglects intermode interactions, which may significantly modify dynamics in multimode regimes.

As a paradigmatic example, we analyze the spatial mode characterization of the hybrid-mode coupler via the wave number $k_m$ and geometric length $l$.
The wave number can be predicted analytically by setting the amplitude equal to zero as described by Eq. \eqref{eq:transcendental}, yielding the wave number solutions indicated by the dashed lines as illustrated in Fig.~\ref{fig:figure2}(a).

Then, using Eqs. \eqref{eq:wave equation_solution2}, \eqref{eq:Bm}, and \eqref{eq:Am},
we compute the spatial envelope function for the five lowest-frequency normal modes, 
employing the parameters from Table \ref{table:designed_params} which are deduced through fitting the experimental spectroscopy with introducing the capacitive parameters determined by electromagnetic simulation.
These results are presented in Fig.~\ref{fig:figure2}(b).
The upper (lower) panel corresponds to flux biases $\Phi/\Phi_0 = 0$ ($\Phi/\Phi_0 = 0.5$), where Josephson inductance modulation induces shifts in the antinode positions.
The position of the Josephson junction influences these shifts, which correspond to changes in the electric potential and affect the coupling between the coupler and qubit, as described by Eq. \eqref{eq:H_QM5}, to be derived in the next section.

The spectrum of the coupler is obtained from the full-circuit model in Eq.~\eqref{eq:H_QM2}, with the resulting dashed lines shown in Fig.~\ref{fig:figure2}(c).
Experimentally, we verify the flux-tunable resonator spectrum for the two lowest modes, presenting the transmission magnitude as a function of flux bias and drive frequency.
The fundamental mode tunes from $5.02$ GHz at $\Phi/\Phi_0 = 0$ to below $3$ GHz at $\Phi/\Phi_0 = 0.5$, while the second mode shifts from $7.95$ GHz to $6.622$ GHz across the same flux range.
Given the large number of data points needed a long time, the measurements spanned two nights, allowing us to observe the full energy spectrum of the coupler as the magnetic flux varies.

Furthermore, analysis of anharmonicity via Eq.~\eqref{eq:anharmocity} reveals weak, flux-tunable anharmonicity in the lower-order modes.
As shown in Fig.~\ref{fig:figure2}(d), we first examine the proportion of charge energy in the coupler spectrum (upper panel).
The theoretically calculated Kerr coefficient $\mathcal{K}_{mm} = \frac{E_{Jc}}{E_{L,m}}E_{C,m}$ is plotted against the applied flux, alongside experimentally measured anharmonicity, as indicated by diamond markers.
These results align well with theoretical predictions Eq. \eqref{eq:anharmocity}, also confirming the validity of the single-mode approximation.
Additionally, the results show that the tunable resonator exhibits weak anharmonicity in the lower-order modes, which can be adjusted by the external flux bias.
Anharmonicity magnitude decreases with increasing mode order, demonstrating reduced junction influence on higher-frequency modes.

\subsection{Coupling between two distant transmons}\label{sec2b}

Next, we model the coupler-qubit system via the Lagrangian formalism.
Basis vectors transform as $\phi_m = \psi \Delta u_m$, substituting into Eq. \eqref{eq:wave equation_solution1}, yielding the Lagrangian $\mathcal{L}=\mathcal{T} - \mathcal{V}$ with
the total kinetic energy
\begin{subequations}\label{eq:lagrangian}
    \begin{equation}\label{eq:lagrangian_T}
    \begin{aligned}
        \mathcal{T} = &\sum_{i=1,2} \frac{C_i}{2} \dot{\phi}_{i}^2 + \sum_m \int_{x} \frac{C_0}{2 (\Delta u_m)^2} \dot{\phi}_{m}^2 dx + \frac{C_J}{2 (\Delta u_m)^2}\dot{\phi}_{m}^2\\
                    &+ \frac{C_{g_1}}{2} \left[\frac{u_m(-l)}{\Delta u_m}\dot{\phi}_{m}-\dot{\phi}_{1}\right]^2 + \frac{C_{g_2}}{2} \left[\dot{\phi}_{2} - \frac{u_m(+l)}{\Delta u_m}\dot{\phi}_{m}\right]^2,\\
                & = \sum_{j=1,2}  \frac{1}{2}  C_{\Sigma j}    \dot{\phi}_{Q_j}^2(t) +  \frac{C_m^{\prime}}{2}\dot{\phi}_m^2-\frac{{\phi}_m^2}{2L_m^{\prime}}\\
                &-G_{m,1}\dot{\phi}_m(t) \dot{\phi}_{Q_1}(t)-G_{m,2}\dot{\phi}_m(t) \dot{\phi}_{Q_2}(t),                 
    \end{aligned}
    \end{equation}
	and the potential energy
    \begin{equation}\label{eq:lagrangian_V}
    \begin{aligned}
            \mathcal{V} & = -\cos \left(\frac{2\pi {\phi}_{Q_j}}{\Phi_0}\right) +\sum_m  U_{nl}(\delta_C),           
    \end{aligned}
    \end{equation}
\end{subequations}
where $C_{\Sigma j}= C_{Q_j}+C_{gj}$, $G_{m,1}= \frac{C_{g1}u_m(-l)}{\Delta u_m}$, and $G_{m,2} = \frac{C_{g2}u_m(+l)}{\Delta u_m}$.
Expressing fluxes as $\mathbf{\phi}=[{\phi}_{Q_1},{\phi}_m,{\phi}_{Q_2}]^T$, the kinetic energy can be written as $\mathcal{T}=\frac{1}{2} \dot{\mathbf{\phi}}^T \mathbf{C} \dot{\mathbf{\phi}}$ with capacitance matrix
\begin{equation}
	\label{eq:C}
	\begin{aligned}
		\mathbf{C} = \begin{bmatrix} C_{\Sigma1} & -G_{m,1} & 0 \\ -G_{m,1} & C_m^{\prime} & -G_{m,2} \\  
			0 & -G_{m,2} & C_{\Sigma2} \end{bmatrix}.
	\end{aligned}
\end{equation}
Conjugate momenta $Q=\partial \mathcal{L} / \partial \dot{\mathbf{\phi}} = \mathbf{C} \dot{\mathbf{\phi}}$ yield the classical Hamiltonian
\begin{equation}
	\label{eq:H_QM4}
	\begin{aligned}
		\mathcal{H} = \frac{1}{2} \mathbf{Q}^T \mathbf{C}^{-1} \mathbf{Q} + \mathcal{V},
	\end{aligned}
\end{equation}
where $\mathbf{C}^{-1}$ is explicitly given by
\begin{equation}
	\label{eq:C_inv}
	\begin{aligned}
		\mathbf{C}^{-1}
  		& = \left[\begin{array}{lll}A_{11} & A_{12} & A_{13} \\ A_{21} & A_{22} & A_{23} \\ A_{31} & A_{32} & A_{33}\end{array}\right]\\
		& \!=\! \left( C_m^{\prime} C_{\Sigma1}C_{\Sigma2}  -  C_{\Sigma2}G_{m,1}^2 - C_{\Sigma1}G_{m,2}^2 \right)^{-1} \\
		& \begin{bmatrix}
			C_{{\Sigma2}} C_{m}^{\prime} - G_{m,2}^2 & C_{{\Sigma2}}\,G_{m,1} & G_{m,1} G_{m,2} \\ 
			C_{{\Sigma2}} G_{m,1} & C_{\Sigma1} C_{{\Sigma2}} & C_{{\Sigma1}} G_{m,2} \\ 
			G_{m,1} G_{m,2} & C_{{\Sigma1}}G_{m,2} & C_{{\Sigma1}} C_{m}^{\prime} \!-\!G_{m,1}^2 
		\end{bmatrix}
	\end{aligned}
\end{equation}
Using canonical quantization, that is $\hat{n}_{j}=Q_j/2e$ and $\hat{\varphi}_{j} = {2\pi} \hat{\phi}_{Q_j}/{\Phi_0}$, the quantum Hamiltonian of the whole system can be written as
\begin{equation}
	\label{eq:H_QM5}
	\begin{aligned}
		\hat{\mathcal{H}}_m &=\sum_{j=1,2}  4E_{C,Q_j}\hat{n}_{j}^2 - E_{Jj}\cos\hat{\varphi}_{j}\\
		& + 4E^{\prime}_{C,m}\hat{n}_m^2+\frac{1}{2}E_{L,m}\hat{\varphi}_m^2 -\frac{1}{24}E_{Jc}\hat{\varphi}_m^4\\
		& + 4 E_{1m} \hat{n}_{Q_1} \hat{n}_{m} + 4 E_{2m} \hat{n}_{Q_2} \hat{n}_{m} + 4 E_{12} \hat{n}_{Q_1} \hat{n}_{Q_2}
	\end{aligned}
\end{equation}
with charge energies $E^{\prime}_{C,m} = e^2 A_{33}/2,\, E_{C,Q_j}=e^2 A_{jj}/2$, qubit-coupler coupling energies
$E_{1m}= \mathrm{sign} (u_m(-l))e^2 |A_{12}|,\, E_{2m}=\mathrm{sign}(u_m(+l))e^2 |A _{23}|$.
Sign conventions reflect voltage polarities at coupler boundaries.
Here, the term $E_{13}=e^2 |A_{13}|$
does not correspond to the direct coupling arising from the static capacitive interaction.
The effective model in the $m$-th mode captures only dynamic interactions mediated by the coupler.
The direct coupling predicted by the capacitive network is approximately $0.3$ MHz.
Given its negligible contribution to the coupling between qubits, it is omitted in the subsequent analysis.
Determining the sign of direct qubit-qubit coupling strength, induced by parasitic capacitance networks in distributed-element models, presents an inherent challenge. Unlike floating or grounded couplers \cite{yan2018,Sete2021}, our qubit-coupler system decomposes exclusively into common-mode and differential-mode components \cite{longSuperconductingQuantumCircuits}. Crucially, the capacitance matrix formulation excludes the common-mode basis vector, a primary source of parasitic direct coupling. We therefore remain this ambiguity in our analysis, as the direct coupling term is negligible when qubit-coupler coupling capacitances substantially exceed the intrinsic qubit-qubit coupling capacitance.

\begin{figure}[hbtp!]
	\centering	
	\includegraphics[width=0.5\textwidth]{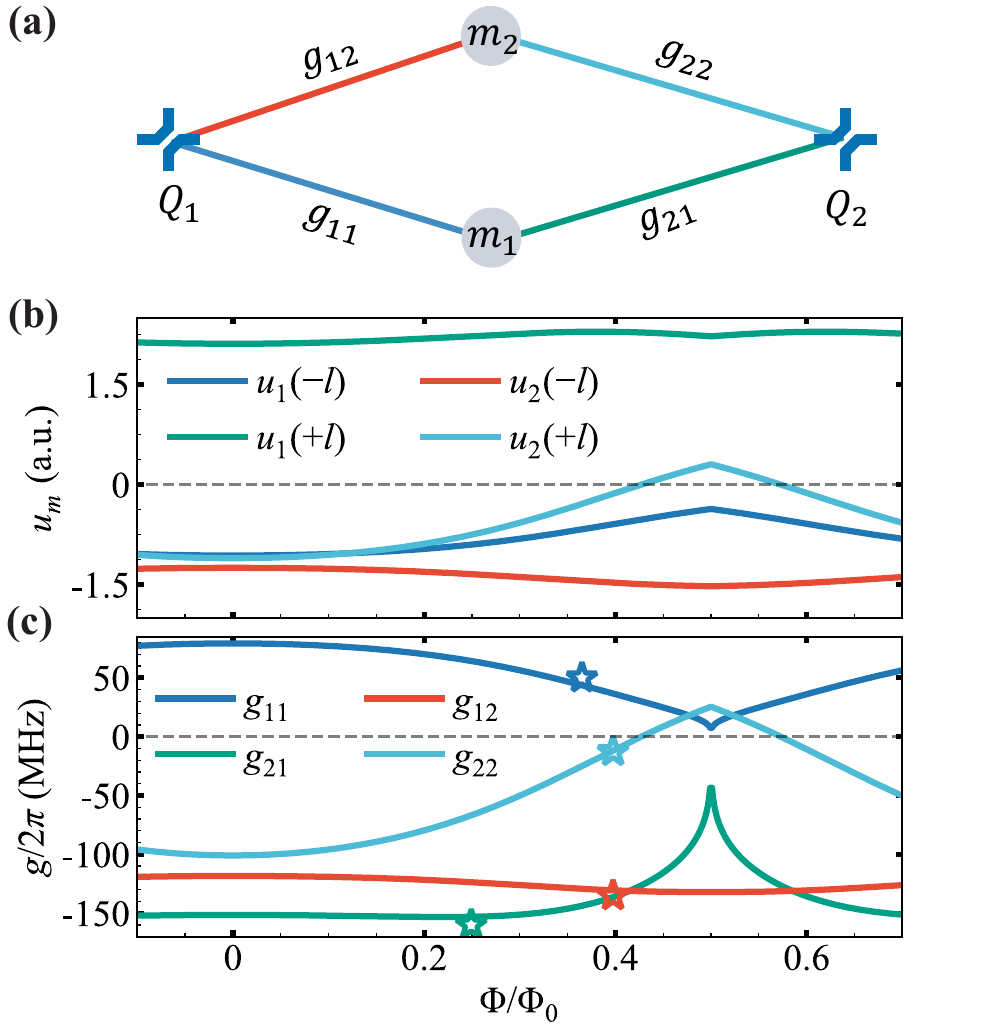}	
    \caption{\justifying
    (a) Schematic diagram of the coupling architecture between two transmon qubits. The 1 cm coupling length suppresses higher-order mode interference, enabling an effective two-path coupling model between the qubits.  
	(b) Electric field profile in the coupling port as a function of flux bias $\Phi/\Phi_0$.
	The amplitude and phase of the field, modulated by the embedded Josephson junction, directly determine the coupling strength between the transmon and the mode. Flux engineering enables precise control over field synchronization.  
	(c) Theoretical predictions of the mode-transmon coupling strengths by using Eq. \eqref{eq:g_jc}.
	Star makers represent the result inffered from Fig. \ref{fig:figure3}.
    }
    \label{fig:figure4}
\end{figure}

Similarly, we introduce the qubit number and phase operators via creation and annihilation operators
\begin{equation}
	\label{eq:a+a_operator_q}
	\begin{aligned}
		\hat{n}_{j} = \frac{i}{\sqrt{2}} \left(\frac{E_{Jj}}{8E_{C,Q_j}}\right)^{\frac{1}{4}} ( \hat{b}^{\dagger}_{j} - \hat{b}_{j}) ,\\
		\hat{\varphi}_{j}  = \frac{1}{\sqrt{2}} \left(\frac{8E_{C,{Q_j}}}{E_{Jj}}\right)^{\frac{1}{4}} (\hat{b}^{\dagger}_{j} + \hat{b}_{j}).
	\end{aligned}
\end{equation}
Truncating the qubit potential to the fourth order and nelecting double-excitation terms ($\hat{b}^{\dagger}_{j} \hat{a}^{\dagger}_{m} + \hat{b}_{j} \hat{a}_{m}$), the capacitive-coupling Hamiltonian takes the form
\begin{equation}
	\label{eq:H_QM6}
	\begin{aligned}
		\hat{\mathcal{H}}_m = &\sum_{j=1,2} \hbar \omega_{j} \hat{b}^{\dagger}_{j} \hat{b}_{j}-\frac{E_{C,Q_j}}{2}\hat{b}^{\dagger}_{j} \hat{b}^{\dagger}_{j} \hat{b}_{j} \hat{b}_{j} \\
		+ & \hbar \omega_{C,m}\hat{a}^{\dagger}_m \hat{a}_m \!- \! \frac{\mathcal{K}_{mm}}{2}\hat{a}^{\dagger}_m \hat{a}^{\dagger}_m \hat{a}_m \hat{a}_m  \\
		+ & g_{1m}\left(\hat{b}^{\dagger}_{1} \hat{a}_m+\hat{b}_{1}\hat{a}^{\dagger}_m \right)+ g_{2m} \left(\hat{b}^{\dagger}_{2} \hat{a}_m + \hat{b}_{2} \hat{a}^{\dagger}_m \right)\\
		+ & g_{q1q2}\left(\hat{b}^{\dagger}_{1} \hat{b}_{2}+\hat{b}_{2} \hat{b}^{\dagger}_{1} \right)	
	\end{aligned}
\end{equation}
where $\hbar \omega_{j}=\sqrt{8E_{C,Q_j}E_{Jj}}-E_{C,Q_j}$ is the qubit transition frequency.
The coupling strengths are given by
\begin{equation}
	\label{eq:g_jc}
	\begin{aligned}
		g_{jm} = \frac{E_{jm}}{\sqrt{2}} \left(\frac{E_{Jj}E_{L,m}}{E_{C,Q_j} E^{'}_{C,m}}\right)^{\frac{1}{4}}.\\
	\end{aligned}
\end{equation}

We now turn to the experimental sample.
The qubits are composed of two identical Josephson junctions shunted by a capacitor $C_J$.
At sweet spots, they exhibit transition frequencies of $\omega_1/2\pi = 6.505$ GHz and $\omega_2/2\pi = 6.322$ GHz,
with Charge energies of $E_{C,Q_1}/2\pi = 0.222$ GHz and $E_{C,Q_2}/2\pi = 0.196$ GHz.
The measured relaxation times are $T_{1,Q_1} = 2.1~\mu\mathrm{s}$ and $T_{1,Q_2} = 5.3~\mu\mathrm{s}$,
while Ramsey interference yields dephasing times of $T_{2,Q_1}^{\star} = 1.2~\mu\mathrm{s}$ and $T_{2,Q_2}^{\star} = 0.8~\mu\mathrm{s}$ under $\omega_1/2\pi = 6.334$ GHz and $\omega_2/2\pi = 6.145$ GHz.
The limited coherence may be attributed to high transition frequencies, which enhance sensitivity to both the Purcell effect and stray photon noise.

First, we experimentally determine the coupling term $g_{jm}$ through spectroscopy measurments, fitting the avoided crossing results between the transmon transition frequency $\omega_{j}$ and the coupler modes $\nu_m$ near resonance, as shown in Fig. \ref{fig:figure3}(a) and (b).
The solid black lines represent the fitting curves used to extract the coupling strengths. 
Thus, we obtain $|g_{21}|/2\pi = 160$ MHz and $|g_{11}|/2\pi = 50$ MHz.
The coupling $g_{12}$ is derived from the expression $J_{m2r1} \simeq g_{12} g_{q1r1}[1/(\omega_{r1}-\omega_{1})+1/(\nu_{2}-\omega_{1})]$ in the dispersive regime.
By fitting the avoided crossing results between the $Q_1$ readout resonator and $\nu_2$ at the resonant frequency $6.954$ GHz, as shown in Fig. \ref{fig:figure3} (c), we obtain $J_{m2r1}/2\pi = 21.04$ MHz.
The coupling $g_{q1r1}$ is derived from the dispersive shift $\chi = g_{q1r1}/(\omega_{1}-\omega_{r1})$ of the $Q_1$ readout resonator.
Considering the shift $\chi/2\pi = 10.9$ MHz directly measured in experiment, we find $|g_{12}|/2\pi = 135.18$ MHz, under the condition $\nu_{2}/2\pi = 6.954$ GHz and $\omega_1/2\pi = 6.504$ GHz.
Here, $\omega_{r1}$ is the bare frequency of the $Q_1$ readout resonator.
Meanwhile, the analysis of Eqs. \eqref{eq:H_QM5} and \eqref{eq:g_jc}, together with the results shown in Fig.~\ref{fig:figure2}(b), further indicates $|g_{22}|/2\pi = 12.6$ MHz.

\begin{figure}[hbtp!]
	\centering
	\includegraphics[width=0.45\textwidth]{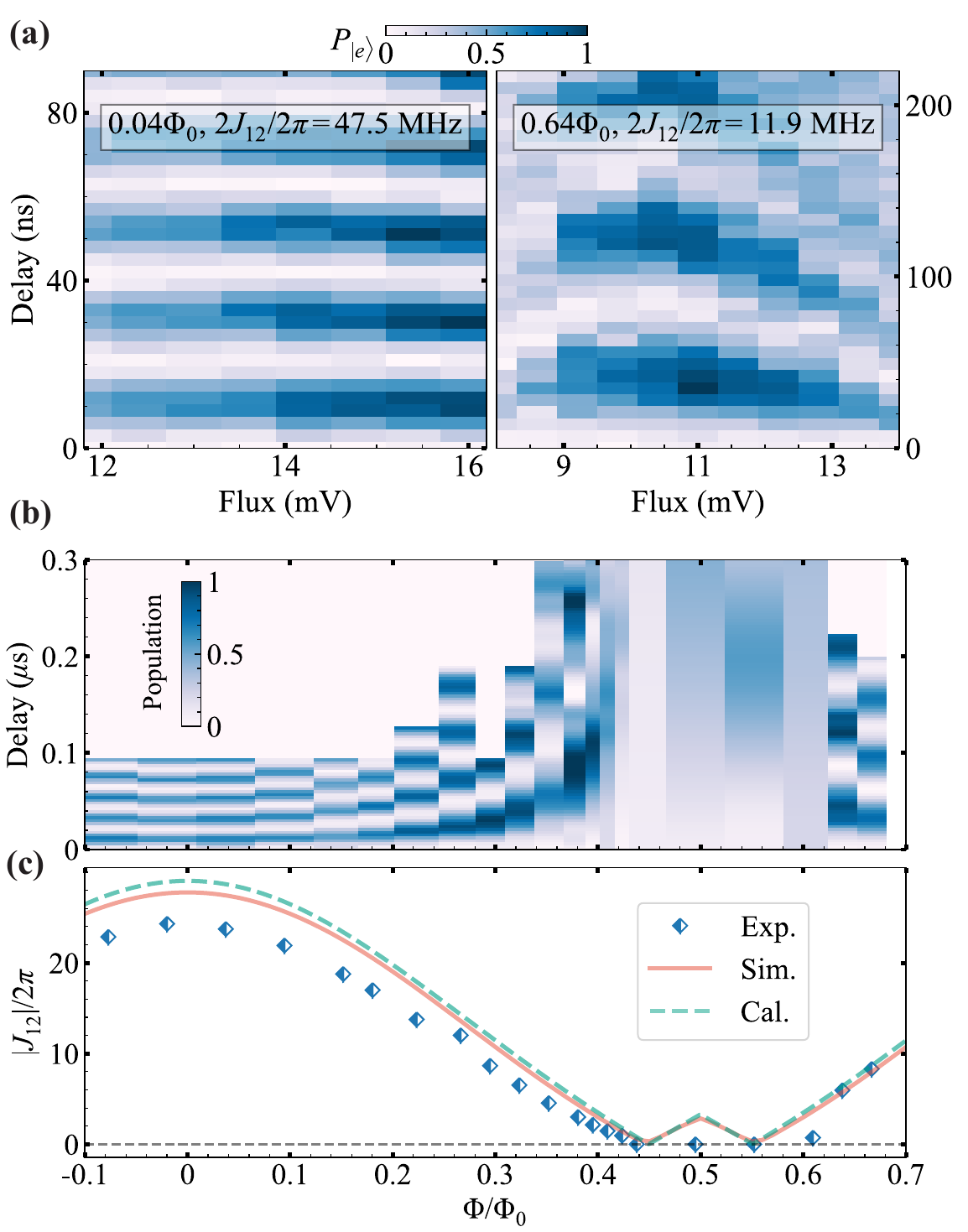}
    \caption{\justifying
    Experiemtally measured $XX$ interactions between qubits.
	(a) Chevron pattern of vacuum Rabi oscillations. The left qubit is fixed at $\omega_1/2\pi = 5.5$ GHz, while the right qubit frequency is tuned via its flux bias.
    	Left panel: The coupler is biased at $\Phi/\Phi_0 = 0.04$, with mode frequencies $\nu_1/2\pi = 5.012$ GHz and $\nu_2/2\pi = 7.937$ GHz.
    	Right panel: The coupler is biased at $\Phi/\Phi_0 = 0.64$, with mode frequencies $\nu_1/2\pi = 3.907$ GHz and $\nu_2/2\pi = 7.054$ GHz.
	The trajectory of the slowest oscillation in the Chevron pattern, plotted versus $\Phi$, is shown in (b) to clarify the distinct exchange speeds.
  (b) The $XX$ strength is extracted by fitting the vacuum Rabi oscillations.
     A strong coupling strength reaching $24.3$ MHz ($27.8$ MHz) is achieved in experiment (simulation), while the minimum coupling is $0.0$ MHz (-2.8 MHz) at $\Phi/\Phi_0 = 0.5$, yielding a switching ratio.
    Markers represent experimental data,
    the dashed line denotes calculations based on the two-mode coupling model Eq. \eqref{eq:H_eff} in the dispersive regime, and the pink line corresponds to numerical simulations.
  }
    \label{fig:figure5}
\end{figure}

It is noteworthy that the higher-order modes in a $1$-cm coupling length characterize higher frequencies, such as $\omega_{C,3}/2\pi>13$ GHz in simulations, thus restricting our description to the two lowest modes and establishing an effective dual-path coupling model.
The model we used in experiment can be equivalent to the schematic diagram of the coupling architecture between two transmon qubits, as shown in Fig. \ref{fig:figure4}(a).
For theoretical modeling, we approximate all elements as two-level systems via the truncated Hamiltonian
\begin{equation}\label{eq:H_2level}
    \begin{aligned}    
      H =  &\sum_{j=1,2}\frac{\omega_j}{2}\sigma_j^z + \sum_{m=1,2}\frac{\nu_m}{2}\sigma_m^z + \sum_{\substack{j=1,2 \\ m=1,2}} g_{jm}(\sigma_j^{+}\sigma_m^{-} + \mathrm{H.c.})
    \end{aligned}
\end{equation}
where H.c. denotes Hermitian conjungate, the $\sigma_{\lambda}^{z}$, $\sigma_{\lambda}^{+}$, and $\sigma_{\lambda}^{-}$ ($\lambda \in \{j,m\}$) denote the Pauli-$Z$, raising, and lowering operators.
The coupler's electric field spatial envelope versus flux bias $\Phi/\Phi_0$ as shown in Fig.~\ref{fig:figure4}(b).
Josephson junction modulation of field amplitude and phase determines transmon-mode coupling strength, with flux engineering providing precise field synchronization control.
Theoretical coupling strengths, as shown in Fig.~\ref{fig:figure4}(c), derived from Eq.~\eqref{eq:g_jc} and the field-phase relation in (b), exhibit flux-tunable modulation.

\section{Demonstration of the tunable coupling strength}\label{sec3}
In our configuration, the qubits exhibit opposite-sign detunings relative to the coupler modes, such that $\Delta_{j1} \equiv \omega_j - \nu_1 > 0$ and $\Delta_{j2} \equiv \omega_j - \nu_2 < 0$.
Both couplings operate in the dispersive regime, where $g_{jm} \ll |\Delta_{jm}|$.
Under the Schrieffer-Wolff transformation
$
U = \exp \left[ \sum_{\substack{j=1,2 \\ m=1,2}} \frac{g_{jm}}{\Delta_{jm}} \left( \sigma_j^{+} \sigma_m^{-} - \sigma_j^{-} \sigma_m^{+}\right) \right]
$,
the effective two-qubit Hamiltonian can be written as
\begin{equation}\label{eq:H_eff}
H_\mathrm{eff} = \sum_{j=1,2} \frac{\tilde{\omega}_j}{2} \sigma_j^z + J_\mathrm{12} \left( \sigma_1^+ \sigma_2^- + \mathrm{H.c.} \right)\!,
\end{equation}
where $J_\mathrm{12} = \sum_{m=1,2}g_{1m} g_{2m} \left( \Delta_{1m}^{-1} + \Delta_{2m}^{-1} \right)$ denotes the total effective qubit-qubit coupling, and $\tilde{\omega}_j = \omega_j + \sum_{m=1,2} g_{jm}^2 /(\Delta_{1m}+ \Delta_{2m})$ is the Lamb-shifted qubit frequency. Throughout, we assume the coupler mode remains in its ground state.

Unlike conventional tunable couplers, $J_\mathrm{12}$ depends on the external flux $\Phi$ through both the coupler's frequency $\omega_m(\Phi)$ and the qubit-coupler couplings $g_{jm}(\Phi)$. Crucially, while $\Delta_{1m} > 0$ and $\Delta_{2m} < 0$ yield opposing signs for the individual coupling terms, the condition $g_{1m}g_{2m} > 0$ ensures $J_\mathrm{12} > 0$ across all flux. Notably, the qubit-coupler coupling strength can vary significantly and even undergo sign reversal due to abrupt electric-field changes at the coupler Josephson junction, which modulates the port voltage. Consequently, adjusting $\Phi$ enables a high contrast ratio for $J_\mathrm{12}$, analogous to flux-tunable coupling schemes \cite{yan2018,Sete2021,li2024}.

\begin{figure}[hbtp!]
	\centering
	\includegraphics[width=0.475\textwidth]{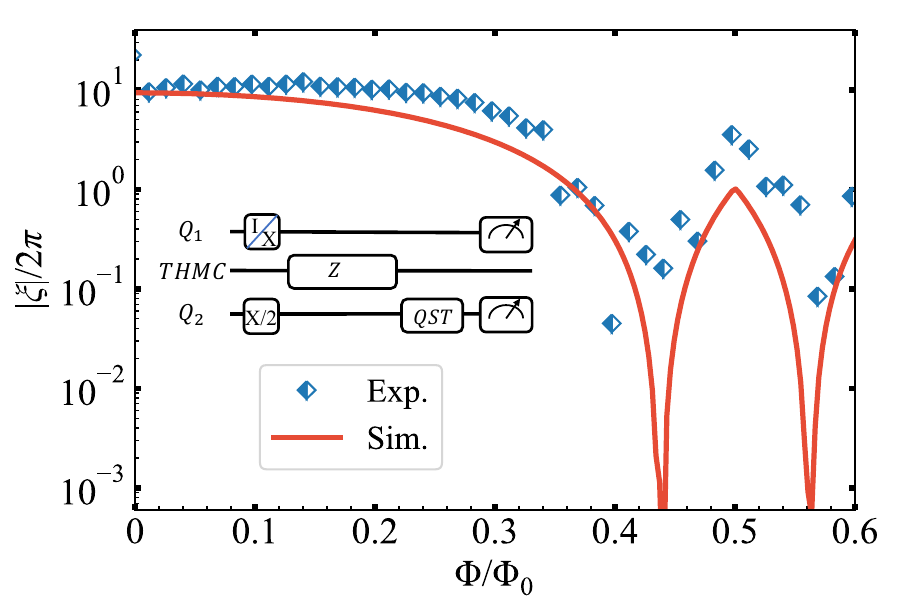}
    \caption{\justifying
    Experimentally measured $ZZ$ interaction strength $\xi$ using the conventional method, where the controlled qubit $Q_1$ is prepared in either ground or excited state, while the target qubit $Q_2$ is in a superpostion state.
			The controlled phase was then measured via quantum state tomography ($QST$).
			The experimental condition is $\omega_1/2\pi=6.293$ GHz, $\omega_2/2\pi=6.152$ GHz, and $\Delta_{12}/2\pi=141$ MHz.
			The red line shows simulation results that are in agreement with the experimental data.
	}
    \label{fig:figure6_exp}
\end{figure}

To characterize the tunability of the coupler, we measured the effective coupling strength $J_{12}$ as a function of the coupler flux bias.
The coupler enables continuous tuning of $J_{12}$ from a maximum value exceeding 23 (27) MHz down to 0 (-2.8) MHz in experiment (simulation), demonstrating performance comparable to conventional tunable couplers.
For instance, at a coupler bias of $\Phi/\Phi_0 = 0.04$, Fig. \ref{fig:figure5}(a) shows Chevron pattern between the excited states of the qubits, resulting in resonant vacuum Rabi oscillations. The $XX$ interaction strength $2J_{12}/2\pi = 47.5$ MHz is extracted by fitting the oscillation pattern, where the frequency follows $\sqrt{\Delta_{12}^2 + J_{12}^2}$, with $\Delta_{12}$ denoting the detuning between the transmon qubits. 
Similarly, when the coupler bias is set to $\Phi/\Phi_0 = 0.64$, we obtain $2J_{12}/2\pi = 11.9$ MHz.
Theoretical calculations of $J_{12}$, based on Eq.~\eqref{eq:H_eff}, exhibit in agreement with the experimental data, as shown by the close correspondence between simulation and measurement in Fig. \ref{fig:figure5}(b), thus validating the accuracy of the underlying model.
We note that conventional tunable couplers, such as those discussed in previous works \cite{yan2018, Sete2021},
typically achieve a zero $XX$ interaction, a feature that our scheme does fully reach.
In pratical, the residual $XX$ interaction can also be mitigated by introducing detuning between qubits,
which supresses unwanted couplings.

Next, we experimentally demonstrated that the $ZZ$ interaction can be obtained via the adiabatic evolution method.
The full Hamiltonian $H$ is considered in its instantaneous energy eigenbasis, corresponding to the adiabatic states that depend on the coupler modes.
Thus, the $ZZ$ interaction $\xi$ can be defined as $\xi = E_{|\overline{1001}\rangle}+E_{|\overline{0000}\rangle}-E_{|\overline{1000}\rangle}-E_{|\overline{0001}\rangle}$
where $|\overline{q_1 m_1 m_2 q_2}\rangle$ denotes the addressed state corresponding to the product state $|q_1 m_1 m_2 q_2 \rangle$ of the coupler and qubits,
$E_{\overline{|q_1 m_1 m_2 q_2 \rangle}}$ is the corresponding eigenenergy under adiabatic evolution.
These energies and interactions are parameterized by the coupler flux $\Phi$.

As a conventional method for calibrating the $ZZ$ interaction, as shown in Fig. \ref{fig:figure6_exp}, the target qubit $Q_1$ is prepared in a superposition state via $X/2$ gate, while the control qubit $Q_2$ is set in either the ground ($I$ gate) or excited state ($X$ gate).
The phase accumulated during a fixed evolution time is quantified by measuring the phase shift using quantum state tomography, which enables the extraction of the controlled phase by comparing results across different states of $Q_1$.
The experimental parameters are $\omega_1/2\pi = 6.293$ GHz, $\omega_2/2\pi = 6.152$ GHz, and $\Delta_{12}/2\pi = 141$ MHz, less than the qubit anharmonicities, with $E_{C,Q_1}/2\pi = 0.222$ GHz and $E_{C,Q_2}/2\pi = 0.196$ GHz.
Simultaneously, simulations under these experimental conditions, represented by the red line, are in agreement with the experimental data, confirming that the theoretical scheme can describe the $ZZ$ interaction strength.

It is noteworthy that the results obtained under these conditions are suboptimal,
with the residual $ZZ$ interaction remaining relatively high.
For instance, when $\Phi \notin (0.42, 0.46)$, the residual interaction is more than 0.1 MHz.
This performance likely reflects limitations of the current experimental setup, which is unable to achieve more optimal conditions, primarily due to significant decoherence during qubit biasing.
The decoherence is attributeed to flux noise in the experimental environment and fabrication-induced deviations in the qubit Josephson energy.
The latter necessitates biasing the qubit to the condition, corresponding to a noise-sensitive point, thereby constraining the ability to perform controlled phase manipulations.
To address these challenges, we proceed with a theoretical analysis in Section~\ref{sec4},
aimed at identifying improved conditions that would enhance the $ZZ$ interaction while simultaneously reducing the residual interaction.

\begin{figure}[hbtp!]
	\centering
	\includegraphics[width=0.45\textwidth]{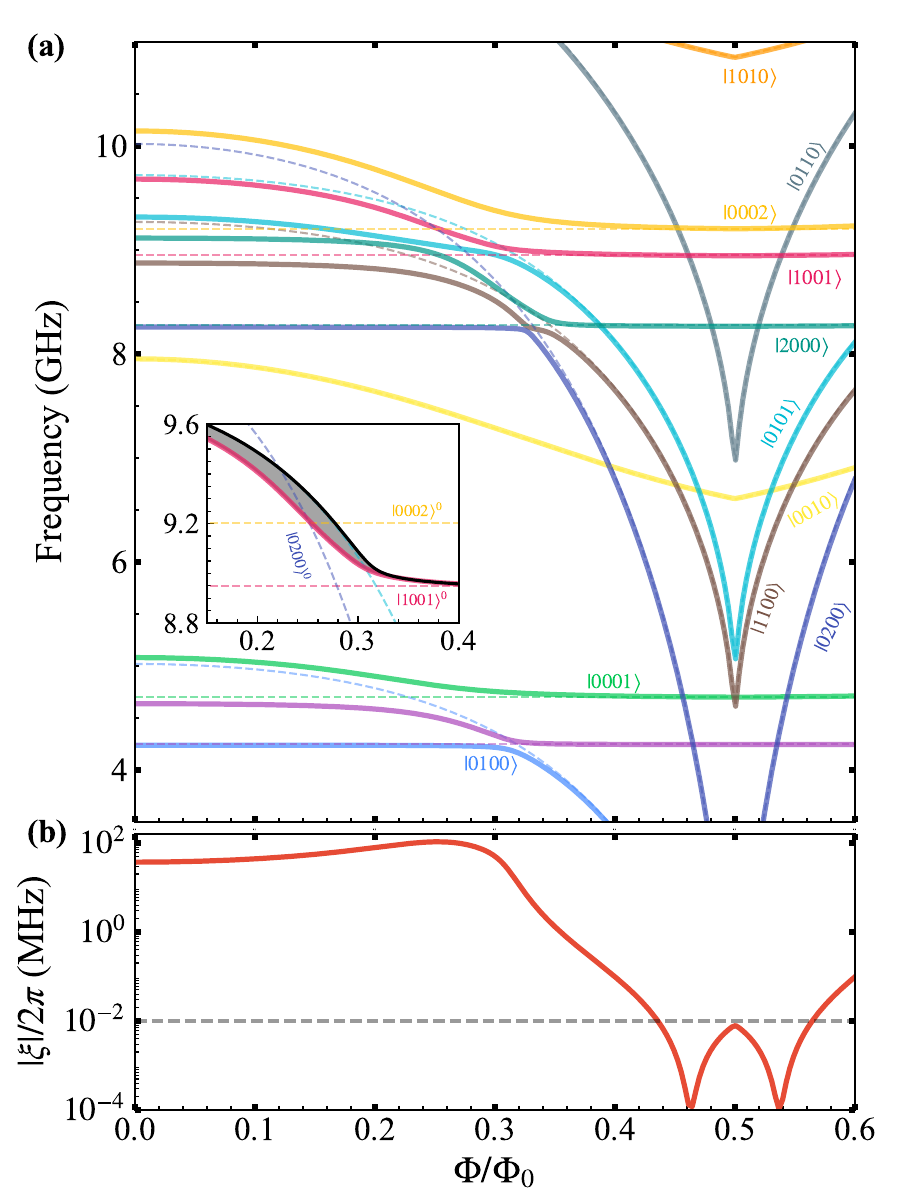}
    \caption{\justifying
    (a) Energy spectrum of the truncated Hamiltonian.
	The eigenenergies calculated from Eq. \eqref{eq:H_QM6} are plotted as a function of external flux. Solid curves represent the energy spectrum of adiabatic states as the first mode frequency of the coupler increases, while dashed curves denote non-adiabatic states.
	Multiple avoided crossings within the 2-subspace exhibit large $ZZ$ interactions.
	Inset: Comparison of the adiabatic $|1001\rangle$ state (red solid curve) and the sum frequency of $|1000\rangle$ and $|0001\rangle$ states (black solid curve).
	The shaded region highlights the interesting working range in Control-Z gate.
	The qubits are biased at $\omega_1/2\pi=4.25$ GHz, $\omega_2/2\pi=4.7$ GHz, and $\Delta_{12}/2\pi=450$ MHz.
    (b) $ZZ$ interaction strength $\xi$ as a function of $\Phi/\Phi_0$, where circuit parameters are identical to (a).
    A small interaction ($|\xi|/2\pi \approx 10^{-4}$ MHz) at $\Phi/\Phi_0=0.4643$ and maximum interaction strength ($|\xi|/2\pi \approx 106$ MHz) at $\Phi/\Phi_0=0.2533$, confirming a high contrast in our scheme.
}
    \label{fig:figure6}
\end{figure}

\begin{figure}[hbtp!]
	\centering
	\includegraphics[width=0.475\textwidth]{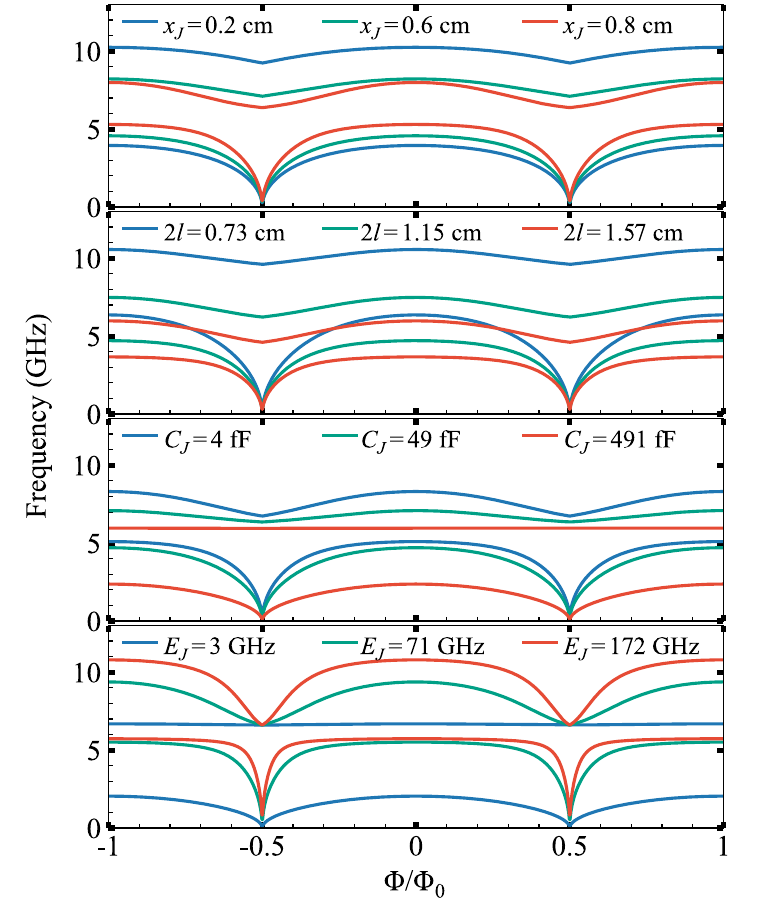}
    \caption{\justifying
    The two lowest spectra of the coupler are shown as a function of position $x_J$, length $l$, shunt capacitance $C_J$, and Josephson energy $E_{J}$ in the panels from top to bottom.
    Three distinct parameter sets, highlighted by blue, green, and red, respectively, are selected to illustrate the variation of spectral features with flux.
    These results emphasize the need to optimize the parameters for different coupler lengths to achieve more optimal conditions in high-performance quantum manipulation.
  }
    \label{fig:figure7}
\end{figure}

\section{discussions}\label{sec4}
Here, we present the scheme for engineering appreciable $ZZ$ interactions.
Fig. \ref{fig:figure6}(a) shows the energy spectrum versus the coupler flux under the Hamiltonian of Eq. \eqref{eq:H_2level}.
The higher-order mode exhibits a smaller tuning range from 6.62 GHz to 7.95 GHz and remains in the dispersive coupling regime with the qubits. Consequently, its contribution to the effective coupling is equivalent to the direct coupling $g_{q1q2}$ in a conventional coupler scheme.
Simultaneously, when the external flux exceeds $\Phi/\Phi_0 = 0.4$, the lower-order mode is dispersively coupled to the qubits.

However, in the coupler flux range of $0.2 \Phi_0$ to $0.4 \Phi_0$, the lower-order mode have multiple avoided crossings with the qubits.
This strong interaction causes a significant deviation between the energy of the adiabatic state $|1001\rangle$ and the sum of the energies of the adiabatic states $|1000\rangle$ and $|0001\rangle$, as indicated by the shaded region in the inset.
Fig. \ref{fig:figure6}(b) directly plots the deviation, that is the $ZZ$ interaction strength $|\xi|/2\pi$, versus the coupler flux, where the adiabatic state $|0000\rangle$ is taken as the zero energy reference.
We observe that although there is no direct qubit-qubit coupling $g_{q1q2}$, the system still exhibits $ZZ$ suppression points due to the influence of the higher-order mode. At $\Phi/\Phi_0 = 0.4372$, the $ZZ$ coupling reaches an extremely low level $\mathcal{O}(10^{-2})$.
Conversely, at $\Phi/\Phi_0 = 0.2533$, the $ZZ$ coupling reaches a high level $\mathcal{O}(10^{2})$, primarily dominated by the lower-order mode.
In addition, the simultaneous suppression of both the $XX$ and $ZZ$ couplings is essential in experiemts.
To simultaneously eliminate these subtle residual interactions, a widely employed approach is to optimize the detuning between qubits, thereby suppressing the $XX$ interaction at the $ZZ$ minimum.

Therefore, 
regarding the $ZZ$ interaction, our work can achieve a tunability range comparable to conventional schemes.
For instance, in Ref. \cite{chu2021}, the $ZZ$ strength varies from approximately 10 kHz to above 100 MHz, offering a contrast of roughly $10^4$.
While it is possible to set $ZZ=0$ under specific conditions, achieving high-fidelity control remains challenging in such regimes.
In contrast, fixed-frequency couplers, as discussed in \cite{Kandala2021},
suffer from the lack of tunability.
Despite this, they can still enable high-fidelity quantum control through relatively high coherence times and coupling strengths in the MHz range, with a coupling-to-undesired-coupling ratio typically around 130.

An alternative approach for fixed-frequency couplers involves the use of resonator-induced phase gates \cite{Cross2015,Paik2016,Puri2016}.
This scheme is impressive for leveraging resonators, which naturally extend to connect distant qubits, thus facilitating strong $ZZ$ interactions.
This concept gave rise to our previous work, which demonstrated that the coupling distance can be extended to the sub-meter scale \cite{deng2025}.
A key feature of this scheme is that the photon lifetime in the resonator significantly impacts control performance. 
As discussed in this work, it is important to note that the present study does not rely on driving the coupler.
Instead, it functions like a conventional coupler, just with more tunable modes and lower anharmonicity.

Although these modes could be manipulated to below 3 GHz, corresponding to such as $\Phi/\Phi_0>0.4372$ with $|\xi| < 10^{-2}$ in Fig. 7(b), the thermal occupation remains below 0.01 at a 20 mK environment, which is a typical condition in superconducting qubit platforms.
The system behaves like a normal coupler, with thermal photons similarly acting as they would in standard coupler setups.
Therefore, we do not consider this limitation to be a significant concern for achieving the desired performance.

We next examine the influence of sample parameters on the coupler modes.
As shown in Fig.~\ref{fig:figure7}, the two lowest spectral modes of the coupler are plotted as a function of position $x_J$, length $l$, shunt capacitance $C_J$, and Josephson energy $E_{J}$, from top to bottom panels.
With increasing length, the mode gap, analogous to FSR, decreases, while the position $x_J$ influences the gap increase as approaching the coulpler center position.
The shunt capacitance mitigates nonlinear effects on higher-order modes, while the Josephson energy enhances the nonlinearity.
These results highlight the scalability, particularly with respect to crosstalk and frequency collisions in superconducting processors. 
The spatial extent of the coupler leads to a reduced mode gap, which can exacerbate crosstalk and frequency collisions between neighboring qubits and couplers.
Nevertheless, we believe that feasible solutions exist for addressing these challenges in various application scenarios.

For example, to mitigate frequency crowding in the manipulation regime, careful coordination of relevant parameters is required, as illustrated in Fig. \ref{fig:figure7}.
Specifically, appropriately selecting the junction position, shunt capacitance, and Josephson energy can alleviate this issue.
Additionally, unwanted coupling channels can be suppressed by introducing additional design elements \cite{qiu2025, zhong2019, niu2023, zhong2021}.
For instance, in our previous work \cite{deng2025}, we demonstrated how introducing two driving resonators could selectively address specific long-range coupler modes, thus mitigating leakages from low-mode resonances that might otherwise interfere with qubit controls.
Another strategy involves improving control techniques \cite{Martinis2014,zheng2022accelerated,xu2024} and leveraging the frequency distribution characteristics of long-range couplers to tailor control pulses more effectively for high-performance manipulations \cite{mollenhauer2024,song2024}.

It is note that the resonator-like coupler design itself is relatively simple and reproducible. This is important for scalability in fabrication.
Consequently, employing a variety of resonator-like couplers could mitigate the limited connectivity in superconducting qubits, enabling the construction of hardware-efficient processors capable of executing quantum tasks with improved efficiency.

\section{conclusions}\label{sec5}
To conclude, we have introduced a hybrid-mode coupler that enables flux-tunable coupling between transmons spaced at a centimeter-scale distance.
This architecture mediates strong $XX$ and $ZZ$ couplings, with experimental achieved 23 MHz and 10 MHz, respectively, in agreement with an effective two-channel model.
Theretical prediction suggests that $ZZ$ interactions can reach up to 100 MHz, with modulation contrasts exceeding $10^4$.
Our scheme thus promises an important component for modular superconducting quantum processors, where resonator-mediated coupling across different interaction ranges supports hardware-efficient quantum tasks.

\section*{Acknowledgements}
We are grateful to anonymous referees for their valuable comments, which greatly enhance the quality of this work.
We thank Wei Fang, Jiayu Ding and Orkesh Nurbolat for technical supports.
Additionally, we extend our appreciation to Peng Zhao for his insightful contributions during the discussions.
This work was partially supported by
the Innovation Program for Quantum Science and Technology (Grant Nos. 2021ZD0301702, 2024ZD0302000),
NSF of Jiangsu Province (Grant No. BK20232002),
NSFC (Grant Nos. U21A20436 and 12074179),
and
Natural Science Foundation of Shandong Province (Grant No. ZR2023LZH002).


\begin{thebibliography}{88}%
\makeatletter
\providecommand \@ifxundefined [1]{%
 \@ifx{#1\undefined}
}%
\providecommand \@ifnum [1]{%
 \ifnum #1\expandafter \@firstoftwo
 \else \expandafter \@secondoftwo
 \fi
}%
\providecommand \@ifx [1]{%
 \ifx #1\expandafter \@firstoftwo
 \else \expandafter \@secondoftwo
 \fi
}%
\providecommand \natexlab [1]{#1}%
\providecommand \enquote  [1]{``#1''}%
\providecommand \bibnamefont  [1]{#1}%
\providecommand \bibfnamefont [1]{#1}%
\providecommand \citenamefont [1]{#1}%
\providecommand \href@noop [0]{\@secondoftwo}%
\providecommand \href [0]{\begingroup \@sanitize@url \@href}%
\providecommand \@href[1]{\@@startlink{#1}\@@href}%
\providecommand \@@href[1]{\endgroup#1\@@endlink}%
\providecommand \@sanitize@url [0]{\catcode `\\12\catcode `\$12\catcode `\&12\catcode `\#12\catcode `\^12\catcode `\_12\catcode `\%12\relax}%
\providecommand \@@startlink[1]{}%
\providecommand \@@endlink[0]{}%
\providecommand \url  [0]{\begingroup\@sanitize@url \@url }%
\providecommand \@url [1]{\endgroup\@href {#1}{\urlprefix }}%
\providecommand \urlprefix  [0]{URL }%
\providecommand \Eprint [0]{\href }%
\providecommand \doibase [0]{https://doi.org/}%
\providecommand \selectlanguage [0]{\@gobble}%
\providecommand \bibinfo  [0]{\@secondoftwo}%
\providecommand \bibfield  [0]{\@secondoftwo}%
\providecommand \translation [1]{[#1]}%
\providecommand \BibitemOpen [0]{}%
\providecommand \bibitemStop [0]{}%
\providecommand \bibitemNoStop [0]{.\EOS\space}%
\providecommand \EOS [0]{\spacefactor3000\relax}%
\providecommand \BibitemShut  [1]{\csname bibitem#1\endcsname}%
\let\auto@bib@innerbib\@empty
\bibitem [{\citenamefont {You}\ and\ \citenamefont {Nori}(2011)}]{you2011}%
  \BibitemOpen
  \bibfield  {author} {\bibinfo {author} {\bibfnamefont {J.~Q.}\ \bibnamefont {You}}\ and\ \bibinfo {author} {\bibfnamefont {F.}~\bibnamefont {Nori}},\ }\bibfield  {title} {\bibinfo {title} {Atomic physics and quantum optics using superconducting circuits},\ }\href {https://doi.org/10.1038/nature10122} {\bibfield  {journal} {\bibinfo  {journal} {Nature}\ }\textbf {\bibinfo {volume} {474}},\ \bibinfo {pages} {589} (\bibinfo {year} {2011})}\BibitemShut {NoStop}%
\bibitem [{\citenamefont {Krantz}\ \emph {et~al.}(2019)\citenamefont {Krantz}, \citenamefont {Kjaergaard}, \citenamefont {Yan}, \citenamefont {Orlando}, \citenamefont {Gustavsson},\ and\ \citenamefont {Oliver}}]{krantz2019}%
  \BibitemOpen
  \bibfield  {author} {\bibinfo {author} {\bibfnamefont {P.}~\bibnamefont {Krantz}}, \bibinfo {author} {\bibfnamefont {M.}~\bibnamefont {Kjaergaard}}, \bibinfo {author} {\bibfnamefont {F.}~\bibnamefont {Yan}}, \bibinfo {author} {\bibfnamefont {T.~P.}\ \bibnamefont {Orlando}}, \bibinfo {author} {\bibfnamefont {S.}~\bibnamefont {Gustavsson}},\ and\ \bibinfo {author} {\bibfnamefont {W.~D.}\ \bibnamefont {Oliver}},\ }\bibfield  {title} {\bibinfo {title} {A quantum engineer's guide to superconducting qubits},\ }\bibfield  {journal} {\bibinfo  {journal} {Appl. Phys. Rev.}\ }\textbf {\bibinfo {volume} {6}},\ \href {https://doi.org/10.1063/1.5089550} {10.1063/1.5089550} (\bibinfo {year} {2019})\BibitemShut {NoStop}%
\bibitem [{\citenamefont {Blais}\ \emph {et~al.}(2021)\citenamefont {Blais}, \citenamefont {Grimsmo}, \citenamefont {Girvin},\ and\ \citenamefont {Wallraff}}]{Blais2021}%
  \BibitemOpen
  \bibfield  {author} {\bibinfo {author} {\bibfnamefont {A.}~\bibnamefont {Blais}}, \bibinfo {author} {\bibfnamefont {A.~L.}\ \bibnamefont {Grimsmo}}, \bibinfo {author} {\bibfnamefont {S.~M.}\ \bibnamefont {Girvin}},\ and\ \bibinfo {author} {\bibfnamefont {A.}~\bibnamefont {Wallraff}},\ }\bibfield  {title} {\bibinfo {title} {Circuit quantum electrodynamics},\ }\href {https://doi.org/10.1103/RevModPhys.93.025005} {\bibfield  {journal} {\bibinfo  {journal} {Rev. Mod. Phys.}\ }\textbf {\bibinfo {volume} {93}},\ \bibinfo {pages} {025005} (\bibinfo {year} {2021})}\BibitemShut {NoStop}%
\bibitem [{\citenamefont {Koch}\ \emph {et~al.}(2007)\citenamefont {Koch}, \citenamefont {Yu}, \citenamefont {Gambetta}, \citenamefont {Houck}, \citenamefont {Schuster}, \citenamefont {Majer}, \citenamefont {Blais}, \citenamefont {Devoret}, \citenamefont {Girvin},\ and\ \citenamefont {Schoelkopf}}]{Koch2007}%
  \BibitemOpen
  \bibfield  {author} {\bibinfo {author} {\bibfnamefont {J.}~\bibnamefont {Koch}}, \bibinfo {author} {\bibfnamefont {T.~M.}\ \bibnamefont {Yu}}, \bibinfo {author} {\bibfnamefont {J.}~\bibnamefont {Gambetta}}, \bibinfo {author} {\bibfnamefont {A.~A.}\ \bibnamefont {Houck}}, \bibinfo {author} {\bibfnamefont {D.~I.}\ \bibnamefont {Schuster}}, \bibinfo {author} {\bibfnamefont {J.}~\bibnamefont {Majer}}, \bibinfo {author} {\bibfnamefont {A.}~\bibnamefont {Blais}}, \bibinfo {author} {\bibfnamefont {M.~H.}\ \bibnamefont {Devoret}}, \bibinfo {author} {\bibfnamefont {S.~M.}\ \bibnamefont {Girvin}},\ and\ \bibinfo {author} {\bibfnamefont {R.~J.}\ \bibnamefont {Schoelkopf}},\ }\bibfield  {title} {\bibinfo {title} {Charge-insensitive qubit design derived from the cooper pair box},\ }\href {https://doi.org/10.1103/PhysRevA.76.042319} {\bibfield  {journal} {\bibinfo  {journal} {Phys. Rev. A}\ }\textbf {\bibinfo {volume} {76}},\ \bibinfo {pages} {042319} (\bibinfo {year} {2007})}\BibitemShut {NoStop}%
\bibitem [{\citenamefont {Chen}\ \emph {et~al.}(2014)\citenamefont {Chen}, \citenamefont {Neill}, \citenamefont {Roushan}, \citenamefont {Leung}, \citenamefont {Fang}, \citenamefont {Barends}, \citenamefont {Kelly}, \citenamefont {Campbell}, \citenamefont {Chen}, \citenamefont {Chiaro}, \citenamefont {Dunsworth}, \citenamefont {Jeffrey}, \citenamefont {Megrant}, \citenamefont {Mutus}, \citenamefont {O'Malley}, \citenamefont {Quintana}, \citenamefont {Sank}, \citenamefont {Vainsencher}, \citenamefont {Wenner}, \citenamefont {White}, \citenamefont {Geller}, \citenamefont {Cleland},\ and\ \citenamefont {Martinis}}]{chen2014}%
  \BibitemOpen
  \bibfield  {author} {\bibinfo {author} {\bibfnamefont {Y.}~\bibnamefont {Chen}}, \bibinfo {author} {\bibfnamefont {C.}~\bibnamefont {Neill}}, \bibinfo {author} {\bibfnamefont {P.}~\bibnamefont {Roushan}}, \bibinfo {author} {\bibfnamefont {N.}~\bibnamefont {Leung}}, \bibinfo {author} {\bibfnamefont {M.}~\bibnamefont {Fang}}, \bibinfo {author} {\bibfnamefont {R.}~\bibnamefont {Barends}}, \bibinfo {author} {\bibfnamefont {J.}~\bibnamefont {Kelly}}, \bibinfo {author} {\bibfnamefont {B.}~\bibnamefont {Campbell}}, \bibinfo {author} {\bibfnamefont {Z.}~\bibnamefont {Chen}}, \bibinfo {author} {\bibfnamefont {B.}~\bibnamefont {Chiaro}}, \bibinfo {author} {\bibfnamefont {A.}~\bibnamefont {Dunsworth}}, \bibinfo {author} {\bibfnamefont {E.}~\bibnamefont {Jeffrey}}, \bibinfo {author} {\bibfnamefont {A.}~\bibnamefont {Megrant}}, \bibinfo {author} {\bibfnamefont {J.~Y.}\ \bibnamefont {Mutus}}, \bibinfo {author} {\bibfnamefont {P.~J.~J.}\ \bibnamefont {O'Malley}}, \bibinfo {author} {\bibfnamefont {C.~M.}\ \bibnamefont {Quintana}}, \bibinfo {author} {\bibfnamefont {D.}~\bibnamefont {Sank}}, \bibinfo {author} {\bibfnamefont {A.}~\bibnamefont {Vainsencher}}, \bibinfo {author} {\bibfnamefont {J.}~\bibnamefont {Wenner}}, \bibinfo {author} {\bibfnamefont {T.~C.}\ \bibnamefont {White}}, \bibinfo {author} {\bibfnamefont {M.~R.}\ \bibnamefont {Geller}}, \bibinfo {author} {\bibfnamefont {A.~N.}\ \bibnamefont {Cleland}},\ and\ \bibinfo {author} {\bibfnamefont {J.~M.}\ \bibnamefont {Martinis}},\ }\bibfield  {title} {\bibinfo {title} {Qubit architecture with high coherence and fast tunable coupling},\ }\href {https://doi.org/10.1103/PhysRevLett.113.220502} {\bibfield  {journal} {\bibinfo  {journal} {Phys. Rev. Lett.}\ }\textbf {\bibinfo {volume} {113}},\ \bibinfo {pages} {220502} (\bibinfo {year} {2014})},\ \Eprint {https://arxiv.org/abs/1402.7367} {arXiv:1402.7367 [cond-mat, physics:quant-ph]} \BibitemShut {NoStop}%
\bibitem [{\citenamefont {Kurpiers}\ \emph {et~al.}(2018)\citenamefont {Kurpiers}, \citenamefont {Magnard}, \citenamefont {Walter}, \citenamefont {Royer}, \citenamefont {Pechal}, \citenamefont {Heinsoo}, \citenamefont {Salath{\'e}}, \citenamefont {Akin}, \citenamefont {Storz}, \citenamefont {Besse} \emph {et~al.}}]{kurpiers2018}%
  \BibitemOpen
  \bibfield  {author} {\bibinfo {author} {\bibfnamefont {P.}~\bibnamefont {Kurpiers}}, \bibinfo {author} {\bibfnamefont {P.}~\bibnamefont {Magnard}}, \bibinfo {author} {\bibfnamefont {T.}~\bibnamefont {Walter}}, \bibinfo {author} {\bibfnamefont {B.}~\bibnamefont {Royer}}, \bibinfo {author} {\bibfnamefont {M.}~\bibnamefont {Pechal}}, \bibinfo {author} {\bibfnamefont {J.}~\bibnamefont {Heinsoo}}, \bibinfo {author} {\bibfnamefont {Y.}~\bibnamefont {Salath{\'e}}}, \bibinfo {author} {\bibfnamefont {A.}~\bibnamefont {Akin}}, \bibinfo {author} {\bibfnamefont {S.}~\bibnamefont {Storz}}, \bibinfo {author} {\bibfnamefont {J.-C.}\ \bibnamefont {Besse}}, \emph {et~al.},\ }\bibfield  {title} {\bibinfo {title} {Deterministic quantum state transfer and remote entanglement using microwave photons},\ }\href {https://doi.org/https://doi.org/10.1038/s41586-018-0195-y} {\bibfield  {journal} {\bibinfo  {journal} {Nature}\ }\textbf {\bibinfo {volume} {558}},\ \bibinfo {pages} {264} (\bibinfo {year} {2018})}\BibitemShut {NoStop}%
\bibitem [{\citenamefont {Zhou}\ \emph {et~al.}(2023)\citenamefont {Zhou}, \citenamefont {Lu}, \citenamefont {Praquin}, \citenamefont {Chien}, \citenamefont {Kaufman}, \citenamefont {Cao}, \citenamefont {Xia}, \citenamefont {Mong}, \citenamefont {Pfaff}, \citenamefont {Pekker} \emph {et~al.}}]{zhou2023}%
  \BibitemOpen
  \bibfield  {author} {\bibinfo {author} {\bibfnamefont {C.}~\bibnamefont {Zhou}}, \bibinfo {author} {\bibfnamefont {P.}~\bibnamefont {Lu}}, \bibinfo {author} {\bibfnamefont {M.}~\bibnamefont {Praquin}}, \bibinfo {author} {\bibfnamefont {T.-C.}\ \bibnamefont {Chien}}, \bibinfo {author} {\bibfnamefont {R.}~\bibnamefont {Kaufman}}, \bibinfo {author} {\bibfnamefont {X.}~\bibnamefont {Cao}}, \bibinfo {author} {\bibfnamefont {M.}~\bibnamefont {Xia}}, \bibinfo {author} {\bibfnamefont {R.~S.}\ \bibnamefont {Mong}}, \bibinfo {author} {\bibfnamefont {W.}~\bibnamefont {Pfaff}}, \bibinfo {author} {\bibfnamefont {D.}~\bibnamefont {Pekker}}, \emph {et~al.},\ }\bibfield  {title} {\bibinfo {title} {Realizing all-to-all couplings among detachable quantum modules using a microwave quantum state router},\ }\href {https://doi.org/https://doi.org/10.1038/s41534-023-00723-7} {\bibfield  {journal} {\bibinfo  {journal} {npj Quantum Inf.}\ }\textbf {\bibinfo {volume} {9}},\ \bibinfo {pages} {54} (\bibinfo {year} {2023})}\BibitemShut {NoStop}%
\bibitem [{\citenamefont {Abrams}\ \emph {et~al.}(2020)\citenamefont {Abrams}, \citenamefont {Didier}, \citenamefont {Johnson}, \citenamefont {Silva},\ and\ \citenamefont {Ryan}}]{abrams2020}%
  \BibitemOpen
  \bibfield  {author} {\bibinfo {author} {\bibfnamefont {D.~M.}\ \bibnamefont {Abrams}}, \bibinfo {author} {\bibfnamefont {N.}~\bibnamefont {Didier}}, \bibinfo {author} {\bibfnamefont {B.~R.}\ \bibnamefont {Johnson}}, \bibinfo {author} {\bibfnamefont {M.~P.~D.}\ \bibnamefont {Silva}},\ and\ \bibinfo {author} {\bibfnamefont {C.~A.}\ \bibnamefont {Ryan}},\ }\bibfield  {title} {\bibinfo {title} {Implementation of {XY} entangling gates with a single calibrated pulse},\ }\href {https://doi.org/10.1038/s41928-020-00498-1} {\bibfield  {journal} {\bibinfo  {journal} {Nat. Electron.}\ }\textbf {\bibinfo {volume} {3}},\ \bibinfo {pages} {744} (\bibinfo {year} {2020})}\BibitemShut {NoStop}%
\bibitem [{\citenamefont {Nguyen}\ \emph {et~al.}(2024)\citenamefont {Nguyen}, \citenamefont {Kim}, \citenamefont {Hashim}, \citenamefont {Goss}, \citenamefont {Marinelli}, \citenamefont {Bhandari}, \citenamefont {Das}, \citenamefont {Naik}, \citenamefont {Kreikebaum}, \citenamefont {Jordan}, \citenamefont {Santiago},\ and\ \citenamefont {Siddiqi}}]{nguyen2024}%
  \BibitemOpen
  \bibfield  {author} {\bibinfo {author} {\bibfnamefont {L.~B.}\ \bibnamefont {Nguyen}}, \bibinfo {author} {\bibfnamefont {Y.}~\bibnamefont {Kim}}, \bibinfo {author} {\bibfnamefont {A.}~\bibnamefont {Hashim}}, \bibinfo {author} {\bibfnamefont {N.}~\bibnamefont {Goss}}, \bibinfo {author} {\bibfnamefont {B.}~\bibnamefont {Marinelli}}, \bibinfo {author} {\bibfnamefont {B.}~\bibnamefont {Bhandari}}, \bibinfo {author} {\bibfnamefont {D.}~\bibnamefont {Das}}, \bibinfo {author} {\bibfnamefont {R.~K.}\ \bibnamefont {Naik}}, \bibinfo {author} {\bibfnamefont {J.~M.}\ \bibnamefont {Kreikebaum}}, \bibinfo {author} {\bibfnamefont {A.~N.}\ \bibnamefont {Jordan}}, \bibinfo {author} {\bibfnamefont {D.~I.}\ \bibnamefont {Santiago}},\ and\ \bibinfo {author} {\bibfnamefont {I.}~\bibnamefont {Siddiqi}},\ }\bibfield  {title} {\bibinfo {title} {Programmable {Heisenberg} interactions between {Floquet} qubits},\ }\href {https://doi.org/10.1038/s41567-023-02326-7} {\bibfield  {journal} {\bibinfo  {journal} {Nat. Phys.}\ }\textbf {\bibinfo {volume} {20}},\ \bibinfo {pages} {240} (\bibinfo {year} {2024})}\BibitemShut {NoStop}%
\bibitem [{\citenamefont {Qiu}\ \emph {et~al.}(2025{\natexlab{a}})\citenamefont {Qiu}, \citenamefont {Zhang}, \citenamefont {Wang}, \citenamefont {Zhang}, \citenamefont {Zhou}, \citenamefont {Sun}, \citenamefont {Zhang}, \citenamefont {Linpeng}, \citenamefont {Liu}, \citenamefont {Niu}, \citenamefont {Zhong},\ and\ \citenamefont {Yu}}]{qiu2025a}%
  \BibitemOpen
  \bibfield  {author} {\bibinfo {author} {\bibfnamefont {J.}~\bibnamefont {Qiu}}, \bibinfo {author} {\bibfnamefont {Z.}~\bibnamefont {Zhang}}, \bibinfo {author} {\bibfnamefont {Z.}~\bibnamefont {Wang}}, \bibinfo {author} {\bibfnamefont {L.}~\bibnamefont {Zhang}}, \bibinfo {author} {\bibfnamefont {Y.}~\bibnamefont {Zhou}}, \bibinfo {author} {\bibfnamefont {X.}~\bibnamefont {Sun}}, \bibinfo {author} {\bibfnamefont {J.}~\bibnamefont {Zhang}}, \bibinfo {author} {\bibfnamefont {X.}~\bibnamefont {Linpeng}}, \bibinfo {author} {\bibfnamefont {S.}~\bibnamefont {Liu}}, \bibinfo {author} {\bibfnamefont {J.}~\bibnamefont {Niu}}, \bibinfo {author} {\bibfnamefont {Y.}~\bibnamefont {Zhong}},\ and\ \bibinfo {author} {\bibfnamefont {D.}~\bibnamefont {Yu}},\ }\href {https://doi.org/10.48550/arXiv.2503.01133} {\bibinfo {title} {A thermal-noise-resilient microwave quantum network traversing 4 {K}}} (\bibinfo {year} {2025}{\natexlab{a}}),\ \bibinfo {note} {arXiv:2503.01133 [quant-ph]}\BibitemShut {NoStop}%
\bibitem [{\citenamefont {Zhao}\ \emph {et~al.}(2020)\citenamefont {Zhao}, \citenamefont {Xu}, \citenamefont {Lan}, \citenamefont {Chu}, \citenamefont {Tan}, \citenamefont {Yu},\ and\ \citenamefont {Yu}}]{zhao2020}%
  \BibitemOpen
  \bibfield  {author} {\bibinfo {author} {\bibfnamefont {P.}~\bibnamefont {Zhao}}, \bibinfo {author} {\bibfnamefont {P.}~\bibnamefont {Xu}}, \bibinfo {author} {\bibfnamefont {D.}~\bibnamefont {Lan}}, \bibinfo {author} {\bibfnamefont {J.}~\bibnamefont {Chu}}, \bibinfo {author} {\bibfnamefont {X.}~\bibnamefont {Tan}}, \bibinfo {author} {\bibfnamefont {H.}~\bibnamefont {Yu}},\ and\ \bibinfo {author} {\bibfnamefont {Y.}~\bibnamefont {Yu}},\ }\bibfield  {title} {\bibinfo {title} {High-contrast $zz$ interaction using superconducting qubits with opposite-sign anharmonicity},\ }\href {https://doi.org/10.1103/PhysRevLett.125.200503} {\bibfield  {journal} {\bibinfo  {journal} {Phys. Rev. Lett.}\ }\textbf {\bibinfo {volume} {125}},\ \bibinfo {pages} {200503} (\bibinfo {year} {2020})}\BibitemShut {NoStop}%
\bibitem [{\citenamefont {Yan}\ \emph {et~al.}(2018)\citenamefont {Yan}, \citenamefont {Krantz}, \citenamefont {Sung}, \citenamefont {Kjaergaard}, \citenamefont {Campbell}, \citenamefont {Orlando}, \citenamefont {Gustavsson},\ and\ \citenamefont {Oliver}}]{yan2018}%
  \BibitemOpen
  \bibfield  {author} {\bibinfo {author} {\bibfnamefont {F.}~\bibnamefont {Yan}}, \bibinfo {author} {\bibfnamefont {P.}~\bibnamefont {Krantz}}, \bibinfo {author} {\bibfnamefont {Y.}~\bibnamefont {Sung}}, \bibinfo {author} {\bibfnamefont {M.}~\bibnamefont {Kjaergaard}}, \bibinfo {author} {\bibfnamefont {D.~L.}\ \bibnamefont {Campbell}}, \bibinfo {author} {\bibfnamefont {T.~P.}\ \bibnamefont {Orlando}}, \bibinfo {author} {\bibfnamefont {S.}~\bibnamefont {Gustavsson}},\ and\ \bibinfo {author} {\bibfnamefont {W.~D.}\ \bibnamefont {Oliver}},\ }\bibfield  {title} {\bibinfo {title} {Tunable coupling scheme for implementing high-fidelity two-qubit gates},\ }\href {https://doi.org/10.1103/PhysRevApplied.10.054062} {\bibfield  {journal} {\bibinfo  {journal} {Phys. Rev. Appl.}\ }\textbf {\bibinfo {volume} {10}},\ \bibinfo {pages} {054062} (\bibinfo {year} {2018})}\BibitemShut {NoStop}%
\bibitem [{\citenamefont {Sete}\ \emph {et~al.}(2021)\citenamefont {Sete}, \citenamefont {Chen}, \citenamefont {Manenti}, \citenamefont {Kulshreshtha},\ and\ \citenamefont {Poletto}}]{Sete2021}%
  \BibitemOpen
  \bibfield  {author} {\bibinfo {author} {\bibfnamefont {E.~A.}\ \bibnamefont {Sete}}, \bibinfo {author} {\bibfnamefont {A.~Q.}\ \bibnamefont {Chen}}, \bibinfo {author} {\bibfnamefont {R.}~\bibnamefont {Manenti}}, \bibinfo {author} {\bibfnamefont {S.}~\bibnamefont {Kulshreshtha}},\ and\ \bibinfo {author} {\bibfnamefont {S.}~\bibnamefont {Poletto}},\ }\bibfield  {title} {\bibinfo {title} {Floating tunable coupler for scalable quantum computing architectures},\ }\href {https://doi.org/10.1103/PhysRevApplied.15.064063} {\bibfield  {journal} {\bibinfo  {journal} {Phys. Rev. Appl.}\ }\textbf {\bibinfo {volume} {15}},\ \bibinfo {pages} {064063} (\bibinfo {year} {2021})}\BibitemShut {NoStop}%
\bibitem [{\citenamefont {Li}\ \emph {et~al.}(2024)\citenamefont {Li}, \citenamefont {Kubo}, \citenamefont {Ho}, \citenamefont {Yan}, \citenamefont {Nakamura},\ and\ \citenamefont {Goto}}]{li2024}%
  \BibitemOpen
  \bibfield  {author} {\bibinfo {author} {\bibfnamefont {R.}~\bibnamefont {Li}}, \bibinfo {author} {\bibfnamefont {K.}~\bibnamefont {Kubo}}, \bibinfo {author} {\bibfnamefont {Y.}~\bibnamefont {Ho}}, \bibinfo {author} {\bibfnamefont {Z.}~\bibnamefont {Yan}}, \bibinfo {author} {\bibfnamefont {Y.}~\bibnamefont {Nakamura}},\ and\ \bibinfo {author} {\bibfnamefont {H.}~\bibnamefont {Goto}},\ }\bibfield  {title} {\bibinfo {title} {Realization of high-fidelity cz gate based on a double-transmon coupler},\ }\href {https://doi.org/10.1103/PhysRevX.14.041050} {\bibfield  {journal} {\bibinfo  {journal} {Phys. Rev. X}\ }\textbf {\bibinfo {volume} {14}},\ \bibinfo {pages} {041050} (\bibinfo {year} {2024})}\BibitemShut {NoStop}%
\bibitem [{\citenamefont {Mundada}\ \emph {et~al.}(2019)\citenamefont {Mundada}, \citenamefont {Zhang}, \citenamefont {Hazard},\ and\ \citenamefont {Houck}}]{Mundada2019}%
  \BibitemOpen
  \bibfield  {author} {\bibinfo {author} {\bibfnamefont {P.}~\bibnamefont {Mundada}}, \bibinfo {author} {\bibfnamefont {G.}~\bibnamefont {Zhang}}, \bibinfo {author} {\bibfnamefont {T.}~\bibnamefont {Hazard}},\ and\ \bibinfo {author} {\bibfnamefont {A.}~\bibnamefont {Houck}},\ }\bibfield  {title} {\bibinfo {title} {Suppression of qubit crosstalk in a tunable coupling superconducting circuit},\ }\href {https://doi.org/10.1103/PhysRevApplied.12.054023} {\bibfield  {journal} {\bibinfo  {journal} {Phys. Rev. Appl.}\ }\textbf {\bibinfo {volume} {12}},\ \bibinfo {pages} {054023} (\bibinfo {year} {2019})}\BibitemShut {NoStop}%
\bibitem [{\citenamefont {Xu}\ \emph {et~al.}(2020)\citenamefont {Xu}, \citenamefont {Chu}, \citenamefont {Yuan}, \citenamefont {Qiu}, \citenamefont {Zhou}, \citenamefont {Zhang}, \citenamefont {Tan}, \citenamefont {Yu}, \citenamefont {Liu}, \citenamefont {Li} \emph {et~al.}}]{Xu2020}%
  \BibitemOpen
  \bibfield  {author} {\bibinfo {author} {\bibfnamefont {Y.}~\bibnamefont {Xu}}, \bibinfo {author} {\bibfnamefont {J.}~\bibnamefont {Chu}}, \bibinfo {author} {\bibfnamefont {J.}~\bibnamefont {Yuan}}, \bibinfo {author} {\bibfnamefont {J.}~\bibnamefont {Qiu}}, \bibinfo {author} {\bibfnamefont {Y.}~\bibnamefont {Zhou}}, \bibinfo {author} {\bibfnamefont {L.}~\bibnamefont {Zhang}}, \bibinfo {author} {\bibfnamefont {X.}~\bibnamefont {Tan}}, \bibinfo {author} {\bibfnamefont {Y.}~\bibnamefont {Yu}}, \bibinfo {author} {\bibfnamefont {S.}~\bibnamefont {Liu}}, \bibinfo {author} {\bibfnamefont {J.}~\bibnamefont {Li}}, \emph {et~al.},\ }\bibfield  {title} {\bibinfo {title} {High-fidelity, high-scalability two-qubit gate scheme for superconducting qubits},\ }\href {https://doi.org/10.1103/PhysRevLett.125.240503} {\bibfield  {journal} {\bibinfo  {journal} {Phys. Rev. Lett.}\ }\textbf {\bibinfo {volume} {125}},\ \bibinfo {pages} {240503} (\bibinfo {year} {2020})}\BibitemShut {NoStop}%
\bibitem [{\citenamefont {Sung}\ \emph {et~al.}(2021)\citenamefont {Sung}, \citenamefont {Ding}, \citenamefont {Braum\"uller}, \citenamefont {Veps\"al\"ainen}, \citenamefont {Kannan}, \citenamefont {Kjaergaard}, \citenamefont {Greene}, \citenamefont {Samach}, \citenamefont {McNally}, \citenamefont {Kim}, \citenamefont {Melville}, \citenamefont {Niedzielski}, \citenamefont {Schwartz}, \citenamefont {Yoder}, \citenamefont {Orlando}, \citenamefont {Gustavsson},\ and\ \citenamefont {Oliver}}]{Sung2021}%
  \BibitemOpen
  \bibfield  {author} {\bibinfo {author} {\bibfnamefont {Y.}~\bibnamefont {Sung}}, \bibinfo {author} {\bibfnamefont {L.}~\bibnamefont {Ding}}, \bibinfo {author} {\bibfnamefont {J.}~\bibnamefont {Braum\"uller}}, \bibinfo {author} {\bibfnamefont {A.}~\bibnamefont {Veps\"al\"ainen}}, \bibinfo {author} {\bibfnamefont {B.}~\bibnamefont {Kannan}}, \bibinfo {author} {\bibfnamefont {M.}~\bibnamefont {Kjaergaard}}, \bibinfo {author} {\bibfnamefont {A.}~\bibnamefont {Greene}}, \bibinfo {author} {\bibfnamefont {G.~O.}\ \bibnamefont {Samach}}, \bibinfo {author} {\bibfnamefont {C.}~\bibnamefont {McNally}}, \bibinfo {author} {\bibfnamefont {D.}~\bibnamefont {Kim}}, \bibinfo {author} {\bibfnamefont {A.}~\bibnamefont {Melville}}, \bibinfo {author} {\bibfnamefont {B.~M.}\ \bibnamefont {Niedzielski}}, \bibinfo {author} {\bibfnamefont {M.~E.}\ \bibnamefont {Schwartz}}, \bibinfo {author} {\bibfnamefont {J.~L.}\ \bibnamefont {Yoder}}, \bibinfo {author} {\bibfnamefont {T.~P.}\ \bibnamefont {Orlando}}, \bibinfo {author} {\bibfnamefont {S.}~\bibnamefont {Gustavsson}},\ and\ \bibinfo {author} {\bibfnamefont {W.~D.}\ \bibnamefont {Oliver}},\ }\bibfield  {title} {\bibinfo {title} {Realization of high-fidelity cz and $zz$-free iswap gates with a tunable coupler},\ }\href {https://doi.org/10.1103/PhysRevX.11.021058} {\bibfield  {journal} {\bibinfo  {journal} {Phys. Rev. X}\ }\textbf {\bibinfo {volume} {11}},\ \bibinfo {pages} {021058} (\bibinfo {year} {2021})}\BibitemShut {NoStop}%
\bibitem [{\citenamefont {Stehlik}\ \emph {et~al.}(2021)\citenamefont {Stehlik}, \citenamefont {Zajac}, \citenamefont {Underwood}, \citenamefont {Phung}, \citenamefont {Blair}, \citenamefont {Carnevale}, \citenamefont {Klaus}, \citenamefont {Keefe}, \citenamefont {Carniol}, \citenamefont {Kumph}, \citenamefont {Steffen},\ and\ \citenamefont {Dial}}]{Stehlik2021}%
  \BibitemOpen
  \bibfield  {author} {\bibinfo {author} {\bibfnamefont {J.}~\bibnamefont {Stehlik}}, \bibinfo {author} {\bibfnamefont {D.~M.}\ \bibnamefont {Zajac}}, \bibinfo {author} {\bibfnamefont {D.~L.}\ \bibnamefont {Underwood}}, \bibinfo {author} {\bibfnamefont {T.}~\bibnamefont {Phung}}, \bibinfo {author} {\bibfnamefont {J.}~\bibnamefont {Blair}}, \bibinfo {author} {\bibfnamefont {S.}~\bibnamefont {Carnevale}}, \bibinfo {author} {\bibfnamefont {D.}~\bibnamefont {Klaus}}, \bibinfo {author} {\bibfnamefont {G.~A.}\ \bibnamefont {Keefe}}, \bibinfo {author} {\bibfnamefont {A.}~\bibnamefont {Carniol}}, \bibinfo {author} {\bibfnamefont {M.}~\bibnamefont {Kumph}}, \bibinfo {author} {\bibfnamefont {M.}~\bibnamefont {Steffen}},\ and\ \bibinfo {author} {\bibfnamefont {O.~E.}\ \bibnamefont {Dial}},\ }\bibfield  {title} {\bibinfo {title} {Tunable coupling architecture for fixed-frequency transmon superconducting qubits},\ }\href {https://doi.org/10.1103/PhysRevLett.127.080505} {\bibfield  {journal} {\bibinfo  {journal} {Phys. Rev. Lett.}\ }\textbf {\bibinfo {volume} {127}},\ \bibinfo {pages} {080505} (\bibinfo {year} {2021})}\BibitemShut {NoStop}%
\bibitem [{\citenamefont {Gong}\ \emph {et~al.}(2021)\citenamefont {Gong}, \citenamefont {Wang}, \citenamefont {Zha}, \citenamefont {Chen}, \citenamefont {Huang}, \citenamefont {Wu}, \citenamefont {Zhu}, \citenamefont {Zhao}, \citenamefont {Li}, \citenamefont {Guo} \emph {et~al.}}]{gong2021}%
  \BibitemOpen
  \bibfield  {author} {\bibinfo {author} {\bibfnamefont {M.}~\bibnamefont {Gong}}, \bibinfo {author} {\bibfnamefont {S.}~\bibnamefont {Wang}}, \bibinfo {author} {\bibfnamefont {C.}~\bibnamefont {Zha}}, \bibinfo {author} {\bibfnamefont {M.-C.}\ \bibnamefont {Chen}}, \bibinfo {author} {\bibfnamefont {H.-L.}\ \bibnamefont {Huang}}, \bibinfo {author} {\bibfnamefont {Y.}~\bibnamefont {Wu}}, \bibinfo {author} {\bibfnamefont {Q.}~\bibnamefont {Zhu}}, \bibinfo {author} {\bibfnamefont {Y.}~\bibnamefont {Zhao}}, \bibinfo {author} {\bibfnamefont {S.}~\bibnamefont {Li}}, \bibinfo {author} {\bibfnamefont {S.}~\bibnamefont {Guo}}, \emph {et~al.},\ }\bibfield  {title} {\bibinfo {title} {Quantum walks on a programmable two-dimensional 62-qubit superconducting processor},\ }\href {https://doi.org/10.1126/science.abg7812} {\bibfield  {journal} {\bibinfo  {journal} {Science}\ }\textbf {\bibinfo {volume} {372}},\ \bibinfo {pages} {948} (\bibinfo {year} {2021})}\BibitemShut {NoStop}%
\bibitem [{\citenamefont {Chu}\ \emph {et~al.}(2023)\citenamefont {Chu}, \citenamefont {He}, \citenamefont {Zhou}, \citenamefont {Yuan}, \citenamefont {Zhang}, \citenamefont {Guo}, \citenamefont {Hai}, \citenamefont {Han}, \citenamefont {Hu}, \citenamefont {Huang} \emph {et~al.}}]{chu2023}%
  \BibitemOpen
  \bibfield  {author} {\bibinfo {author} {\bibfnamefont {J.}~\bibnamefont {Chu}}, \bibinfo {author} {\bibfnamefont {X.}~\bibnamefont {He}}, \bibinfo {author} {\bibfnamefont {Y.}~\bibnamefont {Zhou}}, \bibinfo {author} {\bibfnamefont {J.}~\bibnamefont {Yuan}}, \bibinfo {author} {\bibfnamefont {L.}~\bibnamefont {Zhang}}, \bibinfo {author} {\bibfnamefont {Q.}~\bibnamefont {Guo}}, \bibinfo {author} {\bibfnamefont {Y.}~\bibnamefont {Hai}}, \bibinfo {author} {\bibfnamefont {Z.}~\bibnamefont {Han}}, \bibinfo {author} {\bibfnamefont {C.-K.}\ \bibnamefont {Hu}}, \bibinfo {author} {\bibfnamefont {W.}~\bibnamefont {Huang}}, \emph {et~al.},\ }\bibfield  {title} {\bibinfo {title} {Scalable algorithm simplification using quantum and logic},\ }\href {https://doi.org/https://doi.org/10.1038/s41567-022-01813-7} {\bibfield  {journal} {\bibinfo  {journal} {Nat. Phys.}\ }\textbf {\bibinfo {volume} {19}},\ \bibinfo {pages} {126} (\bibinfo {year} {2023})}\BibitemShut {NoStop}%
\bibitem [{\citenamefont {Zhang}\ \emph {et~al.}(2022)\citenamefont {Zhang}, \citenamefont {Jiang}, \citenamefont {Deng}, \citenamefont {Wang}, \citenamefont {Chen}, \citenamefont {Zhang}, \citenamefont {Ren}, \citenamefont {Dong}, \citenamefont {Xu}, \citenamefont {Gao}, \citenamefont {Jin}, \citenamefont {Zhu}, \citenamefont {Guo}, \citenamefont {Li}, \citenamefont {Song}, \citenamefont {Gorshkov}, \citenamefont {Iadecola}, \citenamefont {Liu}, \citenamefont {Gong}, \citenamefont {Wang}, \citenamefont {Deng},\ and\ \citenamefont {Wang}}]{zhang2023}%
  \BibitemOpen
  \bibfield  {author} {\bibinfo {author} {\bibfnamefont {X.}~\bibnamefont {Zhang}}, \bibinfo {author} {\bibfnamefont {W.}~\bibnamefont {Jiang}}, \bibinfo {author} {\bibfnamefont {J.}~\bibnamefont {Deng}}, \bibinfo {author} {\bibfnamefont {K.}~\bibnamefont {Wang}}, \bibinfo {author} {\bibfnamefont {J.}~\bibnamefont {Chen}}, \bibinfo {author} {\bibfnamefont {P.}~\bibnamefont {Zhang}}, \bibinfo {author} {\bibfnamefont {W.}~\bibnamefont {Ren}}, \bibinfo {author} {\bibfnamefont {H.}~\bibnamefont {Dong}}, \bibinfo {author} {\bibfnamefont {S.}~\bibnamefont {Xu}}, \bibinfo {author} {\bibfnamefont {Y.}~\bibnamefont {Gao}}, \bibinfo {author} {\bibfnamefont {F.}~\bibnamefont {Jin}}, \bibinfo {author} {\bibfnamefont {X.}~\bibnamefont {Zhu}}, \bibinfo {author} {\bibfnamefont {Q.}~\bibnamefont {Guo}}, \bibinfo {author} {\bibfnamefont {H.}~\bibnamefont {Li}}, \bibinfo {author} {\bibfnamefont {C.}~\bibnamefont {Song}}, \bibinfo {author} {\bibfnamefont {A.~V.}\ \bibnamefont {Gorshkov}}, \bibinfo {author} {\bibfnamefont {T.}~\bibnamefont {Iadecola}}, \bibinfo {author} {\bibfnamefont {F.}~\bibnamefont {Liu}}, \bibinfo {author} {\bibfnamefont {Z.-X.}\ \bibnamefont {Gong}}, \bibinfo {author} {\bibfnamefont {Z.}~\bibnamefont {Wang}}, \bibinfo {author} {\bibfnamefont {D.-L.}\ \bibnamefont {Deng}},\ and\ \bibinfo {author} {\bibfnamefont {H.}~\bibnamefont {Wang}},\ }\bibfield  {title} {\bibinfo {title} {Digital quantum simulation of {Floquet} symmetry-protected topological phases},\ }\href {https://doi.org/10.1038/s41586-022-04854-3} {\bibfield  {journal} {\bibinfo  {journal} {Nature}\ }\textbf {\bibinfo {volume} {607}},\ \bibinfo {pages} {468} (\bibinfo {year} {2022})}\BibitemShut {NoStop}%
\bibitem [{\citenamefont {Arute}\ \emph {et~al.}(2019)\citenamefont {Arute}, \citenamefont {Arya}, \citenamefont {Babbush}, \citenamefont {Bacon}, \citenamefont {Bardin}, \citenamefont {Barends}, \citenamefont {Biswas}, \citenamefont {Boixo}, \citenamefont {Brandao}, \citenamefont {Buell} \emph {et~al.}}]{arute2019}%
  \BibitemOpen
  \bibfield  {author} {\bibinfo {author} {\bibfnamefont {F.}~\bibnamefont {Arute}}, \bibinfo {author} {\bibfnamefont {K.}~\bibnamefont {Arya}}, \bibinfo {author} {\bibfnamefont {R.}~\bibnamefont {Babbush}}, \bibinfo {author} {\bibfnamefont {D.}~\bibnamefont {Bacon}}, \bibinfo {author} {\bibfnamefont {J.~C.}\ \bibnamefont {Bardin}}, \bibinfo {author} {\bibfnamefont {R.}~\bibnamefont {Barends}}, \bibinfo {author} {\bibfnamefont {R.}~\bibnamefont {Biswas}}, \bibinfo {author} {\bibfnamefont {S.}~\bibnamefont {Boixo}}, \bibinfo {author} {\bibfnamefont {F.~G.}\ \bibnamefont {Brandao}}, \bibinfo {author} {\bibfnamefont {D.~A.}\ \bibnamefont {Buell}}, \emph {et~al.},\ }\bibfield  {title} {\bibinfo {title} {Quantum supremacy using a programmable superconducting processor},\ }\href {https://doi.org/10.1038/s41586-019-1666-5} {\bibfield  {journal} {\bibinfo  {journal} {Nature}\ }\textbf {\bibinfo {volume} {574}},\ \bibinfo {pages} {505} (\bibinfo {year} {2019})}\BibitemShut {NoStop}%
\bibitem [{\citenamefont {Wu}\ \emph {et~al.}(2021)\citenamefont {Wu}, \citenamefont {Bao}, \citenamefont {Cao}, \citenamefont {Chen}, \citenamefont {Chen}, \citenamefont {Chen}, \citenamefont {Chung}, \citenamefont {Deng}, \citenamefont {Du}, \citenamefont {Fan} \emph {et~al.}}]{Wu2021}%
  \BibitemOpen
  \bibfield  {author} {\bibinfo {author} {\bibfnamefont {Y.}~\bibnamefont {Wu}}, \bibinfo {author} {\bibfnamefont {W.-S.}\ \bibnamefont {Bao}}, \bibinfo {author} {\bibfnamefont {S.}~\bibnamefont {Cao}}, \bibinfo {author} {\bibfnamefont {F.}~\bibnamefont {Chen}}, \bibinfo {author} {\bibfnamefont {M.-C.}\ \bibnamefont {Chen}}, \bibinfo {author} {\bibfnamefont {X.}~\bibnamefont {Chen}}, \bibinfo {author} {\bibfnamefont {T.-H.}\ \bibnamefont {Chung}}, \bibinfo {author} {\bibfnamefont {H.}~\bibnamefont {Deng}}, \bibinfo {author} {\bibfnamefont {Y.}~\bibnamefont {Du}}, \bibinfo {author} {\bibfnamefont {D.}~\bibnamefont {Fan}}, \emph {et~al.},\ }\bibfield  {title} {\bibinfo {title} {Strong quantum computational advantage using a superconducting quantum processor},\ }\href {https://doi.org/10.1103/PhysRevLett.127.180501} {\bibfield  {journal} {\bibinfo  {journal} {Phys. Rev. Lett.}\ }\textbf {\bibinfo {volume} {127}},\ \bibinfo {pages} {180501} (\bibinfo {year} {2021})}\BibitemShut {NoStop}%
\bibitem [{\citenamefont {AI}\ and\ \citenamefont {Collaborators}(2025)}]{Google2025}%
  \BibitemOpen
  \bibfield  {author} {\bibinfo {author} {\bibfnamefont {G.~Q.}\ \bibnamefont {AI}}\ and\ \bibinfo {author} {\bibnamefont {Collaborators}},\ }\bibfield  {title} {\bibinfo {title} {Quantum error correction below the surface code threshold},\ }\href {https://doi.org/10.1038/s41586-024-08449-y} {\bibfield  {journal} {\bibinfo  {journal} {Nature}\ }\textbf {\bibinfo {volume} {638}},\ \bibinfo {pages} {920} (\bibinfo {year} {2025})}\BibitemShut {NoStop}%
\bibitem [{\citenamefont {Gao}\ \emph {et~al.}(2025)\citenamefont {Gao}, \citenamefont {Fan}, \citenamefont {Zha}, \citenamefont {Bei}, \citenamefont {Cai}, \citenamefont {Cai}, \citenamefont {Cao}, \citenamefont {Chen}, \citenamefont {Chen}, \citenamefont {Chen}, \citenamefont {Chen}, \citenamefont {Chen}, \citenamefont {Chen}, \citenamefont {Chen}, \citenamefont {Chen}, \citenamefont {Chu}, \citenamefont {Deng}, \citenamefont {Deng}, \citenamefont {Ding}, \citenamefont {Ding}, \citenamefont {Ding}, \citenamefont {Dong}, \citenamefont {Dong}, \citenamefont {Fan}, \citenamefont {Fu}, \citenamefont {Gao}, \citenamefont {Ge}, \citenamefont {Gong}, \citenamefont {Gui}, \citenamefont {Guo}, \citenamefont {Guo}, \citenamefont {Guo}, \citenamefont {Han}, \citenamefont {He}, \citenamefont {Hong}, \citenamefont {Hu}, \citenamefont {Huang}, \citenamefont {Huo}, \citenamefont {Jiang}, \citenamefont {Jiang}, \citenamefont {Jin}, \citenamefont {Leng}, \citenamefont {Li}, \citenamefont {Li}, \citenamefont {Li}, \citenamefont {Li}, \citenamefont {Li}, \citenamefont {Li}, \citenamefont {Li}, \citenamefont {Li}, \citenamefont {Li}, \citenamefont {Li}, \citenamefont {Li}, \citenamefont {Li}, \citenamefont {Liang}, \citenamefont {Liang}, \citenamefont {Liao}, \citenamefont {Lin}, \citenamefont {Lin}, \citenamefont {Liu}, \citenamefont {Liu}, \citenamefont {Liu}, \citenamefont {Liu}, \citenamefont {Liu}, \citenamefont {Liu}, \citenamefont {Lou}, \citenamefont {Ma}, \citenamefont {Meng}, \citenamefont {Mou}, \citenamefont {Nan}, \citenamefont {Nie}, \citenamefont {Nie}, \citenamefont {Ning}, \citenamefont {Niu}, \citenamefont {Peng}, \citenamefont {Qian}, \citenamefont {Rong}, \citenamefont {Rong}, \citenamefont {Shen}, \citenamefont {Shen}, \citenamefont {Su}, \citenamefont {Su}, \citenamefont {Sun}, \citenamefont {Sun}, \citenamefont {Sun}, \citenamefont {Sun}, \citenamefont {Tan}, \citenamefont {Tan}, \citenamefont {Tang}, \citenamefont {Tu}, \citenamefont {Wan}, \citenamefont {Wang}, \citenamefont {Wang}, \citenamefont {Wang}, \citenamefont {Wang}, \citenamefont {Wang}, \citenamefont {Wang}, \citenamefont {Wang}, \citenamefont {Wang}, \citenamefont {Wang}, \citenamefont {Wang}, \citenamefont {Wang}, \citenamefont {Wang}, \citenamefont {Wang}, \citenamefont {Wei}, \citenamefont {Wei}, \citenamefont {Wu}, \citenamefont {Wu}, \citenamefont {Wu}, \citenamefont {Wu}, \citenamefont {Wu}, \citenamefont {Xie}, \citenamefont {Xin}, \citenamefont {Xu}, \citenamefont {Xue}, \citenamefont {Yan}, \citenamefont {Yang}, \citenamefont {Yang}, \citenamefont {Yang}, \citenamefont {Ye}, \citenamefont {Ye}, \citenamefont {Ying}, \citenamefont {Yu}, \citenamefont {Yu}, \citenamefont {Yu}, \citenamefont {Zeng}, \citenamefont {Zhan}, \citenamefont {Zhang}, \citenamefont {Zhang}, \citenamefont {Zhang}, \citenamefont {Zhang}, \citenamefont {Zhang}, \citenamefont {Zhang}, \citenamefont {Zhang}, \citenamefont {Zhang}, \citenamefont {Zhao}, \citenamefont {Zhao}, \citenamefont {Zhao}, \citenamefont {Zhao}, \citenamefont {Zhao}, \citenamefont {Zhao}, \citenamefont {Zheng}, \citenamefont {Zhou}, \citenamefont {Zhou}, \citenamefont {Zhou}, \citenamefont {Zhou}, \citenamefont {Zhou}, \citenamefont {Zhou}, \citenamefont {Zhou}, \citenamefont {Zhu}, \citenamefont {Zhu}, \citenamefont {Zou}, \citenamefont {Zou}, \citenamefont {Zhang}, \citenamefont {Lu}, \citenamefont {Peng}, \citenamefont {Zhu},\ and\ \citenamefont {Pan}}]{Gao2025}%
  \BibitemOpen
  \bibfield  {author} {\bibinfo {author} {\bibfnamefont {D.}~\bibnamefont {Gao}}, \bibinfo {author} {\bibfnamefont {D.}~\bibnamefont {Fan}}, \bibinfo {author} {\bibfnamefont {C.}~\bibnamefont {Zha}}, \bibinfo {author} {\bibfnamefont {J.}~\bibnamefont {Bei}}, \bibinfo {author} {\bibfnamefont {G.}~\bibnamefont {Cai}}, \bibinfo {author} {\bibfnamefont {J.}~\bibnamefont {Cai}}, \bibinfo {author} {\bibfnamefont {S.}~\bibnamefont {Cao}}, \bibinfo {author} {\bibfnamefont {F.}~\bibnamefont {Chen}}, \bibinfo {author} {\bibfnamefont {J.}~\bibnamefont {Chen}}, \bibinfo {author} {\bibfnamefont {K.}~\bibnamefont {Chen}}, \bibinfo {author} {\bibfnamefont {X.}~\bibnamefont {Chen}}, \bibinfo {author} {\bibfnamefont {X.}~\bibnamefont {Chen}}, \bibinfo {author} {\bibfnamefont {Z.}~\bibnamefont {Chen}}, \bibinfo {author} {\bibfnamefont {Z.}~\bibnamefont {Chen}}, \bibinfo {author} {\bibfnamefont {Z.}~\bibnamefont {Chen}}, \bibinfo {author} {\bibfnamefont {W.}~\bibnamefont {Chu}}, \bibinfo {author} {\bibfnamefont {H.}~\bibnamefont {Deng}}, \bibinfo {author} {\bibfnamefont {Z.}~\bibnamefont {Deng}}, \bibinfo {author} {\bibfnamefont {P.}~\bibnamefont {Ding}}, \bibinfo {author} {\bibfnamefont {X.}~\bibnamefont {Ding}}, \bibinfo {author} {\bibfnamefont {Z.}~\bibnamefont {Ding}}, \bibinfo {author} {\bibfnamefont {S.}~\bibnamefont {Dong}}, \bibinfo {author} {\bibfnamefont {Y.}~\bibnamefont {Dong}}, \bibinfo {author} {\bibfnamefont {B.}~\bibnamefont {Fan}}, \bibinfo {author} {\bibfnamefont {Y.}~\bibnamefont {Fu}}, \bibinfo {author} {\bibfnamefont {S.}~\bibnamefont {Gao}}, \bibinfo {author} {\bibfnamefont {L.}~\bibnamefont {Ge}}, \bibinfo {author} {\bibfnamefont {M.}~\bibnamefont {Gong}}, \bibinfo {author} {\bibfnamefont {J.}~\bibnamefont {Gui}}, \bibinfo {author} {\bibfnamefont {C.}~\bibnamefont {Guo}}, \bibinfo {author} {\bibfnamefont {S.}~\bibnamefont {Guo}}, \bibinfo {author} {\bibfnamefont {X.}~\bibnamefont {Guo}}, \bibinfo {author} {\bibfnamefont {L.}~\bibnamefont {Han}}, \bibinfo {author} {\bibfnamefont {T.}~\bibnamefont {He}}, \bibinfo {author} {\bibfnamefont {L.}~\bibnamefont {Hong}}, \bibinfo {author} {\bibfnamefont {Y.}~\bibnamefont {Hu}}, \bibinfo {author} {\bibfnamefont {H.-L.}\ \bibnamefont {Huang}}, \bibinfo {author} {\bibfnamefont {Y.-H.}\ \bibnamefont {Huo}}, \bibinfo {author} {\bibfnamefont {T.}~\bibnamefont {Jiang}}, \bibinfo {author} {\bibfnamefont {Z.}~\bibnamefont {Jiang}}, \bibinfo {author} {\bibfnamefont {H.}~\bibnamefont {Jin}}, \bibinfo {author} {\bibfnamefont {Y.}~\bibnamefont {Leng}}, \bibinfo {author} {\bibfnamefont {D.}~\bibnamefont {Li}}, \bibinfo {author} {\bibfnamefont {D.}~\bibnamefont {Li}}, \bibinfo {author} {\bibfnamefont {F.}~\bibnamefont {Li}}, \bibinfo {author} {\bibfnamefont {J.}~\bibnamefont {Li}}, \bibinfo {author} {\bibfnamefont {J.}~\bibnamefont {Li}}, \bibinfo {author} {\bibfnamefont {J.}~\bibnamefont {Li}}, \bibinfo {author} {\bibfnamefont {J.}~\bibnamefont {Li}}, \bibinfo {author} {\bibfnamefont {N.}~\bibnamefont {Li}}, \bibinfo {author} {\bibfnamefont {S.}~\bibnamefont {Li}}, \bibinfo {author} {\bibfnamefont {W.}~\bibnamefont {Li}}, \bibinfo {author} {\bibfnamefont {Y.}~\bibnamefont {Li}}, \bibinfo {author} {\bibfnamefont {Y.}~\bibnamefont {Li}}, \bibinfo {author} {\bibfnamefont {F.}~\bibnamefont {Liang}}, \bibinfo {author} {\bibfnamefont {X.}~\bibnamefont {Liang}}, \bibinfo {author} {\bibfnamefont {N.}~\bibnamefont {Liao}}, \bibinfo {author} {\bibfnamefont {J.}~\bibnamefont {Lin}}, \bibinfo {author} {\bibfnamefont {W.}~\bibnamefont {Lin}}, \bibinfo {author} {\bibfnamefont {D.}~\bibnamefont {Liu}}, \bibinfo {author} {\bibfnamefont {H.}~\bibnamefont {Liu}}, \bibinfo {author} {\bibfnamefont {M.}~\bibnamefont {Liu}}, \bibinfo {author} {\bibfnamefont {X.}~\bibnamefont {Liu}}, \bibinfo {author} {\bibfnamefont {X.}~\bibnamefont {Liu}}, \bibinfo {author} {\bibfnamefont {Y.}~\bibnamefont {Liu}}, \bibinfo {author} {\bibfnamefont {H.}~\bibnamefont {Lou}}, \bibinfo {author} {\bibfnamefont {Y.}~\bibnamefont {Ma}}, \bibinfo {author} {\bibfnamefont {L.}~\bibnamefont {Meng}}, \bibinfo {author} {\bibfnamefont {H.}~\bibnamefont {Mou}}, \bibinfo {author} {\bibfnamefont {K.}~\bibnamefont {Nan}}, \bibinfo {author} {\bibfnamefont {B.}~\bibnamefont {Nie}}, \bibinfo {author} {\bibfnamefont {M.}~\bibnamefont {Nie}}, \bibinfo {author} {\bibfnamefont {J.}~\bibnamefont {Ning}}, \bibinfo {author} {\bibfnamefont {L.}~\bibnamefont {Niu}}, \bibinfo {author} {\bibfnamefont {W.}~\bibnamefont {Peng}}, \bibinfo {author} {\bibfnamefont {H.}~\bibnamefont {Qian}}, \bibinfo {author} {\bibfnamefont {H.}~\bibnamefont {Rong}}, \bibinfo {author} {\bibfnamefont {T.}~\bibnamefont {Rong}}, \bibinfo {author} {\bibfnamefont {H.}~\bibnamefont {Shen}}, \bibinfo {author} {\bibfnamefont {Q.}~\bibnamefont {Shen}}, \bibinfo {author} {\bibfnamefont {H.}~\bibnamefont {Su}}, \bibinfo {author} {\bibfnamefont {F.}~\bibnamefont {Su}}, \bibinfo {author} {\bibfnamefont {C.}~\bibnamefont {Sun}}, \bibinfo {author} {\bibfnamefont {L.}~\bibnamefont {Sun}}, \bibinfo {author} {\bibfnamefont {T.}~\bibnamefont {Sun}}, \bibinfo {author} {\bibfnamefont {Y.}~\bibnamefont {Sun}}, \bibinfo {author} {\bibfnamefont {Y.}~\bibnamefont {Tan}}, \bibinfo {author} {\bibfnamefont {J.}~\bibnamefont {Tan}}, \bibinfo {author} {\bibfnamefont {L.}~\bibnamefont {Tang}}, \bibinfo {author} {\bibfnamefont {W.}~\bibnamefont {Tu}}, \bibinfo {author} {\bibfnamefont {C.}~\bibnamefont {Wan}}, \bibinfo {author} {\bibfnamefont {J.}~\bibnamefont {Wang}}, \bibinfo {author} {\bibfnamefont {B.}~\bibnamefont {Wang}}, \bibinfo {author} {\bibfnamefont {C.}~\bibnamefont {Wang}}, \bibinfo {author} {\bibfnamefont {C.}~\bibnamefont {Wang}}, \bibinfo {author} {\bibfnamefont {C.}~\bibnamefont {Wang}}, \bibinfo {author} {\bibfnamefont {J.}~\bibnamefont {Wang}}, \bibinfo {author} {\bibfnamefont {L.}~\bibnamefont {Wang}}, \bibinfo {author} {\bibfnamefont {R.}~\bibnamefont {Wang}}, \bibinfo {author} {\bibfnamefont {S.}~\bibnamefont {Wang}}, \bibinfo {author} {\bibfnamefont {X.}~\bibnamefont {Wang}}, \bibinfo {author} {\bibfnamefont {X.}~\bibnamefont {Wang}}, \bibinfo {author} {\bibfnamefont {X.}~\bibnamefont {Wang}}, \bibinfo {author} {\bibfnamefont {Y.}~\bibnamefont {Wang}}, \bibinfo {author} {\bibfnamefont {Z.}~\bibnamefont {Wei}}, \bibinfo {author} {\bibfnamefont {J.}~\bibnamefont {Wei}}, \bibinfo {author} {\bibfnamefont {D.}~\bibnamefont {Wu}}, \bibinfo {author} {\bibfnamefont {G.}~\bibnamefont {Wu}}, \bibinfo {author} {\bibfnamefont {J.}~\bibnamefont {Wu}}, \bibinfo {author} {\bibfnamefont {S.}~\bibnamefont {Wu}}, \bibinfo {author} {\bibfnamefont {Y.}~\bibnamefont {Wu}}, \bibinfo {author} {\bibfnamefont {S.}~\bibnamefont {Xie}}, \bibinfo {author} {\bibfnamefont {L.}~\bibnamefont {Xin}}, \bibinfo {author} {\bibfnamefont {Y.}~\bibnamefont {Xu}}, \bibinfo {author} {\bibfnamefont {C.}~\bibnamefont {Xue}}, \bibinfo {author} {\bibfnamefont {K.}~\bibnamefont {Yan}}, \bibinfo {author} {\bibfnamefont {W.}~\bibnamefont {Yang}}, \bibinfo {author} {\bibfnamefont {X.}~\bibnamefont {Yang}}, \bibinfo {author} {\bibfnamefont {Y.}~\bibnamefont {Yang}}, \bibinfo {author} {\bibfnamefont {Y.}~\bibnamefont {Ye}}, \bibinfo {author} {\bibfnamefont {Z.}~\bibnamefont {Ye}}, \bibinfo {author} {\bibfnamefont {C.}~\bibnamefont {Ying}}, \bibinfo {author} {\bibfnamefont {J.}~\bibnamefont {Yu}}, \bibinfo {author} {\bibfnamefont {Q.}~\bibnamefont {Yu}}, \bibinfo {author} {\bibfnamefont {W.}~\bibnamefont {Yu}}, \bibinfo {author} {\bibfnamefont {X.}~\bibnamefont {Zeng}}, \bibinfo {author} {\bibfnamefont {S.}~\bibnamefont {Zhan}}, \bibinfo {author} {\bibfnamefont {F.}~\bibnamefont {Zhang}}, \bibinfo {author} {\bibfnamefont {H.}~\bibnamefont {Zhang}}, \bibinfo {author} {\bibfnamefont {K.}~\bibnamefont {Zhang}}, \bibinfo {author} {\bibfnamefont {P.}~\bibnamefont {Zhang}}, \bibinfo {author} {\bibfnamefont {W.}~\bibnamefont {Zhang}}, \bibinfo {author} {\bibfnamefont {Y.}~\bibnamefont {Zhang}}, \bibinfo {author} {\bibfnamefont {Y.}~\bibnamefont {Zhang}}, \bibinfo {author} {\bibfnamefont {L.}~\bibnamefont {Zhang}}, \bibinfo {author} {\bibfnamefont {G.}~\bibnamefont {Zhao}}, \bibinfo {author} {\bibfnamefont {P.}~\bibnamefont {Zhao}}, \bibinfo {author} {\bibfnamefont {X.}~\bibnamefont {Zhao}}, \bibinfo {author} {\bibfnamefont {X.}~\bibnamefont {Zhao}}, \bibinfo {author} {\bibfnamefont {Y.}~\bibnamefont {Zhao}}, \bibinfo {author} {\bibfnamefont {Z.}~\bibnamefont {Zhao}}, \bibinfo {author} {\bibfnamefont {L.}~\bibnamefont {Zheng}}, \bibinfo {author} {\bibfnamefont {F.}~\bibnamefont {Zhou}}, \bibinfo {author} {\bibfnamefont {L.}~\bibnamefont {Zhou}}, \bibinfo {author} {\bibfnamefont {N.}~\bibnamefont {Zhou}}, \bibinfo {author} {\bibfnamefont {N.}~\bibnamefont {Zhou}}, \bibinfo {author} {\bibfnamefont {S.}~\bibnamefont {Zhou}}, \bibinfo {author} {\bibfnamefont {S.}~\bibnamefont {Zhou}}, \bibinfo {author} {\bibfnamefont {Z.}~\bibnamefont {Zhou}}, \bibinfo {author} {\bibfnamefont {C.}~\bibnamefont {Zhu}}, \bibinfo {author} {\bibfnamefont {Q.}~\bibnamefont {Zhu}}, \bibinfo {author} {\bibfnamefont {G.}~\bibnamefont {Zou}}, \bibinfo {author} {\bibfnamefont {H.}~\bibnamefont {Zou}}, \bibinfo {author} {\bibfnamefont {Q.}~\bibnamefont {Zhang}}, \bibinfo {author} {\bibfnamefont {C.-Y.}\ \bibnamefont {Lu}}, \bibinfo {author} {\bibfnamefont {C.-Z.}\ \bibnamefont {Peng}}, \bibinfo {author} {\bibfnamefont {X.}~\bibnamefont {Zhu}},\ and\ \bibinfo {author} {\bibfnamefont {J.-W.}\ \bibnamefont {Pan}},\ }\bibfield  {title} {\bibinfo {title} {Establishing a new benchmark in quantum computational advantage with 105-qubit zuchongzhi 3.0 processor},\ }\href {https://doi.org/10.1103/PhysRevLett.134.090601} {\bibfield  {journal} {\bibinfo  {journal} {Phys. Rev. Lett.}\ }\textbf {\bibinfo {volume} {134}},\ \bibinfo {pages} {090601} (\bibinfo {year} {2025})}\BibitemShut {NoStop}%
\bibitem [{\citenamefont {McKay}\ \emph {et~al.}(2016)\citenamefont {McKay}, \citenamefont {Filipp}, \citenamefont {Mezzacapo}, \citenamefont {Magesan}, \citenamefont {Chow},\ and\ \citenamefont {Gambetta}}]{McKay2016}%
  \BibitemOpen
  \bibfield  {author} {\bibinfo {author} {\bibfnamefont {D.~C.}\ \bibnamefont {McKay}}, \bibinfo {author} {\bibfnamefont {S.}~\bibnamefont {Filipp}}, \bibinfo {author} {\bibfnamefont {A.}~\bibnamefont {Mezzacapo}}, \bibinfo {author} {\bibfnamefont {E.}~\bibnamefont {Magesan}}, \bibinfo {author} {\bibfnamefont {J.~M.}\ \bibnamefont {Chow}},\ and\ \bibinfo {author} {\bibfnamefont {J.~M.}\ \bibnamefont {Gambetta}},\ }\bibfield  {title} {\bibinfo {title} {Universal gate for fixed-frequency qubits via a tunable bus},\ }\href {https://doi.org/10.1103/PhysRevApplied.6.064007} {\bibfield  {journal} {\bibinfo  {journal} {Phys. Rev. Appl.}\ }\textbf {\bibinfo {volume} {6}},\ \bibinfo {pages} {064007} (\bibinfo {year} {2016})}\BibitemShut {NoStop}%
\bibitem [{\citenamefont {Foxen}\ \emph {et~al.}(2020)\citenamefont {Foxen}, \citenamefont {Neill}, \citenamefont {Dunsworth}, \citenamefont {Roushan}, \citenamefont {Chiaro}, \citenamefont {Megrant}, \citenamefont {Kelly}, \citenamefont {Chen}, \citenamefont {Satzinger}, \citenamefont {Barends}, \citenamefont {Arute}, \citenamefont {Arya}, \citenamefont {Babbush}, \citenamefont {Bacon}, \citenamefont {Bardin}, \citenamefont {Boixo}, \citenamefont {Buell}, \citenamefont {Burkett}, \citenamefont {Chen}, \citenamefont {Collins}, \citenamefont {Farhi}, \citenamefont {Fowler}, \citenamefont {Gidney}, \citenamefont {Giustina}, \citenamefont {Graff}, \citenamefont {Harrigan}, \citenamefont {Huang}, \citenamefont {Isakov}, \citenamefont {Jeffrey}, \citenamefont {Jiang}, \citenamefont {Kafri}, \citenamefont {Kechedzhi}, \citenamefont {Klimov}, \citenamefont {Korotkov}, \citenamefont {Kostritsa}, \citenamefont {Landhuis}, \citenamefont {Lucero}, \citenamefont {McClean}, \citenamefont {McEwen}, \citenamefont {Mi}, \citenamefont {Mohseni}, \citenamefont {Mutus}, \citenamefont {Naaman}, \citenamefont {Neeley}, \citenamefont {Niu}, \citenamefont {Petukhov}, \citenamefont {Quintana}, \citenamefont {Rubin}, \citenamefont {Sank}, \citenamefont {Smelyanskiy}, \citenamefont {Vainsencher}, \citenamefont {White}, \citenamefont {Yao}, \citenamefont {Yeh}, \citenamefont {Zalcman}, \citenamefont {Neven},\ and\ \citenamefont {Martinis}}]{Foxen2020}%
  \BibitemOpen
  \bibfield  {author} {\bibinfo {author} {\bibfnamefont {B.}~\bibnamefont {Foxen}}, \bibinfo {author} {\bibfnamefont {C.}~\bibnamefont {Neill}}, \bibinfo {author} {\bibfnamefont {A.}~\bibnamefont {Dunsworth}}, \bibinfo {author} {\bibfnamefont {P.}~\bibnamefont {Roushan}}, \bibinfo {author} {\bibfnamefont {B.}~\bibnamefont {Chiaro}}, \bibinfo {author} {\bibfnamefont {A.}~\bibnamefont {Megrant}}, \bibinfo {author} {\bibfnamefont {J.}~\bibnamefont {Kelly}}, \bibinfo {author} {\bibfnamefont {Z.}~\bibnamefont {Chen}}, \bibinfo {author} {\bibfnamefont {K.}~\bibnamefont {Satzinger}}, \bibinfo {author} {\bibfnamefont {R.}~\bibnamefont {Barends}}, \bibinfo {author} {\bibfnamefont {F.}~\bibnamefont {Arute}}, \bibinfo {author} {\bibfnamefont {K.}~\bibnamefont {Arya}}, \bibinfo {author} {\bibfnamefont {R.}~\bibnamefont {Babbush}}, \bibinfo {author} {\bibfnamefont {D.}~\bibnamefont {Bacon}}, \bibinfo {author} {\bibfnamefont {J.~C.}\ \bibnamefont {Bardin}}, \bibinfo {author} {\bibfnamefont {S.}~\bibnamefont {Boixo}}, \bibinfo {author} {\bibfnamefont {D.}~\bibnamefont {Buell}}, \bibinfo {author} {\bibfnamefont {B.}~\bibnamefont {Burkett}}, \bibinfo {author} {\bibfnamefont {Y.}~\bibnamefont {Chen}}, \bibinfo {author} {\bibfnamefont {R.}~\bibnamefont {Collins}}, \bibinfo {author} {\bibfnamefont {E.}~\bibnamefont {Farhi}}, \bibinfo {author} {\bibfnamefont {A.}~\bibnamefont {Fowler}}, \bibinfo {author} {\bibfnamefont {C.}~\bibnamefont {Gidney}}, \bibinfo {author} {\bibfnamefont {M.}~\bibnamefont {Giustina}}, \bibinfo {author} {\bibfnamefont {R.}~\bibnamefont {Graff}}, \bibinfo {author} {\bibfnamefont {M.}~\bibnamefont {Harrigan}}, \bibinfo {author} {\bibfnamefont {T.}~\bibnamefont {Huang}}, \bibinfo {author} {\bibfnamefont {S.~V.}\ \bibnamefont {Isakov}}, \bibinfo {author} {\bibfnamefont {E.}~\bibnamefont {Jeffrey}}, \bibinfo {author} {\bibfnamefont {Z.}~\bibnamefont {Jiang}}, \bibinfo {author} {\bibfnamefont {D.}~\bibnamefont {Kafri}}, \bibinfo {author} {\bibfnamefont {K.}~\bibnamefont {Kechedzhi}}, \bibinfo {author} {\bibfnamefont {P.}~\bibnamefont {Klimov}}, \bibinfo {author} {\bibfnamefont {A.}~\bibnamefont {Korotkov}}, \bibinfo {author} {\bibfnamefont {F.}~\bibnamefont {Kostritsa}}, \bibinfo {author} {\bibfnamefont {D.}~\bibnamefont {Landhuis}}, \bibinfo {author} {\bibfnamefont {E.}~\bibnamefont {Lucero}}, \bibinfo {author} {\bibfnamefont {J.}~\bibnamefont {McClean}}, \bibinfo {author} {\bibfnamefont {M.}~\bibnamefont {McEwen}}, \bibinfo {author} {\bibfnamefont {X.}~\bibnamefont {Mi}}, \bibinfo {author} {\bibfnamefont {M.}~\bibnamefont {Mohseni}}, \bibinfo {author} {\bibfnamefont {J.~Y.}\ \bibnamefont {Mutus}}, \bibinfo {author} {\bibfnamefont {O.}~\bibnamefont {Naaman}}, \bibinfo {author} {\bibfnamefont {M.}~\bibnamefont {Neeley}}, \bibinfo {author} {\bibfnamefont {M.}~\bibnamefont {Niu}}, \bibinfo {author} {\bibfnamefont {A.}~\bibnamefont {Petukhov}}, \bibinfo {author} {\bibfnamefont {C.}~\bibnamefont {Quintana}}, \bibinfo {author} {\bibfnamefont {N.}~\bibnamefont {Rubin}}, \bibinfo {author} {\bibfnamefont {D.}~\bibnamefont {Sank}}, \bibinfo {author} {\bibfnamefont {V.}~\bibnamefont {Smelyanskiy}}, \bibinfo {author} {\bibfnamefont {A.}~\bibnamefont {Vainsencher}}, \bibinfo {author} {\bibfnamefont {T.~C.}\ \bibnamefont {White}}, \bibinfo {author} {\bibfnamefont {Z.}~\bibnamefont {Yao}}, \bibinfo {author} {\bibfnamefont {P.}~\bibnamefont {Yeh}}, \bibinfo {author} {\bibfnamefont {A.}~\bibnamefont {Zalcman}}, \bibinfo {author} {\bibfnamefont {H.}~\bibnamefont {Neven}},\ and\ \bibinfo {author} {\bibfnamefont {J.~M.}\ \bibnamefont {Martinis}} (\bibinfo {collaboration} {Google AI Quantum}),\ }\bibfield  {title} {\bibinfo {title} {Demonstrating a continuous set of two-qubit gates for near-term quantum algorithms},\ }\href {https://doi.org/10.1103/PhysRevLett.125.120504} {\bibfield  {journal} {\bibinfo  {journal} {Phys. Rev. Lett.}\ }\textbf {\bibinfo {volume} {125}},\ \bibinfo {pages} {120504} (\bibinfo {year} {2020})}\BibitemShut {NoStop}%
\bibitem [{\citenamefont {Chen}\ \emph {et~al.}(2025)\citenamefont {Chen}, \citenamefont {Liu}, \citenamefont {Ma}, \citenamefont {Sun}, \citenamefont {Wang}, \citenamefont {Wang}, \citenamefont {Xu}, \citenamefont {Xue}, \citenamefont {Yan}, \citenamefont {Yang}, \citenamefont {Ding}, \citenamefont {Gao}, \citenamefont {Li}, \citenamefont {Zhang}, \citenamefont {Zhang}, \citenamefont {Jin}, \citenamefont {Yu}, \citenamefont {Chen},\ and\ \citenamefont {Yan}}]{chen2025}%
  \BibitemOpen
  \bibfield  {author} {\bibinfo {author} {\bibfnamefont {Z.}~\bibnamefont {Chen}}, \bibinfo {author} {\bibfnamefont {W.}~\bibnamefont {Liu}}, \bibinfo {author} {\bibfnamefont {Y.}~\bibnamefont {Ma}}, \bibinfo {author} {\bibfnamefont {W.}~\bibnamefont {Sun}}, \bibinfo {author} {\bibfnamefont {R.}~\bibnamefont {Wang}}, \bibinfo {author} {\bibfnamefont {H.}~\bibnamefont {Wang}}, \bibinfo {author} {\bibfnamefont {H.}~\bibnamefont {Xu}}, \bibinfo {author} {\bibfnamefont {G.}~\bibnamefont {Xue}}, \bibinfo {author} {\bibfnamefont {H.}~\bibnamefont {Yan}}, \bibinfo {author} {\bibfnamefont {Z.}~\bibnamefont {Yang}}, \bibinfo {author} {\bibfnamefont {J.}~\bibnamefont {Ding}}, \bibinfo {author} {\bibfnamefont {Y.}~\bibnamefont {Gao}}, \bibinfo {author} {\bibfnamefont {F.}~\bibnamefont {Li}}, \bibinfo {author} {\bibfnamefont {Y.}~\bibnamefont {Zhang}}, \bibinfo {author} {\bibfnamefont {Z.}~\bibnamefont {Zhang}}, \bibinfo {author} {\bibfnamefont {Y.}~\bibnamefont {Jin}}, \bibinfo {author} {\bibfnamefont {H.}~\bibnamefont {Yu}}, \bibinfo {author} {\bibfnamefont {J.}~\bibnamefont {Chen}},\ and\ \bibinfo {author} {\bibfnamefont {F.}~\bibnamefont {Yan}},\ }\href {https://doi.org/10.48550/arXiv.2502.03612} {\bibinfo {title} {Efficient {Implementation} of {Arbitrary} {Two}-{Qubit} {Gates} via {Unified} {Control}}} (\bibinfo {year} {2025}),\ \bibinfo {note} {arXiv:2502.03612 [quant-ph]}\BibitemShut {NoStop}%
\bibitem [{\citenamefont {Jiang}\ \emph {et~al.}(2007)\citenamefont {Jiang}, \citenamefont {Taylor}, \citenamefont {S\o{}rensen},\ and\ \citenamefont {Lukin}}]{Jiang2007}%
  \BibitemOpen
  \bibfield  {author} {\bibinfo {author} {\bibfnamefont {L.}~\bibnamefont {Jiang}}, \bibinfo {author} {\bibfnamefont {J.~M.}\ \bibnamefont {Taylor}}, \bibinfo {author} {\bibfnamefont {A.~S.}\ \bibnamefont {S\o{}rensen}},\ and\ \bibinfo {author} {\bibfnamefont {M.~D.}\ \bibnamefont {Lukin}},\ }\bibfield  {title} {\bibinfo {title} {Distributed quantum computation based on small quantum registers},\ }\href {https://doi.org/10.1103/PhysRevA.76.062323} {\bibfield  {journal} {\bibinfo  {journal} {Phys. Rev. A}\ }\textbf {\bibinfo {volume} {76}},\ \bibinfo {pages} {062323} (\bibinfo {year} {2007})}\BibitemShut {NoStop}%
\bibitem [{\citenamefont {Gambetta}\ \emph {et~al.}(2017)\citenamefont {Gambetta}, \citenamefont {Chow},\ and\ \citenamefont {Steffen}}]{gambetta2017}%
  \BibitemOpen
  \bibfield  {author} {\bibinfo {author} {\bibfnamefont {J.~M.}\ \bibnamefont {Gambetta}}, \bibinfo {author} {\bibfnamefont {J.~M.}\ \bibnamefont {Chow}},\ and\ \bibinfo {author} {\bibfnamefont {M.}~\bibnamefont {Steffen}},\ }\bibfield  {title} {\bibinfo {title} {Building logical qubits in a superconducting quantum computing system},\ }\href {https://doi.org/10.1038/s41534-016-0004-0} {\bibfield  {journal} {\bibinfo  {journal} {npj Quantum Inf.}\ }\textbf {\bibinfo {volume} {3}},\ \bibinfo {pages} {2} (\bibinfo {year} {2017})}\BibitemShut {NoStop}%
\bibitem [{\citenamefont {Bravyi}\ \emph {et~al.}(2022)\citenamefont {Bravyi}, \citenamefont {Dial}, \citenamefont {Gambetta}, \citenamefont {Gil},\ and\ \citenamefont {Nazario}}]{bravyi2022}%
  \BibitemOpen
  \bibfield  {author} {\bibinfo {author} {\bibfnamefont {S.}~\bibnamefont {Bravyi}}, \bibinfo {author} {\bibfnamefont {O.}~\bibnamefont {Dial}}, \bibinfo {author} {\bibfnamefont {J.~M.}\ \bibnamefont {Gambetta}}, \bibinfo {author} {\bibfnamefont {D.}~\bibnamefont {Gil}},\ and\ \bibinfo {author} {\bibfnamefont {Z.}~\bibnamefont {Nazario}},\ }\bibfield  {title} {\bibinfo {title} {The future of quantum computing with superconducting qubits},\ }\href {https://doi.org/10.1063/5.0082975} {\bibfield  {journal} {\bibinfo  {journal} {J. Appl. Phys.}\ }\textbf {\bibinfo {volume} {132}},\ \bibinfo {pages} {160902} (\bibinfo {year} {2022})}\BibitemShut {NoStop}%
\bibitem [{\citenamefont {Ang}\ \emph {et~al.}(2022)\citenamefont {Ang}, \citenamefont {Carini}, \citenamefont {Chen}, \citenamefont {Chuang}, \citenamefont {DeMarco}, \citenamefont {Economou}, \citenamefont {Eickbusch}, \citenamefont {Faraon}, \citenamefont {Fu}, \citenamefont {Girvin}, \citenamefont {Hatridge}, \citenamefont {Houck}, \citenamefont {Hilaire}, \citenamefont {Krsulich}, \citenamefont {Li}, \citenamefont {Liu}, \citenamefont {Liu}, \citenamefont {Martonosi}, \citenamefont {McKay}, \citenamefont {Misewich}, \citenamefont {Ritter}, \citenamefont {Schoelkopf}, \citenamefont {Stein}, \citenamefont {Sussman}, \citenamefont {Tang}, \citenamefont {Tang}, \citenamefont {Tomesh}, \citenamefont {Tubman}, \citenamefont {Wang}, \citenamefont {Wiebe}, \citenamefont {Yao}, \citenamefont {Yost},\ and\ \citenamefont {Zhou}}]{ang2022}%
  \BibitemOpen
  \bibfield  {author} {\bibinfo {author} {\bibfnamefont {J.}~\bibnamefont {Ang}}, \bibinfo {author} {\bibfnamefont {G.}~\bibnamefont {Carini}}, \bibinfo {author} {\bibfnamefont {Y.}~\bibnamefont {Chen}}, \bibinfo {author} {\bibfnamefont {I.}~\bibnamefont {Chuang}}, \bibinfo {author} {\bibfnamefont {M.~A.}\ \bibnamefont {DeMarco}}, \bibinfo {author} {\bibfnamefont {S.~E.}\ \bibnamefont {Economou}}, \bibinfo {author} {\bibfnamefont {A.}~\bibnamefont {Eickbusch}}, \bibinfo {author} {\bibfnamefont {A.}~\bibnamefont {Faraon}}, \bibinfo {author} {\bibfnamefont {K.-M.}\ \bibnamefont {Fu}}, \bibinfo {author} {\bibfnamefont {S.~M.}\ \bibnamefont {Girvin}}, \bibinfo {author} {\bibfnamefont {M.}~\bibnamefont {Hatridge}}, \bibinfo {author} {\bibfnamefont {A.}~\bibnamefont {Houck}}, \bibinfo {author} {\bibfnamefont {P.}~\bibnamefont {Hilaire}}, \bibinfo {author} {\bibfnamefont {K.}~\bibnamefont {Krsulich}}, \bibinfo {author} {\bibfnamefont {A.}~\bibnamefont {Li}}, \bibinfo {author} {\bibfnamefont {C.}~\bibnamefont {Liu}}, \bibinfo {author} {\bibfnamefont {Y.}~\bibnamefont {Liu}}, \bibinfo {author} {\bibfnamefont {M.}~\bibnamefont {Martonosi}}, \bibinfo {author} {\bibfnamefont {D.~C.}\ \bibnamefont {McKay}}, \bibinfo {author} {\bibfnamefont {J.}~\bibnamefont {Misewich}}, \bibinfo {author} {\bibfnamefont {M.}~\bibnamefont {Ritter}}, \bibinfo {author} {\bibfnamefont {R.~J.}\ \bibnamefont {Schoelkopf}}, \bibinfo {author} {\bibfnamefont {S.~A.}\ \bibnamefont {Stein}}, \bibinfo {author} {\bibfnamefont {S.}~\bibnamefont {Sussman}}, \bibinfo {author} {\bibfnamefont {H.~X.}\ \bibnamefont {Tang}}, \bibinfo {author} {\bibfnamefont {W.}~\bibnamefont {Tang}}, \bibinfo {author} {\bibfnamefont {T.}~\bibnamefont {Tomesh}}, \bibinfo {author} {\bibfnamefont {N.~M.}\ \bibnamefont {Tubman}}, \bibinfo {author} {\bibfnamefont {C.}~\bibnamefont {Wang}}, \bibinfo {author} {\bibfnamefont {N.}~\bibnamefont {Wiebe}}, \bibinfo {author} {\bibfnamefont {Y.-X.}\ \bibnamefont {Yao}}, \bibinfo {author} {\bibfnamefont {D.~C.}\ \bibnamefont {Yost}},\ and\ \bibinfo {author} {\bibfnamefont {Y.}~\bibnamefont {Zhou}},\ }\bibfield  {title} {\bibinfo {title} {Architectures for multinode superconducting quantum computers},\ }\href {https://arxiv.org/abs/2212.06167} {\  (\bibinfo {year} {2022})},\ \Eprint {https://arxiv.org/abs/2212.06167} {arXiv:2212.06167 [quant-ph]} \BibitemShut {NoStop}%
\bibitem [{\citenamefont {Field}\ \emph {et~al.}(2024)\citenamefont {Field}, \citenamefont {Chen}, \citenamefont {Scharmann}, \citenamefont {Sete}, \citenamefont {Oruc}, \citenamefont {Vu}, \citenamefont {Kosenko}, \citenamefont {Mutus}, \citenamefont {Poletto},\ and\ \citenamefont {Bestwick}}]{field2024}%
  \BibitemOpen
  \bibfield  {author} {\bibinfo {author} {\bibfnamefont {M.}~\bibnamefont {Field}}, \bibinfo {author} {\bibfnamefont {A.~Q.}\ \bibnamefont {Chen}}, \bibinfo {author} {\bibfnamefont {B.}~\bibnamefont {Scharmann}}, \bibinfo {author} {\bibfnamefont {E.~A.}\ \bibnamefont {Sete}}, \bibinfo {author} {\bibfnamefont {F.}~\bibnamefont {Oruc}}, \bibinfo {author} {\bibfnamefont {K.}~\bibnamefont {Vu}}, \bibinfo {author} {\bibfnamefont {V.}~\bibnamefont {Kosenko}}, \bibinfo {author} {\bibfnamefont {J.~Y.}\ \bibnamefont {Mutus}}, \bibinfo {author} {\bibfnamefont {S.}~\bibnamefont {Poletto}},\ and\ \bibinfo {author} {\bibfnamefont {A.}~\bibnamefont {Bestwick}},\ }\bibfield  {title} {\bibinfo {title} {Modular superconducting-qubit architecture with a multichip tunable coupler},\ }\href {https://doi.org/10.1103/PhysRevApplied.21.054063} {\bibfield  {journal} {\bibinfo  {journal} {Phys. Rev. Appl.}\ }\textbf {\bibinfo {volume} {21}},\ \bibinfo {pages} {54063} (\bibinfo {year} {2024})}\BibitemShut {NoStop}%
\bibitem [{\citenamefont {Cohen}\ \emph {et~al.}(2022)\citenamefont {Cohen}, \citenamefont {Kim}, \citenamefont {Bartlett},\ and\ \citenamefont {Brown}}]{cohen2022}%
  \BibitemOpen
  \bibfield  {author} {\bibinfo {author} {\bibfnamefont {L.~Z.}\ \bibnamefont {Cohen}}, \bibinfo {author} {\bibfnamefont {I.~H.}\ \bibnamefont {Kim}}, \bibinfo {author} {\bibfnamefont {S.~D.}\ \bibnamefont {Bartlett}},\ and\ \bibinfo {author} {\bibfnamefont {B.~J.}\ \bibnamefont {Brown}},\ }\bibfield  {title} {\bibinfo {title} {Low-overhead fault-tolerant quantum computing using long-range connectivity},\ }\href {https://doi.org/10.1126/sciadv.abn1717} {\bibfield  {journal} {\bibinfo  {journal} {Sci. Adv.}\ }\textbf {\bibinfo {volume} {8}},\ \bibinfo {pages} {eabn1717} (\bibinfo {year} {2022})}\BibitemShut {NoStop}%
\bibitem [{\citenamefont {Breuckmann}\ and\ \citenamefont {Eberhardt}(2021)}]{Breuckmann2021}%
  \BibitemOpen
  \bibfield  {author} {\bibinfo {author} {\bibfnamefont {N.~P.}\ \bibnamefont {Breuckmann}}\ and\ \bibinfo {author} {\bibfnamefont {J.~N.}\ \bibnamefont {Eberhardt}},\ }\bibfield  {title} {\bibinfo {title} {Quantum low-density parity-check codes},\ }\href {https://doi.org/10.1103/PRXQuantum.2.040101} {\bibfield  {journal} {\bibinfo  {journal} {PRX Quantum}\ }\textbf {\bibinfo {volume} {2}},\ \bibinfo {pages} {040101} (\bibinfo {year} {2021})}\BibitemShut {NoStop}%
\bibitem [{\citenamefont {Bravyi}\ \emph {et~al.}(2024)\citenamefont {Bravyi}, \citenamefont {Cross}, \citenamefont {Gambetta}, \citenamefont {Maslov}, \citenamefont {Rall},\ and\ \citenamefont {Yoder}}]{bravyi2024}%
  \BibitemOpen
  \bibfield  {author} {\bibinfo {author} {\bibfnamefont {S.}~\bibnamefont {Bravyi}}, \bibinfo {author} {\bibfnamefont {A.~W.}\ \bibnamefont {Cross}}, \bibinfo {author} {\bibfnamefont {J.~M.}\ \bibnamefont {Gambetta}}, \bibinfo {author} {\bibfnamefont {D.}~\bibnamefont {Maslov}}, \bibinfo {author} {\bibfnamefont {P.}~\bibnamefont {Rall}},\ and\ \bibinfo {author} {\bibfnamefont {T.~J.}\ \bibnamefont {Yoder}},\ }\bibfield  {title} {\bibinfo {title} {High-threshold and low-overhead fault-tolerant quantum memory},\ }\href {https://doi.org/https://doi.org/10.1038/s41586-024-07107-7} {\bibfield  {journal} {\bibinfo  {journal} {Nature}\ }\textbf {\bibinfo {volume} {627}},\ \bibinfo {pages} {778} (\bibinfo {year} {2024})}\BibitemShut {NoStop}%
\bibitem [{\citenamefont {Gidney}\ \emph {et~al.}(2025)\citenamefont {Gidney}, \citenamefont {Newman}, \citenamefont {Brooks},\ and\ \citenamefont {Jones}}]{gidney2025}%
  \BibitemOpen
  \bibfield  {author} {\bibinfo {author} {\bibfnamefont {C.}~\bibnamefont {Gidney}}, \bibinfo {author} {\bibfnamefont {M.}~\bibnamefont {Newman}}, \bibinfo {author} {\bibfnamefont {P.}~\bibnamefont {Brooks}},\ and\ \bibinfo {author} {\bibfnamefont {C.}~\bibnamefont {Jones}},\ }\bibfield  {title} {\bibinfo {title} {Yoked surface codes},\ }\href {https://doi.org/10.1038/s41467-025-59714-1} {\bibfield  {journal} {\bibinfo  {journal} {Nat. Commun.}\ }\textbf {\bibinfo {volume} {16}},\ \bibinfo {pages} {4498} (\bibinfo {year} {2025})}\BibitemShut {NoStop}%
\bibitem [{\citenamefont {Wang}\ \emph {et~al.}(2025)\citenamefont {Wang}, \citenamefont {Lu}, \citenamefont {Zhang}, \citenamefont {Liu}, \citenamefont {Chen}, \citenamefont {Wang}, \citenamefont {Wu}, \citenamefont {Xu}, \citenamefont {Zhu}, \citenamefont {Jin}, \citenamefont {Gao}, \citenamefont {Tan}, \citenamefont {Cui}, \citenamefont {Wang}, \citenamefont {Zou}, \citenamefont {Zhang}, \citenamefont {Li}, \citenamefont {Shen}, \citenamefont {Zhong}, \citenamefont {Bao}, \citenamefont {Zhu}, \citenamefont {Han}, \citenamefont {He}, \citenamefont {Shen}, \citenamefont {Wang}, \citenamefont {Yang}, \citenamefont {Song}, \citenamefont {Deng}, \citenamefont {Dong}, \citenamefont {Sun}, \citenamefont {Li}, \citenamefont {Ye}, \citenamefont {Jiang}, \citenamefont {Ma}, \citenamefont {Shen}, \citenamefont {Zhang}, \citenamefont {Li}, \citenamefont {Guo}, \citenamefont {Wang}, \citenamefont {Song}, \citenamefont {Wang},\ and\ \citenamefont {Deng}}]{wang2025}%
  \BibitemOpen
  \bibfield  {author} {\bibinfo {author} {\bibfnamefont {K.}~\bibnamefont {Wang}}, \bibinfo {author} {\bibfnamefont {Z.}~\bibnamefont {Lu}}, \bibinfo {author} {\bibfnamefont {C.}~\bibnamefont {Zhang}}, \bibinfo {author} {\bibfnamefont {G.}~\bibnamefont {Liu}}, \bibinfo {author} {\bibfnamefont {J.}~\bibnamefont {Chen}}, \bibinfo {author} {\bibfnamefont {Y.}~\bibnamefont {Wang}}, \bibinfo {author} {\bibfnamefont {Y.}~\bibnamefont {Wu}}, \bibinfo {author} {\bibfnamefont {S.}~\bibnamefont {Xu}}, \bibinfo {author} {\bibfnamefont {X.}~\bibnamefont {Zhu}}, \bibinfo {author} {\bibfnamefont {F.}~\bibnamefont {Jin}}, \bibinfo {author} {\bibfnamefont {Y.}~\bibnamefont {Gao}}, \bibinfo {author} {\bibfnamefont {Z.}~\bibnamefont {Tan}}, \bibinfo {author} {\bibfnamefont {Z.}~\bibnamefont {Cui}}, \bibinfo {author} {\bibfnamefont {N.}~\bibnamefont {Wang}}, \bibinfo {author} {\bibfnamefont {Y.}~\bibnamefont {Zou}}, \bibinfo {author} {\bibfnamefont {A.}~\bibnamefont {Zhang}}, \bibinfo {author} {\bibfnamefont {T.}~\bibnamefont {Li}}, \bibinfo {author} {\bibfnamefont {F.}~\bibnamefont {Shen}}, \bibinfo {author} {\bibfnamefont {J.}~\bibnamefont {Zhong}}, \bibinfo {author} {\bibfnamefont {Z.}~\bibnamefont {Bao}}, \bibinfo {author} {\bibfnamefont {Z.}~\bibnamefont {Zhu}}, \bibinfo {author} {\bibfnamefont {Y.}~\bibnamefont {Han}}, \bibinfo {author} {\bibfnamefont {Y.}~\bibnamefont {He}}, \bibinfo {author} {\bibfnamefont {J.}~\bibnamefont {Shen}}, \bibinfo {author} {\bibfnamefont {H.}~\bibnamefont {Wang}}, \bibinfo {author} {\bibfnamefont {J.-N.}\ \bibnamefont {Yang}}, \bibinfo {author} {\bibfnamefont {Z.}~\bibnamefont {Song}}, \bibinfo {author} {\bibfnamefont {J.}~\bibnamefont {Deng}}, \bibinfo {author} {\bibfnamefont {H.}~\bibnamefont {Dong}}, \bibinfo {author} {\bibfnamefont {Z.-Z.}\ \bibnamefont {Sun}}, \bibinfo {author} {\bibfnamefont {W.}~\bibnamefont {Li}}, \bibinfo {author} {\bibfnamefont {Q.}~\bibnamefont {Ye}}, \bibinfo {author} {\bibfnamefont {S.}~\bibnamefont {Jiang}}, \bibinfo {author} {\bibfnamefont {Y.}~\bibnamefont {Ma}}, \bibinfo {author} {\bibfnamefont {P.-X.}\ \bibnamefont {Shen}}, \bibinfo {author} {\bibfnamefont {P.}~\bibnamefont {Zhang}}, \bibinfo {author} {\bibfnamefont {H.}~\bibnamefont {Li}}, \bibinfo {author} {\bibfnamefont {Q.}~\bibnamefont {Guo}}, \bibinfo {author} {\bibfnamefont {Z.}~\bibnamefont {Wang}}, \bibinfo {author} {\bibfnamefont {C.}~\bibnamefont {Song}}, \bibinfo {author} {\bibfnamefont {H.}~\bibnamefont {Wang}},\ and\ \bibinfo {author} {\bibfnamefont {D.-L.}\ \bibnamefont {Deng}},\ }\href {https://doi.org/10.48550/arXiv.2505.09684} {\bibinfo {title} {Demonstration of low-overhead quantum error correction codes}} (\bibinfo {year} {2025}),\ \bibinfo {note} {arXiv:2505.09684 [quant-ph]}\BibitemShut {NoStop}%
\bibitem [{\citenamefont {Gottesman}\ and\ \citenamefont {Chuang}(1999)}]{gottesman1999}%
  \BibitemOpen
  \bibfield  {author} {\bibinfo {author} {\bibfnamefont {D.}~\bibnamefont {Gottesman}}\ and\ \bibinfo {author} {\bibfnamefont {I.~L.}\ \bibnamefont {Chuang}},\ }\bibfield  {title} {\bibinfo {title} {Demonstrating the viability of universal quantum computation using teleportation and single-qubit operations},\ }\href {https://doi.org/https://doi.org/10.1038/46503} {\bibfield  {journal} {\bibinfo  {journal} {Nature}\ }\textbf {\bibinfo {volume} {402}},\ \bibinfo {pages} {390} (\bibinfo {year} {1999})}\BibitemShut {NoStop}%
\bibitem [{\citenamefont {Main}\ \emph {et~al.}(2025)\citenamefont {Main}, \citenamefont {Drmota}, \citenamefont {Nadlinger}, \citenamefont {Ainley}, \citenamefont {Agrawal}, \citenamefont {Nichol}, \citenamefont {Srinivas}, \citenamefont {Araneda},\ and\ \citenamefont {Lucas}}]{main2025}%
  \BibitemOpen
  \bibfield  {author} {\bibinfo {author} {\bibfnamefont {D.}~\bibnamefont {Main}}, \bibinfo {author} {\bibfnamefont {P.}~\bibnamefont {Drmota}}, \bibinfo {author} {\bibfnamefont {D.~P.}\ \bibnamefont {Nadlinger}}, \bibinfo {author} {\bibfnamefont {E.~M.}\ \bibnamefont {Ainley}}, \bibinfo {author} {\bibfnamefont {A.}~\bibnamefont {Agrawal}}, \bibinfo {author} {\bibfnamefont {B.~C.}\ \bibnamefont {Nichol}}, \bibinfo {author} {\bibfnamefont {R.}~\bibnamefont {Srinivas}}, \bibinfo {author} {\bibfnamefont {G.}~\bibnamefont {Araneda}},\ and\ \bibinfo {author} {\bibfnamefont {D.~M.}\ \bibnamefont {Lucas}},\ }\bibfield  {title} {\bibinfo {title} {Distributed quantum computing across an optical network link},\ }\href {https://doi.org/10.1038/s41586-024-08404-x} {\bibfield  {journal} {\bibinfo  {journal} {Nature}\ }\textbf {\bibinfo {volume} {638}},\ \bibinfo {pages} {383} (\bibinfo {year} {2025})}\BibitemShut {NoStop}%
\bibitem [{\citenamefont {Chou}\ \emph {et~al.}(2018)\citenamefont {Chou}, \citenamefont {Blumoff}, \citenamefont {Wang}, \citenamefont {Reinhold}, \citenamefont {Axline}, \citenamefont {Gao}, \citenamefont {Frunzio}, \citenamefont {Devoret}, \citenamefont {Jiang},\ and\ \citenamefont {Schoelkopf}}]{chou2018}%
  \BibitemOpen
  \bibfield  {author} {\bibinfo {author} {\bibfnamefont {K.~S.}\ \bibnamefont {Chou}}, \bibinfo {author} {\bibfnamefont {J.~Z.}\ \bibnamefont {Blumoff}}, \bibinfo {author} {\bibfnamefont {C.~S.}\ \bibnamefont {Wang}}, \bibinfo {author} {\bibfnamefont {P.~C.}\ \bibnamefont {Reinhold}}, \bibinfo {author} {\bibfnamefont {C.~J.}\ \bibnamefont {Axline}}, \bibinfo {author} {\bibfnamefont {Y.~Y.}\ \bibnamefont {Gao}}, \bibinfo {author} {\bibfnamefont {L.}~\bibnamefont {Frunzio}}, \bibinfo {author} {\bibfnamefont {M.}~\bibnamefont {Devoret}}, \bibinfo {author} {\bibfnamefont {L.}~\bibnamefont {Jiang}},\ and\ \bibinfo {author} {\bibfnamefont {R.}~\bibnamefont {Schoelkopf}},\ }\bibfield  {title} {\bibinfo {title} {Deterministic teleportation of a quantum gate between two logical qubits},\ }\href {https://doi.org/https://doi.org/10.1038/s41586-018-0470-y} {\bibfield  {journal} {\bibinfo  {journal} {Nature}\ }\textbf {\bibinfo {volume} {561}},\ \bibinfo {pages} {368} (\bibinfo {year} {2018})}\BibitemShut {NoStop}%
\bibitem [{\citenamefont {Qiu}\ \emph {et~al.}(2025{\natexlab{b}})\citenamefont {Qiu}, \citenamefont {Liu}, \citenamefont {Hu}, \citenamefont {Wu}, \citenamefont {Niu}, \citenamefont {Zhang}, \citenamefont {Huang}, \citenamefont {Chen}, \citenamefont {Li}, \citenamefont {Liu}, \citenamefont {Zhong}, \citenamefont {Duan},\ and\ \citenamefont {Yu}}]{qiu2025}%
  \BibitemOpen
  \bibfield  {author} {\bibinfo {author} {\bibfnamefont {J.}~\bibnamefont {Qiu}}, \bibinfo {author} {\bibfnamefont {Y.}~\bibnamefont {Liu}}, \bibinfo {author} {\bibfnamefont {L.}~\bibnamefont {Hu}}, \bibinfo {author} {\bibfnamefont {Y.}~\bibnamefont {Wu}}, \bibinfo {author} {\bibfnamefont {J.}~\bibnamefont {Niu}}, \bibinfo {author} {\bibfnamefont {L.}~\bibnamefont {Zhang}}, \bibinfo {author} {\bibfnamefont {W.}~\bibnamefont {Huang}}, \bibinfo {author} {\bibfnamefont {Y.}~\bibnamefont {Chen}}, \bibinfo {author} {\bibfnamefont {J.}~\bibnamefont {Li}}, \bibinfo {author} {\bibfnamefont {S.}~\bibnamefont {Liu}}, \bibinfo {author} {\bibfnamefont {Y.}~\bibnamefont {Zhong}}, \bibinfo {author} {\bibfnamefont {L.}~\bibnamefont {Duan}},\ and\ \bibinfo {author} {\bibfnamefont {D.}~\bibnamefont {Yu}},\ }\bibfield  {title} {\bibinfo {title} {Deterministic quantum state and gate teleportation between distant superconducting chips},\ }\href {https://doi.org/10.1016/j.scib.2024.11.047} {\bibfield  {journal} {\bibinfo  {journal} {Sci. Bull.}\ }\textbf {\bibinfo {volume} {70}},\ \bibinfo {pages} {351} (\bibinfo {year} {2025}{\natexlab{b}})}\BibitemShut {NoStop}%
\bibitem [{\citenamefont {Zhong}\ \emph {et~al.}(2019)\citenamefont {Zhong}, \citenamefont {Chang}, \citenamefont {Satzinger}, \citenamefont {Chou}, \citenamefont {Bienfait}, \citenamefont {Conner}, \citenamefont {Dumur}, \citenamefont {Grebel}, \citenamefont {Peairs}, \citenamefont {Povey} \emph {et~al.}}]{zhong2019}%
  \BibitemOpen
  \bibfield  {author} {\bibinfo {author} {\bibfnamefont {Y.}~\bibnamefont {Zhong}}, \bibinfo {author} {\bibfnamefont {H.-S.}\ \bibnamefont {Chang}}, \bibinfo {author} {\bibfnamefont {K.}~\bibnamefont {Satzinger}}, \bibinfo {author} {\bibfnamefont {M.-H.}\ \bibnamefont {Chou}}, \bibinfo {author} {\bibfnamefont {A.}~\bibnamefont {Bienfait}}, \bibinfo {author} {\bibfnamefont {C.}~\bibnamefont {Conner}}, \bibinfo {author} {\bibfnamefont {{\'E}.}~\bibnamefont {Dumur}}, \bibinfo {author} {\bibfnamefont {J.}~\bibnamefont {Grebel}}, \bibinfo {author} {\bibfnamefont {G.}~\bibnamefont {Peairs}}, \bibinfo {author} {\bibfnamefont {R.}~\bibnamefont {Povey}}, \emph {et~al.},\ }\bibfield  {title} {\bibinfo {title} {Violating bell’s inequality with remotely connected superconducting qubits},\ }\href {https://doi.org/https://doi.org/10.1038/s41567-019-0507-7} {\bibfield  {journal} {\bibinfo  {journal} {Nat. Phys.}\ }\textbf {\bibinfo {volume} {15}},\ \bibinfo {pages} {741} (\bibinfo {year} {2019})}\BibitemShut {NoStop}%
\bibitem [{\citenamefont {Niu}\ \emph {et~al.}(2023)\citenamefont {Niu}, \citenamefont {Zhang}, \citenamefont {Liu}, \citenamefont {Qiu}, \citenamefont {Huang}, \citenamefont {Huang}, \citenamefont {Jia}, \citenamefont {Liu}, \citenamefont {Tao}, \citenamefont {Wei} \emph {et~al.}}]{niu2023}%
  \BibitemOpen
  \bibfield  {author} {\bibinfo {author} {\bibfnamefont {J.}~\bibnamefont {Niu}}, \bibinfo {author} {\bibfnamefont {L.}~\bibnamefont {Zhang}}, \bibinfo {author} {\bibfnamefont {Y.}~\bibnamefont {Liu}}, \bibinfo {author} {\bibfnamefont {J.}~\bibnamefont {Qiu}}, \bibinfo {author} {\bibfnamefont {W.}~\bibnamefont {Huang}}, \bibinfo {author} {\bibfnamefont {J.}~\bibnamefont {Huang}}, \bibinfo {author} {\bibfnamefont {H.}~\bibnamefont {Jia}}, \bibinfo {author} {\bibfnamefont {J.}~\bibnamefont {Liu}}, \bibinfo {author} {\bibfnamefont {Z.}~\bibnamefont {Tao}}, \bibinfo {author} {\bibfnamefont {W.}~\bibnamefont {Wei}}, \emph {et~al.},\ }\bibfield  {title} {\bibinfo {title} {Low-loss interconnects for modular superconducting quantum processors},\ }\href {https://doi.org/https://doi.org/10.1038/s41928-023-00925-z} {\bibfield  {journal} {\bibinfo  {journal} {Nat. Electron.}\ }\textbf {\bibinfo {volume} {6}},\ \bibinfo {pages} {235} (\bibinfo {year} {2023})}\BibitemShut {NoStop}%
\bibitem [{\citenamefont {Zhong}\ \emph {et~al.}(2021)\citenamefont {Zhong}, \citenamefont {Chang}, \citenamefont {Bienfait}, \citenamefont {Dumur}, \citenamefont {Chou}, \citenamefont {Conner}, \citenamefont {Grebel}, \citenamefont {Povey}, \citenamefont {Yan}, \citenamefont {Schuster} \emph {et~al.}}]{zhong2021}%
  \BibitemOpen
  \bibfield  {author} {\bibinfo {author} {\bibfnamefont {Y.}~\bibnamefont {Zhong}}, \bibinfo {author} {\bibfnamefont {H.-S.}\ \bibnamefont {Chang}}, \bibinfo {author} {\bibfnamefont {A.}~\bibnamefont {Bienfait}}, \bibinfo {author} {\bibfnamefont {{\'E}.}~\bibnamefont {Dumur}}, \bibinfo {author} {\bibfnamefont {M.-H.}\ \bibnamefont {Chou}}, \bibinfo {author} {\bibfnamefont {C.~R.}\ \bibnamefont {Conner}}, \bibinfo {author} {\bibfnamefont {J.}~\bibnamefont {Grebel}}, \bibinfo {author} {\bibfnamefont {R.~G.}\ \bibnamefont {Povey}}, \bibinfo {author} {\bibfnamefont {H.}~\bibnamefont {Yan}}, \bibinfo {author} {\bibfnamefont {D.~I.}\ \bibnamefont {Schuster}}, \emph {et~al.},\ }\bibfield  {title} {\bibinfo {title} {Deterministic multi-qubit entanglement in a quantum network},\ }\href {https://doi.org/https://doi.org/10.1038/s41586-021-03288-7} {\bibfield  {journal} {\bibinfo  {journal} {Nature}\ }\textbf {\bibinfo {volume} {590}},\ \bibinfo {pages} {571} (\bibinfo {year} {2021})}\BibitemShut {NoStop}%
\bibitem [{\citenamefont {Campagne-Ibarcq}\ \emph {et~al.}(2018)\citenamefont {Campagne-Ibarcq}, \citenamefont {Zalys-Geller}, \citenamefont {Narla}, \citenamefont {Shankar}, \citenamefont {Reinhold}, \citenamefont {Burkhart}, \citenamefont {Axline}, \citenamefont {Pfaff}, \citenamefont {Frunzio}, \citenamefont {Schoelkopf},\ and\ \citenamefont {Devoret}}]{Campagne2018}%
  \BibitemOpen
  \bibfield  {author} {\bibinfo {author} {\bibfnamefont {P.}~\bibnamefont {Campagne-Ibarcq}}, \bibinfo {author} {\bibfnamefont {E.}~\bibnamefont {Zalys-Geller}}, \bibinfo {author} {\bibfnamefont {A.}~\bibnamefont {Narla}}, \bibinfo {author} {\bibfnamefont {S.}~\bibnamefont {Shankar}}, \bibinfo {author} {\bibfnamefont {P.}~\bibnamefont {Reinhold}}, \bibinfo {author} {\bibfnamefont {L.}~\bibnamefont {Burkhart}}, \bibinfo {author} {\bibfnamefont {C.}~\bibnamefont {Axline}}, \bibinfo {author} {\bibfnamefont {W.}~\bibnamefont {Pfaff}}, \bibinfo {author} {\bibfnamefont {L.}~\bibnamefont {Frunzio}}, \bibinfo {author} {\bibfnamefont {R.~J.}\ \bibnamefont {Schoelkopf}},\ and\ \bibinfo {author} {\bibfnamefont {M.~H.}\ \bibnamefont {Devoret}},\ }\bibfield  {title} {\bibinfo {title} {Deterministic remote entanglement of superconducting circuits through microwave two-photon transitions},\ }\href {https://doi.org/10.1103/PhysRevLett.120.200501} {\bibfield  {journal} {\bibinfo  {journal} {Phys. Rev. Lett.}\ }\textbf {\bibinfo {volume} {120}},\ \bibinfo {pages} {200501} (\bibinfo {year} {2018})}\BibitemShut {NoStop}%
\bibitem [{\citenamefont {Magnard}\ \emph {et~al.}(2020)\citenamefont {Magnard}, \citenamefont {Storz}, \citenamefont {Kurpiers}, \citenamefont {Sch\"ar}, \citenamefont {Marxer}, \citenamefont {L\"utolf}, \citenamefont {Walter}, \citenamefont {Besse}, \citenamefont {Gabureac}, \citenamefont {Reuer}, \citenamefont {Akin}, \citenamefont {Royer}, \citenamefont {Blais},\ and\ \citenamefont {Wallraff}}]{Magnard2020}%
  \BibitemOpen
  \bibfield  {author} {\bibinfo {author} {\bibfnamefont {P.}~\bibnamefont {Magnard}}, \bibinfo {author} {\bibfnamefont {S.}~\bibnamefont {Storz}}, \bibinfo {author} {\bibfnamefont {P.}~\bibnamefont {Kurpiers}}, \bibinfo {author} {\bibfnamefont {J.}~\bibnamefont {Sch\"ar}}, \bibinfo {author} {\bibfnamefont {F.}~\bibnamefont {Marxer}}, \bibinfo {author} {\bibfnamefont {J.}~\bibnamefont {L\"utolf}}, \bibinfo {author} {\bibfnamefont {T.}~\bibnamefont {Walter}}, \bibinfo {author} {\bibfnamefont {J.-C.}\ \bibnamefont {Besse}}, \bibinfo {author} {\bibfnamefont {M.}~\bibnamefont {Gabureac}}, \bibinfo {author} {\bibfnamefont {K.}~\bibnamefont {Reuer}}, \bibinfo {author} {\bibfnamefont {A.}~\bibnamefont {Akin}}, \bibinfo {author} {\bibfnamefont {B.}~\bibnamefont {Royer}}, \bibinfo {author} {\bibfnamefont {A.}~\bibnamefont {Blais}},\ and\ \bibinfo {author} {\bibfnamefont {A.}~\bibnamefont {Wallraff}},\ }\bibfield  {title} {\bibinfo {title} {Microwave quantum link between superconducting circuits housed in spatially separated cryogenic systems},\ }\href {https://doi.org/10.1103/PhysRevLett.125.260502} {\bibfield  {journal} {\bibinfo  {journal} {Phys. Rev. Lett.}\ }\textbf {\bibinfo {volume} {125}},\ \bibinfo {pages} {260502} (\bibinfo {year} {2020})}\BibitemShut {NoStop}%
\bibitem [{\citenamefont {Kannan}\ \emph {et~al.}(2023)\citenamefont {Kannan}, \citenamefont {Almanakly}, \citenamefont {Sung}, \citenamefont {Di~Paolo}, \citenamefont {Rower}, \citenamefont {Braum{\"u}ller}, \citenamefont {Melville}, \citenamefont {Niedzielski}, \citenamefont {Karamlou}, \citenamefont {Serniak} \emph {et~al.}}]{kannan2023}%
  \BibitemOpen
  \bibfield  {author} {\bibinfo {author} {\bibfnamefont {B.}~\bibnamefont {Kannan}}, \bibinfo {author} {\bibfnamefont {A.}~\bibnamefont {Almanakly}}, \bibinfo {author} {\bibfnamefont {Y.}~\bibnamefont {Sung}}, \bibinfo {author} {\bibfnamefont {A.}~\bibnamefont {Di~Paolo}}, \bibinfo {author} {\bibfnamefont {D.~A.}\ \bibnamefont {Rower}}, \bibinfo {author} {\bibfnamefont {J.}~\bibnamefont {Braum{\"u}ller}}, \bibinfo {author} {\bibfnamefont {A.}~\bibnamefont {Melville}}, \bibinfo {author} {\bibfnamefont {B.~M.}\ \bibnamefont {Niedzielski}}, \bibinfo {author} {\bibfnamefont {A.}~\bibnamefont {Karamlou}}, \bibinfo {author} {\bibfnamefont {K.}~\bibnamefont {Serniak}}, \emph {et~al.},\ }\bibfield  {title} {\bibinfo {title} {On-demand directional microwave photon emission using waveguide quantum electrodynamics},\ }\href {https://doi.org/https://doi.org/10.1038/s41567-022-01869-5} {\bibfield  {journal} {\bibinfo  {journal} {Nat. Phys.}\ }\textbf {\bibinfo {volume} {19}},\ \bibinfo {pages} {394} (\bibinfo {year} {2023})}\BibitemShut {NoStop}%
\bibitem [{\citenamefont {Storz}\ \emph {et~al.}(2023)\citenamefont {Storz}, \citenamefont {Sch{\"a}r}, \citenamefont {Kulikov}, \citenamefont {Magnard}, \citenamefont {Kurpiers}, \citenamefont {L{\"u}tolf}, \citenamefont {Walter}, \citenamefont {Copetudo}, \citenamefont {Reuer}, \citenamefont {Akin} \emph {et~al.}}]{storz2023}%
  \BibitemOpen
  \bibfield  {author} {\bibinfo {author} {\bibfnamefont {S.}~\bibnamefont {Storz}}, \bibinfo {author} {\bibfnamefont {J.}~\bibnamefont {Sch{\"a}r}}, \bibinfo {author} {\bibfnamefont {A.}~\bibnamefont {Kulikov}}, \bibinfo {author} {\bibfnamefont {P.}~\bibnamefont {Magnard}}, \bibinfo {author} {\bibfnamefont {P.}~\bibnamefont {Kurpiers}}, \bibinfo {author} {\bibfnamefont {J.}~\bibnamefont {L{\"u}tolf}}, \bibinfo {author} {\bibfnamefont {T.}~\bibnamefont {Walter}}, \bibinfo {author} {\bibfnamefont {A.}~\bibnamefont {Copetudo}}, \bibinfo {author} {\bibfnamefont {K.}~\bibnamefont {Reuer}}, \bibinfo {author} {\bibfnamefont {A.}~\bibnamefont {Akin}}, \emph {et~al.},\ }\bibfield  {title} {\bibinfo {title} {Loophole-free bell inequality violation with superconducting circuits},\ }\href {https://doi.org/https://doi.org/10.1038/s41586-023-05885-0} {\bibfield  {journal} {\bibinfo  {journal} {Nature}\ }\textbf {\bibinfo {volume} {617}},\ \bibinfo {pages} {265} (\bibinfo {year} {2023})}\BibitemShut {NoStop}%
\bibitem [{\citenamefont {Zhao}\ \emph {et~al.}(2022)\citenamefont {Zhao}, \citenamefont {Zhang}, \citenamefont {Xue}, \citenamefont {Jin},\ and\ \citenamefont {Yu}}]{zhao2022}%
  \BibitemOpen
  \bibfield  {author} {\bibinfo {author} {\bibfnamefont {P.}~\bibnamefont {Zhao}}, \bibinfo {author} {\bibfnamefont {Y.}~\bibnamefont {Zhang}}, \bibinfo {author} {\bibfnamefont {G.}~\bibnamefont {Xue}}, \bibinfo {author} {\bibfnamefont {Y.}~\bibnamefont {Jin}},\ and\ \bibinfo {author} {\bibfnamefont {H.}~\bibnamefont {Yu}},\ }\bibfield  {title} {\bibinfo {title} {Tunable coupling of widely separated superconducting qubits: A possible application toward a modular quantum device},\ }\href {https://doi.org/10.1063/5.0097521} {\bibfield  {journal} {\bibinfo  {journal} {Appl. Phys. Lett.}\ }\textbf {\bibinfo {volume} {121}},\ \bibinfo {pages} {32601} (\bibinfo {year} {2022})}\BibitemShut {NoStop}%
\bibitem [{\citenamefont {Mollenhauer}\ \emph {et~al.}(2025)\citenamefont {Mollenhauer}, \citenamefont {Irfan}, \citenamefont {Cao}, \citenamefont {Mandal},\ and\ \citenamefont {Pfaff}}]{mollenhauer2024}%
  \BibitemOpen
  \bibfield  {author} {\bibinfo {author} {\bibfnamefont {M.}~\bibnamefont {Mollenhauer}}, \bibinfo {author} {\bibfnamefont {A.}~\bibnamefont {Irfan}}, \bibinfo {author} {\bibfnamefont {X.}~\bibnamefont {Cao}}, \bibinfo {author} {\bibfnamefont {S.}~\bibnamefont {Mandal}},\ and\ \bibinfo {author} {\bibfnamefont {W.}~\bibnamefont {Pfaff}},\ }\bibfield  {title} {\bibinfo {title} {A high-efficiency elementary network of interchangeable superconducting qubit devices},\ }\href {https://doi.org/10.1038/s41928-025-01404-3} {\bibfield  {journal} {\bibinfo  {journal} {Nat. Electron.}\ }\textbf {\bibinfo {volume} {8}},\ \bibinfo {pages} {610} (\bibinfo {year} {2025})}\BibitemShut {NoStop}%
\bibitem [{\citenamefont {Song}\ \emph {et~al.}(2025)\citenamefont {Song}, \citenamefont {Yang}, \citenamefont {Liu}, \citenamefont {Zhang}, \citenamefont {Xue}, \citenamefont {Mi}, \citenamefont {Zhang}, \citenamefont {Yan}, \citenamefont {Jin},\ and\ \citenamefont {Yu}}]{song2024}%
  \BibitemOpen
  \bibfield  {author} {\bibinfo {author} {\bibfnamefont {J.}~\bibnamefont {Song}}, \bibinfo {author} {\bibfnamefont {S.}~\bibnamefont {Yang}}, \bibinfo {author} {\bibfnamefont {P.}~\bibnamefont {Liu}}, \bibinfo {author} {\bibfnamefont {H.-L.}\ \bibnamefont {Zhang}}, \bibinfo {author} {\bibfnamefont {G.-M.}\ \bibnamefont {Xue}}, \bibinfo {author} {\bibfnamefont {Z.-Y.}\ \bibnamefont {Mi}}, \bibinfo {author} {\bibfnamefont {W.-G.}\ \bibnamefont {Zhang}}, \bibinfo {author} {\bibfnamefont {F.}~\bibnamefont {Yan}}, \bibinfo {author} {\bibfnamefont {Y.-R.}\ \bibnamefont {Jin}},\ and\ \bibinfo {author} {\bibfnamefont {H.-F.}\ \bibnamefont {Yu}},\ }\bibfield  {title} {\bibinfo {title} {Realization of high-fidelity perfect entanglers between remote superconducting quantum processors},\ }\href {https://doi.org/10.1103/npr7-b7kq} {\bibfield  {journal} {\bibinfo  {journal} {Phys. Rev. Lett.}\ }\textbf {\bibinfo {volume} {135}},\ \bibinfo {pages} {050603} (\bibinfo {year} {2025})}\BibitemShut {NoStop}%
\bibitem [{\citenamefont {Deng}\ \emph {et~al.}(2025)\citenamefont {Deng}, \citenamefont {Zheng}, \citenamefont {Liao}, \citenamefont {Zhou}, \citenamefont {Ge}, \citenamefont {Zhao}, \citenamefont {Lan}, \citenamefont {Tan}, \citenamefont {Zhang}, \citenamefont {Li},\ and\ \citenamefont {Yu}}]{deng2025}%
  \BibitemOpen
  \bibfield  {author} {\bibinfo {author} {\bibfnamefont {X.}~\bibnamefont {Deng}}, \bibinfo {author} {\bibfnamefont {W.}~\bibnamefont {Zheng}}, \bibinfo {author} {\bibfnamefont {X.}~\bibnamefont {Liao}}, \bibinfo {author} {\bibfnamefont {H.}~\bibnamefont {Zhou}}, \bibinfo {author} {\bibfnamefont {Y.}~\bibnamefont {Ge}}, \bibinfo {author} {\bibfnamefont {J.}~\bibnamefont {Zhao}}, \bibinfo {author} {\bibfnamefont {D.}~\bibnamefont {Lan}}, \bibinfo {author} {\bibfnamefont {X.}~\bibnamefont {Tan}}, \bibinfo {author} {\bibfnamefont {Y.}~\bibnamefont {Zhang}}, \bibinfo {author} {\bibfnamefont {S.}~\bibnamefont {Li}},\ and\ \bibinfo {author} {\bibfnamefont {Y.}~\bibnamefont {Yu}},\ }\bibfield  {title} {\bibinfo {title} {Long-range $zz$ interaction via resonator-induced phase in superconducting qubits},\ }\href {https://doi.org/10.1103/PhysRevLett.134.020801} {\bibfield  {journal} {\bibinfo  {journal} {Phys. Rev. Lett.}\ }\textbf {\bibinfo {volume} {134}},\ \bibinfo {pages} {020801} (\bibinfo {year} {2025})}\BibitemShut {NoStop}%
\bibitem [{\citenamefont {Xiong}\ \emph {et~al.}(2025)\citenamefont {Xiong}, \citenamefont {Wang}, \citenamefont {Song}, \citenamefont {Yang}, \citenamefont {Bao}, \citenamefont {Li}, \citenamefont {Mi}, \citenamefont {Zhang}, \citenamefont {Yu}, \citenamefont {Song},\ and\ \citenamefont {Duan}}]{xiong2025}%
  \BibitemOpen
  \bibfield  {author} {\bibinfo {author} {\bibfnamefont {H.}~\bibnamefont {Xiong}}, \bibinfo {author} {\bibfnamefont {J.}~\bibnamefont {Wang}}, \bibinfo {author} {\bibfnamefont {J.}~\bibnamefont {Song}}, \bibinfo {author} {\bibfnamefont {J.}~\bibnamefont {Yang}}, \bibinfo {author} {\bibfnamefont {Z.}~\bibnamefont {Bao}}, \bibinfo {author} {\bibfnamefont {Y.}~\bibnamefont {Li}}, \bibinfo {author} {\bibfnamefont {Z.-Y.}\ \bibnamefont {Mi}}, \bibinfo {author} {\bibfnamefont {H.}~\bibnamefont {Zhang}}, \bibinfo {author} {\bibfnamefont {H.-F.}\ \bibnamefont {Yu}}, \bibinfo {author} {\bibfnamefont {Y.}~\bibnamefont {Song}},\ and\ \bibinfo {author} {\bibfnamefont {L.}~\bibnamefont {Duan}},\ }\href {https://arxiv.org/abs/2502.18902} {\bibinfo {title} {Scalable low-overhead superconducting non-local coupler with exponentially enhanced connectivity}} (\bibinfo {year} {2025}),\ \Eprint {https://arxiv.org/abs/2502.18902} {arXiv:2502.18902 [quant-ph]} \BibitemShut {NoStop}%
\bibitem [{\citenamefont {Heya}\ \emph {et~al.}(2025)\citenamefont {Heya}, \citenamefont {Phung}, \citenamefont {Malekakhlagh}, \citenamefont {Steiner}, \citenamefont {Turchetti}, \citenamefont {Shanks}, \citenamefont {Mamin}, \citenamefont {Lu}, \citenamefont {Kandel}, \citenamefont {Sundaresan},\ and\ \citenamefont {Orcutt}}]{heya2025}%
  \BibitemOpen
  \bibfield  {author} {\bibinfo {author} {\bibfnamefont {K.}~\bibnamefont {Heya}}, \bibinfo {author} {\bibfnamefont {T.}~\bibnamefont {Phung}}, \bibinfo {author} {\bibfnamefont {M.}~\bibnamefont {Malekakhlagh}}, \bibinfo {author} {\bibfnamefont {R.}~\bibnamefont {Steiner}}, \bibinfo {author} {\bibfnamefont {M.}~\bibnamefont {Turchetti}}, \bibinfo {author} {\bibfnamefont {W.}~\bibnamefont {Shanks}}, \bibinfo {author} {\bibfnamefont {J.}~\bibnamefont {Mamin}}, \bibinfo {author} {\bibfnamefont {W.-S.}\ \bibnamefont {Lu}}, \bibinfo {author} {\bibfnamefont {Y.~P.}\ \bibnamefont {Kandel}}, \bibinfo {author} {\bibfnamefont {N.}~\bibnamefont {Sundaresan}},\ and\ \bibinfo {author} {\bibfnamefont {J.}~\bibnamefont {Orcutt}},\ }\href {https://arxiv.org/abs/2502.15034} {\bibinfo {title} {Randomized benchmarking of a high-fidelity remote cnot gate over a meter-scale microwave interconnect}} (\bibinfo {year} {2025}),\ \Eprint {https://arxiv.org/abs/2502.15034} {arXiv:2502.15034 [quant-ph]} \BibitemShut {NoStop}%
\bibitem [{\citenamefont {Majer}\ \emph {et~al.}(2007)\citenamefont {Majer}, \citenamefont {Chow}, \citenamefont {Gambetta}, \citenamefont {Koch}, \citenamefont {Johnson}, \citenamefont {Schreier}, \citenamefont {Frunzio}, \citenamefont {Schuster}, \citenamefont {Houck}, \citenamefont {Wallraff} \emph {et~al.}}]{majer2007}%
  \BibitemOpen
  \bibfield  {author} {\bibinfo {author} {\bibfnamefont {J.}~\bibnamefont {Majer}}, \bibinfo {author} {\bibfnamefont {J.}~\bibnamefont {Chow}}, \bibinfo {author} {\bibfnamefont {J.}~\bibnamefont {Gambetta}}, \bibinfo {author} {\bibfnamefont {J.}~\bibnamefont {Koch}}, \bibinfo {author} {\bibfnamefont {B.}~\bibnamefont {Johnson}}, \bibinfo {author} {\bibfnamefont {J.}~\bibnamefont {Schreier}}, \bibinfo {author} {\bibfnamefont {L.}~\bibnamefont {Frunzio}}, \bibinfo {author} {\bibfnamefont {D.}~\bibnamefont {Schuster}}, \bibinfo {author} {\bibfnamefont {A.~A.}\ \bibnamefont {Houck}}, \bibinfo {author} {\bibfnamefont {A.}~\bibnamefont {Wallraff}}, \emph {et~al.},\ }\bibfield  {title} {\bibinfo {title} {Coupling superconducting qubits via a cavity bus},\ }\href {https://doi.org/https://doi.org/10.1038/nature06184} {\bibfield  {journal} {\bibinfo  {journal} {Nature}\ }\textbf {\bibinfo {volume} {449}},\ \bibinfo {pages} {443} (\bibinfo {year} {2007})}\BibitemShut {NoStop}%
\bibitem [{\citenamefont {Sillanp{\"a}{\"a}}\ \emph {et~al.}(2007)\citenamefont {Sillanp{\"a}{\"a}}, \citenamefont {Park},\ and\ \citenamefont {Simmonds}}]{sillanpaa2007}%
  \BibitemOpen
  \bibfield  {author} {\bibinfo {author} {\bibfnamefont {M.~A.}\ \bibnamefont {Sillanp{\"a}{\"a}}}, \bibinfo {author} {\bibfnamefont {J.~I.}\ \bibnamefont {Park}},\ and\ \bibinfo {author} {\bibfnamefont {R.~W.}\ \bibnamefont {Simmonds}},\ }\bibfield  {title} {\bibinfo {title} {Coherent quantum state storage and transfer between two phase qubits via a resonant cavity},\ }\href {https://doi.org/https://doi.org/10.1038/nature06124} {\bibfield  {journal} {\bibinfo  {journal} {Nature}\ }\textbf {\bibinfo {volume} {449}},\ \bibinfo {pages} {438} (\bibinfo {year} {2007})}\BibitemShut {NoStop}%
\bibitem [{\citenamefont {Hazra}\ \emph {et~al.}(2021)\citenamefont {Hazra}, \citenamefont {Bhattacharjee}, \citenamefont {Chand}, \citenamefont {Salunkhe}, \citenamefont {Gopalakrishnan}, \citenamefont {Patankar},\ and\ \citenamefont {Vijay}}]{Hazra2021}%
  \BibitemOpen
  \bibfield  {author} {\bibinfo {author} {\bibfnamefont {S.}~\bibnamefont {Hazra}}, \bibinfo {author} {\bibfnamefont {A.}~\bibnamefont {Bhattacharjee}}, \bibinfo {author} {\bibfnamefont {M.}~\bibnamefont {Chand}}, \bibinfo {author} {\bibfnamefont {K.~V.}\ \bibnamefont {Salunkhe}}, \bibinfo {author} {\bibfnamefont {S.}~\bibnamefont {Gopalakrishnan}}, \bibinfo {author} {\bibfnamefont {M.~P.}\ \bibnamefont {Patankar}},\ and\ \bibinfo {author} {\bibfnamefont {R.}~\bibnamefont {Vijay}},\ }\bibfield  {title} {\bibinfo {title} {Ring-resonator-based coupling architecture for enhanced connectivity in a superconducting multiqubit network},\ }\href {https://doi.org/10.1103/PhysRevApplied.16.024018} {\bibfield  {journal} {\bibinfo  {journal} {Phys. Rev. Appl.}\ }\textbf {\bibinfo {volume} {16}},\ \bibinfo {pages} {024018} (\bibinfo {year} {2021})}\BibitemShut {NoStop}%
\bibitem [{\citenamefont {Wu}\ \emph {et~al.}(2024)\citenamefont {Wu}, \citenamefont {Yan}, \citenamefont {Andersson}, \citenamefont {Anferov}, \citenamefont {Chou}, \citenamefont {Conner}, \citenamefont {Grebel}, \citenamefont {Joshi}, \citenamefont {Li}, \citenamefont {Miller}, \citenamefont {Povey}, \citenamefont {Qiao},\ and\ \citenamefont {Cleland}}]{wu2024}%
  \BibitemOpen
  \bibfield  {author} {\bibinfo {author} {\bibfnamefont {X.}~\bibnamefont {Wu}}, \bibinfo {author} {\bibfnamefont {H.}~\bibnamefont {Yan}}, \bibinfo {author} {\bibfnamefont {G.}~\bibnamefont {Andersson}}, \bibinfo {author} {\bibfnamefont {A.}~\bibnamefont {Anferov}}, \bibinfo {author} {\bibfnamefont {M.-H.}\ \bibnamefont {Chou}}, \bibinfo {author} {\bibfnamefont {C.~R.}\ \bibnamefont {Conner}}, \bibinfo {author} {\bibfnamefont {J.}~\bibnamefont {Grebel}}, \bibinfo {author} {\bibfnamefont {Y.~J.}\ \bibnamefont {Joshi}}, \bibinfo {author} {\bibfnamefont {S.}~\bibnamefont {Li}}, \bibinfo {author} {\bibfnamefont {J.~M.}\ \bibnamefont {Miller}}, \bibinfo {author} {\bibfnamefont {R.~G.}\ \bibnamefont {Povey}}, \bibinfo {author} {\bibfnamefont {H.}~\bibnamefont {Qiao}},\ and\ \bibinfo {author} {\bibfnamefont {A.~N.}\ \bibnamefont {Cleland}},\ }\bibfield  {title} {\bibinfo {title} {Modular quantum processor with an all-to-all reconfigurable router},\ }\href {https://doi.org/10.1103/PhysRevX.14.041030} {\bibfield  {journal} {\bibinfo  {journal} {Phys. Rev. X}\ }\textbf {\bibinfo {volume} {14}},\ \bibinfo {pages} {041030} (\bibinfo {year} {2024})}\BibitemShut {NoStop}%
\bibitem [{\citenamefont {Zhong}\ \emph {et~al.}(2016)\citenamefont {Zhong}, \citenamefont {Xu}, \citenamefont {Wang}, \citenamefont {Song}, \citenamefont {Guo}, \citenamefont {Liu}, \citenamefont {Xu}, \citenamefont {Xia}, \citenamefont {Lu}, \citenamefont {Han}, \citenamefont {Pan},\ and\ \citenamefont {Wang}}]{Zhong2016}%
  \BibitemOpen
  \bibfield  {author} {\bibinfo {author} {\bibfnamefont {Y.~P.}\ \bibnamefont {Zhong}}, \bibinfo {author} {\bibfnamefont {D.}~\bibnamefont {Xu}}, \bibinfo {author} {\bibfnamefont {P.}~\bibnamefont {Wang}}, \bibinfo {author} {\bibfnamefont {C.}~\bibnamefont {Song}}, \bibinfo {author} {\bibfnamefont {Q.~J.}\ \bibnamefont {Guo}}, \bibinfo {author} {\bibfnamefont {W.~X.}\ \bibnamefont {Liu}}, \bibinfo {author} {\bibfnamefont {K.}~\bibnamefont {Xu}}, \bibinfo {author} {\bibfnamefont {B.~X.}\ \bibnamefont {Xia}}, \bibinfo {author} {\bibfnamefont {C.-Y.}\ \bibnamefont {Lu}}, \bibinfo {author} {\bibfnamefont {S.}~\bibnamefont {Han}}, \bibinfo {author} {\bibfnamefont {J.-W.}\ \bibnamefont {Pan}},\ and\ \bibinfo {author} {\bibfnamefont {H.}~\bibnamefont {Wang}},\ }\bibfield  {title} {\bibinfo {title} {Emulating anyonic fractional statistical behavior in a superconducting quantum circuit},\ }\href {https://doi.org/10.1103/PhysRevLett.117.110501} {\bibfield  {journal} {\bibinfo  {journal} {Phys. Rev. Lett.}\ }\textbf {\bibinfo {volume} {117}},\ \bibinfo {pages} {110501} (\bibinfo {year} {2016})}\BibitemShut {NoStop}%
\bibitem [{\citenamefont {Wang}\ \emph {et~al.}(2019)\citenamefont {Wang}, \citenamefont {Song}, \citenamefont {Feng}, \citenamefont {Cai}, \citenamefont {Xu}, \citenamefont {Deng}, \citenamefont {Li}, \citenamefont {Zheng}, \citenamefont {Zhu}, \citenamefont {Wang}, \citenamefont {Zhu},\ and\ \citenamefont {Scully}}]{wang2019}%
  \BibitemOpen
  \bibfield  {author} {\bibinfo {author} {\bibfnamefont {D.-W.}\ \bibnamefont {Wang}}, \bibinfo {author} {\bibfnamefont {C.}~\bibnamefont {Song}}, \bibinfo {author} {\bibfnamefont {W.}~\bibnamefont {Feng}}, \bibinfo {author} {\bibfnamefont {H.}~\bibnamefont {Cai}}, \bibinfo {author} {\bibfnamefont {D.}~\bibnamefont {Xu}}, \bibinfo {author} {\bibfnamefont {H.}~\bibnamefont {Deng}}, \bibinfo {author} {\bibfnamefont {H.}~\bibnamefont {Li}}, \bibinfo {author} {\bibfnamefont {D.}~\bibnamefont {Zheng}}, \bibinfo {author} {\bibfnamefont {X.}~\bibnamefont {Zhu}}, \bibinfo {author} {\bibfnamefont {H.}~\bibnamefont {Wang}}, \bibinfo {author} {\bibfnamefont {S.-Y.}\ \bibnamefont {Zhu}},\ and\ \bibinfo {author} {\bibfnamefont {M.~O.}\ \bibnamefont {Scully}},\ }\bibfield  {title} {\bibinfo {title} {Synthesis of antisymmetric spin exchange interaction and chiral spin clusters in superconducting circuits},\ }\href {https://doi.org/10.1038/s41567-018-0400-9} {\bibfield  {journal} {\bibinfo  {journal} {Nat. Phys.}\ }\textbf {\bibinfo {volume} {15}},\ \bibinfo {pages} {382} (\bibinfo {year} {2019})}\BibitemShut {NoStop}%
\bibitem [{\citenamefont {Song}\ \emph {et~al.}(2017)\citenamefont {Song}, \citenamefont {Xu}, \citenamefont {Liu}, \citenamefont {Yang}, \citenamefont {Zheng}, \citenamefont {Deng}, \citenamefont {Xie}, \citenamefont {Huang}, \citenamefont {Guo}, \citenamefont {Zhang}, \citenamefont {Zhang}, \citenamefont {Xu}, \citenamefont {Zheng}, \citenamefont {Zhu}, \citenamefont {Wang}, \citenamefont {Chen}, \citenamefont {Lu}, \citenamefont {Han},\ and\ \citenamefont {Pan}}]{song2017}%
  \BibitemOpen
  \bibfield  {author} {\bibinfo {author} {\bibfnamefont {C.}~\bibnamefont {Song}}, \bibinfo {author} {\bibfnamefont {K.}~\bibnamefont {Xu}}, \bibinfo {author} {\bibfnamefont {W.}~\bibnamefont {Liu}}, \bibinfo {author} {\bibfnamefont {C.-p.}\ \bibnamefont {Yang}}, \bibinfo {author} {\bibfnamefont {S.-B.}\ \bibnamefont {Zheng}}, \bibinfo {author} {\bibfnamefont {H.}~\bibnamefont {Deng}}, \bibinfo {author} {\bibfnamefont {Q.}~\bibnamefont {Xie}}, \bibinfo {author} {\bibfnamefont {K.}~\bibnamefont {Huang}}, \bibinfo {author} {\bibfnamefont {Q.}~\bibnamefont {Guo}}, \bibinfo {author} {\bibfnamefont {L.}~\bibnamefont {Zhang}}, \bibinfo {author} {\bibfnamefont {P.}~\bibnamefont {Zhang}}, \bibinfo {author} {\bibfnamefont {D.}~\bibnamefont {Xu}}, \bibinfo {author} {\bibfnamefont {D.}~\bibnamefont {Zheng}}, \bibinfo {author} {\bibfnamefont {X.}~\bibnamefont {Zhu}}, \bibinfo {author} {\bibfnamefont {H.}~\bibnamefont {Wang}}, \bibinfo {author} {\bibfnamefont {Y.-A.}\ \bibnamefont {Chen}}, \bibinfo {author} {\bibfnamefont {C.-Y.}\ \bibnamefont {Lu}}, \bibinfo {author} {\bibfnamefont {S.}~\bibnamefont {Han}},\ and\ \bibinfo {author} {\bibfnamefont {J.-W.}\ \bibnamefont {Pan}},\ }\bibfield  {title} {\bibinfo {title} {10-qubit entanglement and parallel logic operations with a superconducting circuit},\ }\href {https://doi.org/10.1103/PhysRevLett.119.180511} {\bibfield  {journal} {\bibinfo  {journal} {Phys. Rev. Lett.}\ }\textbf {\bibinfo {volume} {119}},\ \bibinfo {pages} {180511} (\bibinfo {year} {2017})}\BibitemShut {NoStop}%
\bibitem [{\citenamefont {Song}\ \emph {et~al.}(2019)\citenamefont {Song}, \citenamefont {Xu}, \citenamefont {Li}, \citenamefont {Zhang}, \citenamefont {Zhang}, \citenamefont {Liu}, \citenamefont {Guo}, \citenamefont {Wang}, \citenamefont {Ren}, \citenamefont {Hao}, \citenamefont {Feng}, \citenamefont {Fan}, \citenamefont {Zheng}, \citenamefont {Wang}, \citenamefont {Wang},\ and\ \citenamefont {Zhu}}]{song2019}%
  \BibitemOpen
  \bibfield  {author} {\bibinfo {author} {\bibfnamefont {C.}~\bibnamefont {Song}}, \bibinfo {author} {\bibfnamefont {K.}~\bibnamefont {Xu}}, \bibinfo {author} {\bibfnamefont {H.}~\bibnamefont {Li}}, \bibinfo {author} {\bibfnamefont {Y.-R.}\ \bibnamefont {Zhang}}, \bibinfo {author} {\bibfnamefont {X.}~\bibnamefont {Zhang}}, \bibinfo {author} {\bibfnamefont {W.}~\bibnamefont {Liu}}, \bibinfo {author} {\bibfnamefont {Q.}~\bibnamefont {Guo}}, \bibinfo {author} {\bibfnamefont {Z.}~\bibnamefont {Wang}}, \bibinfo {author} {\bibfnamefont {W.}~\bibnamefont {Ren}}, \bibinfo {author} {\bibfnamefont {J.}~\bibnamefont {Hao}}, \bibinfo {author} {\bibfnamefont {H.}~\bibnamefont {Feng}}, \bibinfo {author} {\bibfnamefont {H.}~\bibnamefont {Fan}}, \bibinfo {author} {\bibfnamefont {D.}~\bibnamefont {Zheng}}, \bibinfo {author} {\bibfnamefont {D.-W.}\ \bibnamefont {Wang}}, \bibinfo {author} {\bibfnamefont {H.}~\bibnamefont {Wang}},\ and\ \bibinfo {author} {\bibfnamefont {S.-Y.}\ \bibnamefont {Zhu}},\ }\bibfield  {title} {\bibinfo {title} {Generation of multicomponent atomic schrödinger cat states of up to 20 qubits},\ }\href {https://doi.org/10.1126/science.aay0600} {\bibfield  {journal} {\bibinfo  {journal} {Science}\ }\textbf {\bibinfo {volume} {365}},\ \bibinfo {pages} {574} (\bibinfo {year} {2019})}\BibitemShut {NoStop}%
\bibitem [{\citenamefont {Palacios-Laloy}\ \emph {et~al.}(2008)\citenamefont {Palacios-Laloy}, \citenamefont {Nguyen}, \citenamefont {Mallet}, \citenamefont {Bertet}, \citenamefont {Vion},\ and\ \citenamefont {Esteve}}]{palacios2008}%
  \BibitemOpen
  \bibfield  {author} {\bibinfo {author} {\bibfnamefont {A.}~\bibnamefont {Palacios-Laloy}}, \bibinfo {author} {\bibfnamefont {F.}~\bibnamefont {Nguyen}}, \bibinfo {author} {\bibfnamefont {F.}~\bibnamefont {Mallet}}, \bibinfo {author} {\bibfnamefont {P.}~\bibnamefont {Bertet}}, \bibinfo {author} {\bibfnamefont {D.}~\bibnamefont {Vion}},\ and\ \bibinfo {author} {\bibfnamefont {D.}~\bibnamefont {Esteve}},\ }\bibfield  {title} {\bibinfo {title} {Tunable {Resonators} for {Quantum} {Circuits}},\ }\href {https://doi.org/10.1007/s10909-008-9774-x} {\bibfield  {journal} {\bibinfo  {journal} {J. Low Temp. Phys.}\ }\textbf {\bibinfo {volume} {151}},\ \bibinfo {pages} {1034} (\bibinfo {year} {2008})}\BibitemShut {NoStop}%
\bibitem [{\citenamefont {Bourassa}\ \emph {et~al.}(2012)\citenamefont {Bourassa}, \citenamefont {Beaudoin}, \citenamefont {Gambetta},\ and\ \citenamefont {Blais}}]{bourassa2012}%
  \BibitemOpen
  \bibfield  {author} {\bibinfo {author} {\bibfnamefont {J.}~\bibnamefont {Bourassa}}, \bibinfo {author} {\bibfnamefont {F.}~\bibnamefont {Beaudoin}}, \bibinfo {author} {\bibfnamefont {J.~M.}\ \bibnamefont {Gambetta}},\ and\ \bibinfo {author} {\bibfnamefont {A.}~\bibnamefont {Blais}},\ }\bibfield  {title} {\bibinfo {title} {Josephson-junction-embedded transmission-line resonators: From kerr medium to in-line transmon},\ }\href {https://doi.org/10.1103/PhysRevA.86.013814} {\bibfield  {journal} {\bibinfo  {journal} {Phys. Rev. A}\ }\textbf {\bibinfo {volume} {86}},\ \bibinfo {pages} {13814} (\bibinfo {year} {2012})}\BibitemShut {NoStop}%
\bibitem [{\citenamefont {Mallet}\ \emph {et~al.}(2009)\citenamefont {Mallet}, \citenamefont {Ong}, \citenamefont {Palacios-Laloy}, \citenamefont {Nguyen}, \citenamefont {Bertet}, \citenamefont {Vion},\ and\ \citenamefont {Esteve}}]{mallet2009}%
  \BibitemOpen
  \bibfield  {author} {\bibinfo {author} {\bibfnamefont {F.}~\bibnamefont {Mallet}}, \bibinfo {author} {\bibfnamefont {F.~R.}\ \bibnamefont {Ong}}, \bibinfo {author} {\bibfnamefont {A.}~\bibnamefont {Palacios-Laloy}}, \bibinfo {author} {\bibfnamefont {F.}~\bibnamefont {Nguyen}}, \bibinfo {author} {\bibfnamefont {P.}~\bibnamefont {Bertet}}, \bibinfo {author} {\bibfnamefont {D.}~\bibnamefont {Vion}},\ and\ \bibinfo {author} {\bibfnamefont {D.}~\bibnamefont {Esteve}},\ }\bibfield  {title} {\bibinfo {title} {Single-shot qubit readout in circuit quantum electrodynamics},\ }\href {https://doi.org/10.1038/nphys1400} {\bibfield  {journal} {\bibinfo  {journal} {Nat. Phys.}\ }\textbf {\bibinfo {volume} {5}},\ \bibinfo {pages} {791} (\bibinfo {year} {2009})}\BibitemShut {NoStop}%
\bibitem [{\citenamefont {Leib}\ \emph {et~al.}(2012)\citenamefont {Leib}, \citenamefont {Deppe}, \citenamefont {Marx}, \citenamefont {Gross},\ and\ \citenamefont {Hartmann}}]{leib2012}%
  \BibitemOpen
  \bibfield  {author} {\bibinfo {author} {\bibfnamefont {M.}~\bibnamefont {Leib}}, \bibinfo {author} {\bibfnamefont {F.}~\bibnamefont {Deppe}}, \bibinfo {author} {\bibfnamefont {A.}~\bibnamefont {Marx}}, \bibinfo {author} {\bibfnamefont {R.}~\bibnamefont {Gross}},\ and\ \bibinfo {author} {\bibfnamefont {M.}~\bibnamefont {Hartmann}},\ }\bibfield  {title} {\bibinfo {title} {Networks of nonlinear superconducting transmission line resonators},\ }\href {https://doi.org/10.1088/1367-2630/14/7/075024} {\bibfield  {journal} {\bibinfo  {journal} {New J. Phys.}\ }\textbf {\bibinfo {volume} {14}},\ \bibinfo {pages} {75024} (\bibinfo {year} {2012})},\ \Eprint {https://arxiv.org/abs/1202.3240} {1202.3240} \BibitemShut {NoStop}%
\bibitem [{\citenamefont {Sandberg}\ \emph {et~al.}(2008)\citenamefont {Sandberg}, \citenamefont {Wilson}, \citenamefont {Persson}, \citenamefont {Bauch}, \citenamefont {Johansson}, \citenamefont {Shumeiko}, \citenamefont {Duty},\ and\ \citenamefont {Delsing}}]{sandberg2008}%
  \BibitemOpen
  \bibfield  {author} {\bibinfo {author} {\bibfnamefont {M.}~\bibnamefont {Sandberg}}, \bibinfo {author} {\bibfnamefont {C.~M.}\ \bibnamefont {Wilson}}, \bibinfo {author} {\bibfnamefont {F.}~\bibnamefont {Persson}}, \bibinfo {author} {\bibfnamefont {T.}~\bibnamefont {Bauch}}, \bibinfo {author} {\bibfnamefont {G.}~\bibnamefont {Johansson}}, \bibinfo {author} {\bibfnamefont {V.}~\bibnamefont {Shumeiko}}, \bibinfo {author} {\bibfnamefont {T.}~\bibnamefont {Duty}},\ and\ \bibinfo {author} {\bibfnamefont {P.}~\bibnamefont {Delsing}},\ }\bibfield  {title} {\bibinfo {title} {Tuning the field in a microwave resonator faster than the photon lifetime},\ }\href {https://doi.org/10.1063/1.2929367} {\bibfield  {journal} {\bibinfo  {journal} {Appl. Phys. Lett.}\ }\textbf {\bibinfo {volume} {92}},\ \bibinfo {pages} {203501} (\bibinfo {year} {2008})}\BibitemShut {NoStop}%
\bibitem [{\citenamefont {Wang}\ \emph {et~al.}(2013)\citenamefont {Wang}, \citenamefont {Zhong}, \citenamefont {He}, \citenamefont {Wang}, \citenamefont {Martinis}, \citenamefont {Cleland},\ and\ \citenamefont {Xie}}]{wang2013}%
  \BibitemOpen
  \bibfield  {author} {\bibinfo {author} {\bibfnamefont {Z.~L.}\ \bibnamefont {Wang}}, \bibinfo {author} {\bibfnamefont {Y.~P.}\ \bibnamefont {Zhong}}, \bibinfo {author} {\bibfnamefont {L.~J.}\ \bibnamefont {He}}, \bibinfo {author} {\bibfnamefont {H.}~\bibnamefont {Wang}}, \bibinfo {author} {\bibfnamefont {J.~M.}\ \bibnamefont {Martinis}}, \bibinfo {author} {\bibfnamefont {A.~N.}\ \bibnamefont {Cleland}},\ and\ \bibinfo {author} {\bibfnamefont {Q.~W.}\ \bibnamefont {Xie}},\ }\bibfield  {title} {\bibinfo {title} {Quantum state characterization of a fast tunable superconducting resonator},\ }\href {https://doi.org/10.1063/1.4802893} {\bibfield  {journal} {\bibinfo  {journal} {Appl. Phys. Lett.}\ }\textbf {\bibinfo {volume} {102}},\ \bibinfo {pages} {163503} (\bibinfo {year} {2013})}\BibitemShut {NoStop}%
\bibitem [{\citenamefont {Tuohino}\ \emph {et~al.}(2024)\citenamefont {Tuohino}, \citenamefont {Vadimov}, \citenamefont {Teixeira}, \citenamefont {Malmelin}, \citenamefont {Silveri},\ and\ \citenamefont {M{\"o}tt{\"o}nen}}]{Tuohino2024}%
  \BibitemOpen
  \bibfield  {author} {\bibinfo {author} {\bibfnamefont {S.}~\bibnamefont {Tuohino}}, \bibinfo {author} {\bibfnamefont {V.}~\bibnamefont {Vadimov}}, \bibinfo {author} {\bibfnamefont {W.}~\bibnamefont {Teixeira}}, \bibinfo {author} {\bibfnamefont {T.}~\bibnamefont {Malmelin}}, \bibinfo {author} {\bibfnamefont {M.}~\bibnamefont {Silveri}},\ and\ \bibinfo {author} {\bibfnamefont {M.}~\bibnamefont {M{\"o}tt{\"o}nen}},\ }\bibfield  {title} {\bibinfo {title} {Multimode physics of the unimon circuit},\ }\href {https://doi.org/10.1103/PhysRevResearch.6.033001} {\bibfield  {journal} {\bibinfo  {journal} {Phys. Rev. Research}\ }\textbf {\bibinfo {volume} {6}},\ \bibinfo {pages} {33001} (\bibinfo {year} {2024})}\BibitemShut {NoStop}%
\bibitem [{\citenamefont {Duda}\ \emph {et~al.}(2025)\citenamefont {Duda}, \citenamefont {Hyypp{\"a}}, \citenamefont {Mukkula}, \citenamefont {Vadimov},\ and\ \citenamefont {M{\"o}tt{\"o}nen}}]{duda2025}%
  \BibitemOpen
  \bibfield  {author} {\bibinfo {author} {\bibfnamefont {R.}~\bibnamefont {Duda}}, \bibinfo {author} {\bibfnamefont {E.}~\bibnamefont {Hyypp{\"a}}}, \bibinfo {author} {\bibfnamefont {O.}~\bibnamefont {Mukkula}}, \bibinfo {author} {\bibfnamefont {V.}~\bibnamefont {Vadimov}},\ and\ \bibinfo {author} {\bibfnamefont {M.}~\bibnamefont {M{\"o}tt{\"o}nen}},\ }\href {https://doi.org/10.48550/arXiv.2504.20205} {\bibinfo {title} {Parameter optimization for the unimon qubit}} (\bibinfo {year} {2025}),\ \Eprint {https://arxiv.org/abs/2504.20205} {arXiv:2504.20205 [quant-ph]} \BibitemShut {NoStop}%
\bibitem [{\citenamefont {Hyypp{\"a}}\ \emph {et~al.}(2022)\citenamefont {Hyypp{\"a}}, \citenamefont {Kundu}, \citenamefont {Chan}, \citenamefont {Gunyh{\'o}}, \citenamefont {Hotari}, \citenamefont {Janzso}, \citenamefont {Juliusson}, \citenamefont {Kiuru}, \citenamefont {Kotilahti}, \citenamefont {Landra}, \citenamefont {Liu}, \citenamefont {Marxer}, \citenamefont {M{\"a}kinen}, \citenamefont {Orgiazzi}, \citenamefont {Palma}, \citenamefont {Savytskyi}, \citenamefont {Tosto}, \citenamefont {Tuorila}, \citenamefont {Vadimov}, \citenamefont {Li}, \citenamefont {{Ockeloen-Korppi}}, \citenamefont {Heinsoo}, \citenamefont {Tan}, \citenamefont {Hassel},\ and\ \citenamefont {M{\"o}tt{\"o}nen}}]{hyyppa2022}%
  \BibitemOpen
  \bibfield  {author} {\bibinfo {author} {\bibfnamefont {E.}~\bibnamefont {Hyypp{\"a}}}, \bibinfo {author} {\bibfnamefont {S.}~\bibnamefont {Kundu}}, \bibinfo {author} {\bibfnamefont {C.~F.}\ \bibnamefont {Chan}}, \bibinfo {author} {\bibfnamefont {A.}~\bibnamefont {Gunyh{\'o}}}, \bibinfo {author} {\bibfnamefont {J.}~\bibnamefont {Hotari}}, \bibinfo {author} {\bibfnamefont {D.}~\bibnamefont {Janzso}}, \bibinfo {author} {\bibfnamefont {K.}~\bibnamefont {Juliusson}}, \bibinfo {author} {\bibfnamefont {O.}~\bibnamefont {Kiuru}}, \bibinfo {author} {\bibfnamefont {J.}~\bibnamefont {Kotilahti}}, \bibinfo {author} {\bibfnamefont {A.}~\bibnamefont {Landra}}, \bibinfo {author} {\bibfnamefont {W.}~\bibnamefont {Liu}}, \bibinfo {author} {\bibfnamefont {F.}~\bibnamefont {Marxer}}, \bibinfo {author} {\bibfnamefont {A.}~\bibnamefont {M{\"a}kinen}}, \bibinfo {author} {\bibfnamefont {J.-L.}\ \bibnamefont {Orgiazzi}}, \bibinfo {author} {\bibfnamefont {M.}~\bibnamefont {Palma}}, \bibinfo {author} {\bibfnamefont {M.}~\bibnamefont {Savytskyi}}, \bibinfo {author} {\bibfnamefont {F.}~\bibnamefont {Tosto}}, \bibinfo {author} {\bibfnamefont {J.}~\bibnamefont {Tuorila}}, \bibinfo {author} {\bibfnamefont {V.}~\bibnamefont {Vadimov}}, \bibinfo {author} {\bibfnamefont {T.}~\bibnamefont {Li}}, \bibinfo {author} {\bibfnamefont {C.}~\bibnamefont {{Ockeloen-Korppi}}}, \bibinfo {author} {\bibfnamefont {J.}~\bibnamefont {Heinsoo}}, \bibinfo {author} {\bibfnamefont {K.~Y.}\ \bibnamefont {Tan}}, \bibinfo {author} {\bibfnamefont {J.}~\bibnamefont {Hassel}},\ and\ \bibinfo {author} {\bibfnamefont {M.}~\bibnamefont {M{\"o}tt{\"o}nen}},\ }\bibfield  {title} {\bibinfo {title} {Unimon qubit},\ }\href {https://doi.org/10.1038/s41467-022-34614-w} {\bibfield  {journal} {\bibinfo  {journal} {Nat. Commun.}\ }\textbf {\bibinfo {volume} {13}},\ \bibinfo {pages} {6895} (\bibinfo {year} {2022})}\BibitemShut {NoStop}%
\bibitem [{\citenamefont {Bao}\ \emph {et~al.}(2021)\citenamefont {Bao}, \citenamefont {Wang}, \citenamefont {Wu}, \citenamefont {Li}, \citenamefont {Ma}, \citenamefont {Song}, \citenamefont {Zhang},\ and\ \citenamefont {Duan}}]{Bao2021}%
  \BibitemOpen
  \bibfield  {author} {\bibinfo {author} {\bibfnamefont {Z.}~\bibnamefont {Bao}}, \bibinfo {author} {\bibfnamefont {Z.}~\bibnamefont {Wang}}, \bibinfo {author} {\bibfnamefont {Y.}~\bibnamefont {Wu}}, \bibinfo {author} {\bibfnamefont {Y.}~\bibnamefont {Li}}, \bibinfo {author} {\bibfnamefont {C.}~\bibnamefont {Ma}}, \bibinfo {author} {\bibfnamefont {Y.}~\bibnamefont {Song}}, \bibinfo {author} {\bibfnamefont {H.}~\bibnamefont {Zhang}},\ and\ \bibinfo {author} {\bibfnamefont {L.}~\bibnamefont {Duan}},\ }\bibfield  {title} {\bibinfo {title} {On-demand storage and retrieval of microwave photons using a superconducting multiresonator quantum memory},\ }\href {https://doi.org/10.1103/PhysRevLett.127.010503} {\bibfield  {journal} {\bibinfo  {journal} {Phys. Rev. Lett.}\ }\textbf {\bibinfo {volume} {127}},\ \bibinfo {pages} {010503} (\bibinfo {year} {2021})}\BibitemShut {NoStop}%
\bibitem [{\citenamefont {Marinelli}\ \emph {et~al.}(2023)\citenamefont {Marinelli}, \citenamefont {Luo}, \citenamefont {Ren}, \citenamefont {Niedzielski}, \citenamefont {Kim}, \citenamefont {Das}, \citenamefont {Schwartz}, \citenamefont {Santiago},\ and\ \citenamefont {Siddiqi}}]{marinelli2023}%
  \BibitemOpen
  \bibfield  {author} {\bibinfo {author} {\bibfnamefont {B.}~\bibnamefont {Marinelli}}, \bibinfo {author} {\bibfnamefont {J.}~\bibnamefont {Luo}}, \bibinfo {author} {\bibfnamefont {H.}~\bibnamefont {Ren}}, \bibinfo {author} {\bibfnamefont {B.~M.}\ \bibnamefont {Niedzielski}}, \bibinfo {author} {\bibfnamefont {D.~K.}\ \bibnamefont {Kim}}, \bibinfo {author} {\bibfnamefont {R.}~\bibnamefont {Das}}, \bibinfo {author} {\bibfnamefont {M.}~\bibnamefont {Schwartz}}, \bibinfo {author} {\bibfnamefont {D.~I.}\ \bibnamefont {Santiago}},\ and\ \bibinfo {author} {\bibfnamefont {I.}~\bibnamefont {Siddiqi}},\ }\href {https://doi.org/10.48550/arXiv.2303.03507} {\bibinfo {title} {Dynamically {Reconfigurable} {Photon} {Exchange} in a {Superconducting} {Quantum} {Processor}}} (\bibinfo {year} {2023}),\ \bibinfo {note} {arXiv:2303.03507 [quant-ph]}\BibitemShut {NoStop}%
\bibitem [{\citenamefont {Barends}\ \emph {et~al.}(2013)\citenamefont {Barends}, \citenamefont {Kelly}, \citenamefont {Megrant}, \citenamefont {Sank}, \citenamefont {Jeffrey}, \citenamefont {Chen}, \citenamefont {Yin}, \citenamefont {Chiaro}, \citenamefont {Mutus}, \citenamefont {Neill}, \citenamefont {O'Malley}, \citenamefont {Roushan}, \citenamefont {Wenner}, \citenamefont {White}, \citenamefont {Cleland},\ and\ \citenamefont {Martinis}}]{Barends2013}%
  \BibitemOpen
  \bibfield  {author} {\bibinfo {author} {\bibfnamefont {R.}~\bibnamefont {Barends}}, \bibinfo {author} {\bibfnamefont {J.}~\bibnamefont {Kelly}}, \bibinfo {author} {\bibfnamefont {A.}~\bibnamefont {Megrant}}, \bibinfo {author} {\bibfnamefont {D.}~\bibnamefont {Sank}}, \bibinfo {author} {\bibfnamefont {E.}~\bibnamefont {Jeffrey}}, \bibinfo {author} {\bibfnamefont {Y.}~\bibnamefont {Chen}}, \bibinfo {author} {\bibfnamefont {Y.}~\bibnamefont {Yin}}, \bibinfo {author} {\bibfnamefont {B.}~\bibnamefont {Chiaro}}, \bibinfo {author} {\bibfnamefont {J.}~\bibnamefont {Mutus}}, \bibinfo {author} {\bibfnamefont {C.}~\bibnamefont {Neill}}, \bibinfo {author} {\bibfnamefont {P.}~\bibnamefont {O'Malley}}, \bibinfo {author} {\bibfnamefont {P.}~\bibnamefont {Roushan}}, \bibinfo {author} {\bibfnamefont {J.}~\bibnamefont {Wenner}}, \bibinfo {author} {\bibfnamefont {T.~C.}\ \bibnamefont {White}}, \bibinfo {author} {\bibfnamefont {A.~N.}\ \bibnamefont {Cleland}},\ and\ \bibinfo {author} {\bibfnamefont {J.~M.}\ \bibnamefont {Martinis}},\ }\bibfield  {title} {\bibinfo {title} {Coherent josephson qubit suitable for scalable quantum integrated circuits},\ }\href {https://doi.org/10.1103/PhysRevLett.111.080502} {\bibfield  {journal} {\bibinfo  {journal} {Phys. Rev. Lett.}\ }\textbf {\bibinfo {volume} {111}},\ \bibinfo {pages} {080502} (\bibinfo {year} {2013})}\BibitemShut {NoStop}%
\bibitem [{\citenamefont {Zheng}\ \emph {et~al.}(2022{\natexlab{a}})\citenamefont {Zheng}, \citenamefont {Xu}, \citenamefont {Ma}, \citenamefont {Li}, \citenamefont {Dong}, \citenamefont {Zhang}, \citenamefont {Wang}, \citenamefont {Sun}, \citenamefont {Wu}, \citenamefont {Zhao}, \citenamefont {Li}, \citenamefont {Lan}, \citenamefont {Tan},\ and\ \citenamefont {Yu}}]{zheng2022}%
  \BibitemOpen
  \bibfield  {author} {\bibinfo {author} {\bibfnamefont {W.}~\bibnamefont {Zheng}}, \bibinfo {author} {\bibfnamefont {J.}~\bibnamefont {Xu}}, \bibinfo {author} {\bibfnamefont {Z.}~\bibnamefont {Ma}}, \bibinfo {author} {\bibfnamefont {Y.}~\bibnamefont {Li}}, \bibinfo {author} {\bibfnamefont {Y.}~\bibnamefont {Dong}}, \bibinfo {author} {\bibfnamefont {Y.}~\bibnamefont {Zhang}}, \bibinfo {author} {\bibfnamefont {X.}~\bibnamefont {Wang}}, \bibinfo {author} {\bibfnamefont {G.}~\bibnamefont {Sun}}, \bibinfo {author} {\bibfnamefont {P.}~\bibnamefont {Wu}}, \bibinfo {author} {\bibfnamefont {J.}~\bibnamefont {Zhao}}, \bibinfo {author} {\bibfnamefont {S.}~\bibnamefont {Li}}, \bibinfo {author} {\bibfnamefont {D.}~\bibnamefont {Lan}}, \bibinfo {author} {\bibfnamefont {X.}~\bibnamefont {Tan}},\ and\ \bibinfo {author} {\bibfnamefont {Y.}~\bibnamefont {Yu}},\ }\bibfield  {title} {\bibinfo {title} {Measuring {Quantum} {Geometric} {Tensor} of {Non}-{Abelian} {System} in {Superconducting} {Circuits}},\ }\href {https://doi.org/10.1088/0256-307X/39/10/100202} {\bibfield  {journal} {\bibinfo  {journal} {Chin. Phys. Lett.}\ }\textbf {\bibinfo {volume} {39}},\ \bibinfo {pages} {100202} (\bibinfo {year} {2022}{\natexlab{a}})}\BibitemShut {NoStop}%
\bibitem [{\citenamefont {Goto}(2022)}]{Goto2022}%
  \BibitemOpen
  \bibfield  {author} {\bibinfo {author} {\bibfnamefont {H.}~\bibnamefont {Goto}},\ }\bibfield  {title} {\bibinfo {title} {Double-transmon coupler: Fast two-qubit gate with no residual coupling for highly detuned superconducting qubits},\ }\href {https://doi.org/10.1103/PhysRevApplied.18.034038} {\bibfield  {journal} {\bibinfo  {journal} {Phys. Rev. Appl.}\ }\textbf {\bibinfo {volume} {18}},\ \bibinfo {pages} {034038} (\bibinfo {year} {2022})}\BibitemShut {NoStop}%
\bibitem [{\citenamefont {Campbell}\ \emph {et~al.}(2023)\citenamefont {Campbell}, \citenamefont {Kamal}, \citenamefont {Ranzani}, \citenamefont {Senatore},\ and\ \citenamefont {LaHaye}}]{Campbell2023}%
  \BibitemOpen
  \bibfield  {author} {\bibinfo {author} {\bibfnamefont {D.~L.}\ \bibnamefont {Campbell}}, \bibinfo {author} {\bibfnamefont {A.}~\bibnamefont {Kamal}}, \bibinfo {author} {\bibfnamefont {L.}~\bibnamefont {Ranzani}}, \bibinfo {author} {\bibfnamefont {M.}~\bibnamefont {Senatore}},\ and\ \bibinfo {author} {\bibfnamefont {M.~D.}\ \bibnamefont {LaHaye}},\ }\bibfield  {title} {\bibinfo {title} {Modular tunable coupler for superconducting circuits},\ }\href {https://doi.org/10.1103/PhysRevApplied.19.064043} {\bibfield  {journal} {\bibinfo  {journal} {Phys. Rev. Appl.}\ }\textbf {\bibinfo {volume} {19}},\ \bibinfo {pages} {064043} (\bibinfo {year} {2023})}\BibitemShut {NoStop}%
\bibitem [{\citenamefont {Li}\ \emph {et~al.}(2025)\citenamefont {Li}, \citenamefont {Kubo}, \citenamefont {Ho}, \citenamefont {Yan}, \citenamefont {Inoue}, \citenamefont {Nakamura},\ and\ \citenamefont {Goto}}]{li2025}%
  \BibitemOpen
  \bibfield  {author} {\bibinfo {author} {\bibfnamefont {R.}~\bibnamefont {Li}}, \bibinfo {author} {\bibfnamefont {K.}~\bibnamefont {Kubo}}, \bibinfo {author} {\bibfnamefont {Y.}~\bibnamefont {Ho}}, \bibinfo {author} {\bibfnamefont {Z.}~\bibnamefont {Yan}}, \bibinfo {author} {\bibfnamefont {S.}~\bibnamefont {Inoue}}, \bibinfo {author} {\bibfnamefont {Y.}~\bibnamefont {Nakamura}},\ and\ \bibinfo {author} {\bibfnamefont {H.}~\bibnamefont {Goto}},\ }\href {https://doi.org/10.48550/arXiv.2503.03053} {\bibinfo {title} {Capacitively {Shunted} {Double}-{Transmon} {Coupler} {Realizing} {Bias}-{Free} {Idling} and {High}-{Fidelity} {CZ} {Gate}}} (\bibinfo {year} {2025}),\ \bibinfo {note} {arXiv:2503.03053 [quant-ph]}\BibitemShut {NoStop}%
\bibitem [{\citenamefont {Long}(2020)}]{longSuperconductingQuantumCircuits}%
  \BibitemOpen
  \bibfield  {author} {\bibinfo {author} {\bibfnamefont {J.}~\bibnamefont {Long}},\ }\emph {\bibinfo {title} {Superconducting Quantum Circuits for Quantum Information Processing}},\ \href@noop {} {Ph.D. thesis},\ \bibinfo  {school} {University of Colorado} (\bibinfo {year} {2020})\BibitemShut {NoStop}%
\bibitem [{\citenamefont {Chu}\ and\ \citenamefont {Yan}(2021)}]{chu2021}%
  \BibitemOpen
  \bibfield  {author} {\bibinfo {author} {\bibfnamefont {J.}~\bibnamefont {Chu}}\ and\ \bibinfo {author} {\bibfnamefont {F.}~\bibnamefont {Yan}},\ }\bibfield  {title} {\bibinfo {title} {Coupler-assisted controlled-phase gate with enhanced adiabaticity},\ }\href {https://doi.org/10.1103/PhysRevApplied.16.054020} {\bibfield  {journal} {\bibinfo  {journal} {Phys. Rev. Appl.}\ }\textbf {\bibinfo {volume} {16}},\ \bibinfo {pages} {054020} (\bibinfo {year} {2021})}\BibitemShut {NoStop}%
\bibitem [{\citenamefont {Kandala}\ \emph {et~al.}(2021)\citenamefont {Kandala}, \citenamefont {Wei}, \citenamefont {Srinivasan}, \citenamefont {Magesan}, \citenamefont {Carnevale}, \citenamefont {Keefe}, \citenamefont {Klaus}, \citenamefont {Dial},\ and\ \citenamefont {McKay}}]{Kandala2021}%
  \BibitemOpen
  \bibfield  {author} {\bibinfo {author} {\bibfnamefont {A.}~\bibnamefont {Kandala}}, \bibinfo {author} {\bibfnamefont {K.~X.}\ \bibnamefont {Wei}}, \bibinfo {author} {\bibfnamefont {S.}~\bibnamefont {Srinivasan}}, \bibinfo {author} {\bibfnamefont {E.}~\bibnamefont {Magesan}}, \bibinfo {author} {\bibfnamefont {S.}~\bibnamefont {Carnevale}}, \bibinfo {author} {\bibfnamefont {G.~A.}\ \bibnamefont {Keefe}}, \bibinfo {author} {\bibfnamefont {D.}~\bibnamefont {Klaus}}, \bibinfo {author} {\bibfnamefont {O.}~\bibnamefont {Dial}},\ and\ \bibinfo {author} {\bibfnamefont {D.~C.}\ \bibnamefont {McKay}},\ }\bibfield  {title} {\bibinfo {title} {Demonstration of a high-fidelity cnot gate for fixed-frequency transmons with engineered $zz$ suppression},\ }\href {https://doi.org/10.1103/PhysRevLett.127.130501} {\bibfield  {journal} {\bibinfo  {journal} {Phys. Rev. Lett.}\ }\textbf {\bibinfo {volume} {127}},\ \bibinfo {pages} {130501} (\bibinfo {year} {2021})}\BibitemShut {NoStop}%
\bibitem [{\citenamefont {Cross}\ and\ \citenamefont {Gambetta}(2015)}]{Cross2015}%
  \BibitemOpen
  \bibfield  {author} {\bibinfo {author} {\bibfnamefont {A.~W.}\ \bibnamefont {Cross}}\ and\ \bibinfo {author} {\bibfnamefont {J.~M.}\ \bibnamefont {Gambetta}},\ }\bibfield  {title} {\bibinfo {title} {Optimized pulse shapes for a resonator-induced phase gate},\ }\href {https://doi.org/10.1103/PhysRevA.91.032325} {\bibfield  {journal} {\bibinfo  {journal} {Phys. Rev. A}\ }\textbf {\bibinfo {volume} {91}},\ \bibinfo {pages} {032325} (\bibinfo {year} {2015})}\BibitemShut {NoStop}%
\bibitem [{\citenamefont {Paik}\ \emph {et~al.}(2016)\citenamefont {Paik}, \citenamefont {Mezzacapo}, \citenamefont {Sandberg}, \citenamefont {McClure}, \citenamefont {Abdo}, \citenamefont {C\'orcoles}, \citenamefont {Dial}, \citenamefont {Bogorin}, \citenamefont {Plourde}, \citenamefont {Steffen}, \citenamefont {Cross}, \citenamefont {Gambetta},\ and\ \citenamefont {Chow}}]{Paik2016}%
  \BibitemOpen
  \bibfield  {author} {\bibinfo {author} {\bibfnamefont {H.}~\bibnamefont {Paik}}, \bibinfo {author} {\bibfnamefont {A.}~\bibnamefont {Mezzacapo}}, \bibinfo {author} {\bibfnamefont {M.}~\bibnamefont {Sandberg}}, \bibinfo {author} {\bibfnamefont {D.~T.}\ \bibnamefont {McClure}}, \bibinfo {author} {\bibfnamefont {B.}~\bibnamefont {Abdo}}, \bibinfo {author} {\bibfnamefont {A.~D.}\ \bibnamefont {C\'orcoles}}, \bibinfo {author} {\bibfnamefont {O.}~\bibnamefont {Dial}}, \bibinfo {author} {\bibfnamefont {D.~F.}\ \bibnamefont {Bogorin}}, \bibinfo {author} {\bibfnamefont {B.~L.~T.}\ \bibnamefont {Plourde}}, \bibinfo {author} {\bibfnamefont {M.}~\bibnamefont {Steffen}}, \bibinfo {author} {\bibfnamefont {A.~W.}\ \bibnamefont {Cross}}, \bibinfo {author} {\bibfnamefont {J.~M.}\ \bibnamefont {Gambetta}},\ and\ \bibinfo {author} {\bibfnamefont {J.~M.}\ \bibnamefont {Chow}},\ }\bibfield  {title} {\bibinfo {title} {Experimental demonstration of a resonator-induced phase gate in a multiqubit circuit-qed system},\ }\href {https://doi.org/10.1103/PhysRevLett.117.250502} {\bibfield  {journal} {\bibinfo  {journal} {Phys. Rev. Lett.}\ }\textbf {\bibinfo {volume} {117}},\ \bibinfo {pages} {250502} (\bibinfo {year} {2016})}\BibitemShut {NoStop}%
\bibitem [{\citenamefont {Puri}\ and\ \citenamefont {Blais}(2016)}]{Puri2016}%
  \BibitemOpen
  \bibfield  {author} {\bibinfo {author} {\bibfnamefont {S.}~\bibnamefont {Puri}}\ and\ \bibinfo {author} {\bibfnamefont {A.}~\bibnamefont {Blais}},\ }\bibfield  {title} {\bibinfo {title} {High-fidelity resonator-induced phase gate with single-mode squeezing},\ }\href {https://doi.org/10.1103/PhysRevLett.116.180501} {\bibfield  {journal} {\bibinfo  {journal} {Phys. Rev. Lett.}\ }\textbf {\bibinfo {volume} {116}},\ \bibinfo {pages} {180501} (\bibinfo {year} {2016})}\BibitemShut {NoStop}%
\bibitem [{\citenamefont {Martinis}\ and\ \citenamefont {Geller}(2014)}]{Martinis2014}%
  \BibitemOpen
  \bibfield  {author} {\bibinfo {author} {\bibfnamefont {J.~M.}\ \bibnamefont {Martinis}}\ and\ \bibinfo {author} {\bibfnamefont {M.~R.}\ \bibnamefont {Geller}},\ }\bibfield  {title} {\bibinfo {title} {Fast adiabatic qubit gates using only ${\ensuremath{\sigma}}_{z}$ control},\ }\href {https://doi.org/10.1103/PhysRevA.90.022307} {\bibfield  {journal} {\bibinfo  {journal} {Phys. Rev. A}\ }\textbf {\bibinfo {volume} {90}},\ \bibinfo {pages} {022307} (\bibinfo {year} {2014})}\BibitemShut {NoStop}%
\bibitem [{\citenamefont {Zheng}\ \emph {et~al.}(2022{\natexlab{b}})\citenamefont {Zheng}, \citenamefont {Xu}, \citenamefont {Wang}, \citenamefont {Dong}, \citenamefont {Lan}, \citenamefont {Tan},\ and\ \citenamefont {Yu}}]{zheng2022accelerated}%
  \BibitemOpen
  \bibfield  {author} {\bibinfo {author} {\bibfnamefont {W.}~\bibnamefont {Zheng}}, \bibinfo {author} {\bibfnamefont {J.}~\bibnamefont {Xu}}, \bibinfo {author} {\bibfnamefont {Z.}~\bibnamefont {Wang}}, \bibinfo {author} {\bibfnamefont {Y.}~\bibnamefont {Dong}}, \bibinfo {author} {\bibfnamefont {D.}~\bibnamefont {Lan}}, \bibinfo {author} {\bibfnamefont {X.}~\bibnamefont {Tan}},\ and\ \bibinfo {author} {\bibfnamefont {Y.}~\bibnamefont {Yu}},\ }\bibfield  {title} {\bibinfo {title} {Accelerated quantum adiabatic transfer in superconducting qubits},\ }\href {https://doi.org/10.1103/PhysRevApplied.18.044014} {\bibfield  {journal} {\bibinfo  {journal} {Phys. Rev. Appl.}\ }\textbf {\bibinfo {volume} {18}},\ \bibinfo {pages} {044014} (\bibinfo {year} {2022}{\natexlab{b}})}\BibitemShut {NoStop}%
\bibitem [{\citenamefont {Xu}\ \emph {et~al.}(2024)\citenamefont {Xu}, \citenamefont {Zhang}, \citenamefont {Zheng}, \citenamefont {Cai}, \citenamefont {Zhou}, \citenamefont {Li}, \citenamefont {Liao}, \citenamefont {Zhang}, \citenamefont {Li}, \citenamefont {Lan}, \citenamefont {Tan},\ and\ \citenamefont {Yu}}]{xu2024}%
  \BibitemOpen
  \bibfield  {author} {\bibinfo {author} {\bibfnamefont {J.}~\bibnamefont {Xu}}, \bibinfo {author} {\bibfnamefont {Y.}~\bibnamefont {Zhang}}, \bibinfo {author} {\bibfnamefont {W.}~\bibnamefont {Zheng}}, \bibinfo {author} {\bibfnamefont {H.}~\bibnamefont {Cai}}, \bibinfo {author} {\bibfnamefont {H.}~\bibnamefont {Zhou}}, \bibinfo {author} {\bibfnamefont {X.}~\bibnamefont {Li}}, \bibinfo {author} {\bibfnamefont {X.}~\bibnamefont {Liao}}, \bibinfo {author} {\bibfnamefont {Y.}~\bibnamefont {Zhang}}, \bibinfo {author} {\bibfnamefont {S.}~\bibnamefont {Li}}, \bibinfo {author} {\bibfnamefont {D.}~\bibnamefont {Lan}}, \bibinfo {author} {\bibfnamefont {X.}~\bibnamefont {Tan}},\ and\ \bibinfo {author} {\bibfnamefont {Y.}~\bibnamefont {Yu}},\ }\bibfield  {title} {\bibinfo {title} {Balancing the {Quantum} {Speed} {Limit} and {Instantaneous} {Energy} {Cost} in {Adiabatic} {Quantum} {Evolution}},\ }\href {https://doi.org/10.1088/0256-307X/41/4/040202} {\bibfield  {journal} {\bibinfo  {journal} {Chin. Phys. Lett.}\ }\textbf {\bibinfo {volume} {41}},\ \bibinfo {pages} {040202} (\bibinfo {year} {2024})}\BibitemShut {NoStop}%
\end{thebibliography}
\providecommand{\noopsort}[1]{}\providecommand{\singleletter}[1]{#1}%

\end{document}